\documentclass[preprint,showpacs,preprintnumbers,amsmath,amssymb]{revtex4-2}
\usepackage{bm}% bold math
\usepackage{amsmath}
\usepackage{amssymb}
\usepackage{amsfonts}
\usepackage{graphicx}
\begin{document}

\title{ \textbf{
    Searching for entanglement in final polarization states of the
    neutron-proton scattering
} }

\author{H.~Wita{\l}a}
\affiliation{M. Smoluchowski Institute of Physics, 
Faculty of Physics, Astronomy and Applied Computer Science,
Jagiellonian University, PL-30348 Krak\'ow, Poland}

\author{J.~Golak}
\affiliation{M. Smoluchowski Institute of Physics, 
Faculty of Physics, Astronomy and Applied Computer Science,
Jagiellonian University, PL-30348 Krak\'ow, Poland}

\author{R.~Skibi\'nski}
\affiliation{M. Smoluchowski Institute of Physics, 
Faculty of Physics, Astronomy and Applied Computer Science,
Jagiellonian University, PL-30348 Krak\'ow, Poland}

\date{\today}

\begin{abstract}
  We investigate  polarization states of the outgoing neutron-proton ($np$) pair
   in elastic polarized 
  neutron and proton  scattering, aiming 
  to find  unambiguous
  evidence for entanglement of their spin states.
  To obtain complete information about these
  states, we calculate, using the high precision nucleon-nucleon
  potential AV18, 
  the final polarizations of the neutron and proton as well as
  their spin correlation coefficients, which unequivocally define the
  corresponding  spin density
  matrix. We compute all terms contributing to polarizations and spin
  correlations, e.g. not only induced polarizations and correlations 
  resulting from unpolarized $np$ scattering, but also contributions
  from single polarization and correlation transfers from individual polarized
  incoming nucleons, and, for the first time, allotment to both  quantities
  stemming  from a doubly spin polarized initial state.
  We find that for the most part the final spin states are statistical
  mixture of states. 
    The only pure states occur for highly polarized incoming neutrons and
    protons with maximal polarizations.
    By quantifying the degree of entanglement through entanglement power
    and concurrence, we observed that the entanglement of impure final
    states increases with energy.
    Among the pure spin states resulting from incoming states with maximal
    neutron and proton polarizations, we found, at $E_{lab}=100$~MeV, cases
    of strongly entangled Bell-type states with only a small admixture of
    entanglement-spoiling contributions.
\end{abstract}

%\pacs{21.45.-v, 21.45.Bc, 25.10.+s, 25.40.Cm}

\maketitle \setcounter{page}{1}

\section{Introduction}
\label{intro}

Studies of reactions with polarized incoming  particles and/or
measured polarizations of
outgoing reactant(s) provide more 
information about transition matrices than investigations of unpolarized
processes~\cite{ohlsen1972,simonius1973}.
 They also offer a unique tool to generate in a systematic way
the final spin states of the outgoing particles by varying 
 polarizations of the incoming reactants.
With a recent growing interest in nuclear entanglement 
and a quest for experimental  footprints of its remnants in final  
nucleon pairs, it seems natural to begin investigating this problem in
the simplest
and best-controlled nuclear process of the neutron-proton ($np$) or
proton-proton ($pp$) scattering. 
    Indeed, the problem of entanglement generation in $np$ scattering has
    been extensively investigated in numerous studies; see,
    e.g., ~\cite{beane2019,beane2021,bai2022,bai2023,liu2023,kirch2,bai2024}
    and references therein.
    By calculating the entanglement power, which measures the degree of
    entanglement induced by the S matrix acting on an unentangled state,
    one obtained valuable information on the entanglement generation
    caused by the NN interaction.
    In the present study, we approach this problem by using the transition
    operator T instead of the S matrix. It directly provides the outgoing
    neutron and proton polarizations and their spin correlations, which
    unequivocally determine the spin density matrix of the outgoing
    $np$ pair.

The nucleon-nucleon ($NN$) elastic scattering reaction has been studied
 extensively in the past decades both experimentally and
 theoretically, leading to an abundant experimental database
 ~\cite{nijmpwa,arndt}
 as well as to the present high precision 
$NN$ potentials, (semi)phe\-no\-me\-no\-lo\-gical ~\cite{mach_adv,nijm,AV18,cdb}
or chiral ones ~\cite{mach,epel2015}, which describe the
collected data with unprecedented precision of $\chi^2/\textrm{datum} \approx 1$
~\cite{mach,reinert2018}. 
Experimentally, the reaction has been investigated at numerous  energies 
with different initial polarizations, starting from the simplest case
of unpolarized incoming nucleons and measurement of the unpolarized
cross section only. Even in this simplest case the outgoing nucleons turn out
to be polarized and this induced polarization is due to the
presence of the spin-orbit term in the $NN$ potential.
A measurement of the cross section when one of the incoming nucleons
is polarized provides the so-called analyzing powers, which turn out
to be equal to the induced polarizations in unpolarized
$NN$ scattering. 
 In this case also the polarizations of the outgoing
 nucleons receive an additional contribution beside the induced
 polarization, and measurement of these final polarizations provides 
 the corresponding single spin transfer coefficient.
 Polarizations of different particles in the initial state lead to different
analyzing powers as well as to different single polarization transfer
coefficients. 

With the advancement of ion source technology and the development of
sophisticated polarization techniques  more complex spin experiments
have become available. Cross section measurements with both beam and target
polarized provide spin correlation coefficients,
 which yield even greater insight into the transition matrix elements. Under
 such initial conditions, with both beam and target particles polarized
 in the initial state,  
 the measurement of the polarization of the outgoing nucleon 
 offers the possibility to determine 
 the polarization transfer from a doubly spin-polarized
initial state, which is  different from single polarization
transfer coefficients when only one of the nucleons in the initial state
is polarized. To the best of our knowledge, measurements of
the polarization transfer from doubly spin-polarized
initial state have not been performed up to now. However, 
the longstanding experimental and theoretical investigations of $NN$
elastic scattering below the pion production threshold allow us to assume with
a high level of certainty, 
that using high precision $NN$ potentials and applying them
in the Lippmann-Schwinger equation would provide predictions for double
spin-polarized transfers and other not yet measured observables,
which very likely would  properly predict results of anticipated
future experiments.

The polarization state of each of the two outgoing
nucleons in $NN$ scattering can be set up in four different ways.
In addition to the fixed induced polarization there are three variable
contributions: two single polarization transfers from each of
the incoming nucleons and one polarization transfer from a doubly
spin-polarized initial state, each of which can be modified by changing
the incoming nucleons' polarizations. The aim of the present investigation is
to look for entanglement signatures within that abundant set of the final
polarization states of the $np$  elastic scattering 
at different scattering angles, generated by varying polarizations
of the incoming nucleons.

In Sec.~\ref{form} we briefly describe the $NN$ spin formalism and provide
 definitions of the required observables. We also discuss the
 properties of the entangled states of two nucleons which can be used
 to ascertain their presence.  
Predictions for different
final polarization observables obtained with different initial states 
follow in Sec.~\ref{results}. Finally, in Sec.~\ref{sumary} summary
and conclusions are given.

\section{Formalism}
\label{form}

We briefly outline, for the reader's convenience, 
the main points of the  formalism for spin description in reaction
$\vec b(\vec a,\vec c)\vec d$ with polarized incoming particles and measured
polarizations of outgoing products. This process is  
described by a transition amplitude $T^{\{m_{in}\}}_{\{m_{out}\}}$
 with corresponding sets of
 spin projections $\{m_{in}\}$ and $\{m_{out}\}$ for the incoming and
 outgoing reactants, respectively.  For details of spin formalism 
  we refer to ~\cite{ohlsen1972,simonius1973}.   
  When describing  polarization, we always use the right-handed
  Cartesian coordinate
  systems  defined according to the Madison Convention~\cite{Madison1971}.

In the following, we focus on $np$ elastic scattering so particles $a$ and $b$
as well as $c$ and $d$ become neutron and proton, respectively.
The question  arises: how to specify the spin density matrix of the outgoing
neutron-proton pair? This $4 \times 4$ Hermitian matrix can be uniquely
written in terms of 15 traceless Hermitian matrices:
$\{ \sigma_i^n \otimes I^p, I^n \otimes  \sigma_i^p, 
\sigma_i^n \otimes \sigma_j^p \}$ and a unit matrix $I=I^n \otimes I^p$, 
with $i,j=1, 2, 3$ and $\sigma_i$ being the Pauli matrices ~\cite{ohlsen1972}:
\begin{eqnarray}
  \rho &=& \frac {1} {4} [ I +
  \sum_{i=1}^3 \langle \sigma_i^n\rangle \sigma_i^n \otimes I^p +
  \sum_{i=1}^3 \langle \sigma_i^p\rangle  I^n \otimes \sigma_i^p +  
  \sum_{i,j=1}^3 \langle \sigma_i^n \sigma_j^p\rangle  \sigma_i^n
  \otimes \sigma_j^p ]  ~,
\label{eq1_a0_0}
\end{eqnarray}
with $\langle O\rangle  \equiv \text{Tr}(\rho O )$ being an average value
of an operator $O$
in spin state described by
the density matrix $\rho$. 
Thus the polarizations of the outgoing neutron  $\langle \sigma_i^n\rangle $
and proton $\langle \sigma_i^p\rangle $ together
with their spin correlation coefficients
$\langle \sigma_i^n \sigma_j^p\rangle $ specify the spin
density matrix of the outgoing $np$ pair. 

This is the reason why we are interested in polarizations of the outgoing
  particles $c$ and $d$ as well as 
  in  their spin correlations for different polarization states of the incoming
  particles $a$ and $b$.  
  To determine polarizations we define single and  double spin-polarization
  transfer
observables for a scattering with one or two polarized particles
in the initial state, respectively, and for measuring polarization
of only one outgoing particle: $\vec b(\vec a,\vec c)d$.
We also define the corresponding observables when spin correlations of the
two outgoing particles are required:  $\vec b(\vec a,\vec c)\vec d$.

In a standard scattering experiment the density matrix of the initial state
$\rho^{in}$ is given in terms of the
polarization tensors of particles $a$, $t_{k_aq_a}$, and $b$, $t_{k_bq_b}$, by
~\cite{ohlsen1972,simonius1973}:
\begin{eqnarray}
  \rho^{in} &=& \frac {1} {(2s_a+1)(2s_b+1)}
  \sum_{k_bq_b} t_{k_bq_b}\tau_{k_bq_b}^{\dagger}
  \sum_{k_aq_a} t_{k_aq_a}\tau_{k_aq_a}^{\dagger} ~,
\label{eq1_a0}
\end{eqnarray}
where $\tau_{kq}$ are spherical tensor operators with $k=0, 1,\dots, 2s$, 
 $q=-k,-k+1, \dots, k$, and $s$ is particle's spin. 

The density matrix $\rho^{in}$ together with the transition operator $T$   
determines the density matrix of the final state 
 $\rho^{out}$
\begin{eqnarray}
\rho^{out} &=& T \rho^{in} T^{\dagger} ~.
\label{eq2_a0}
\end{eqnarray}

The spin correlation tensors of the outgoing particle $c$ and $d$,
$t_{k_cq_c,k_dq_d}(t_{k_aq_a}^a,t_{k_bq_b}^b)$, depend on the polarizations of
particles $a$ and $b$ in the initial state and are given by traces over
spin projections of the outgoing particles:
\begin{eqnarray}
  t_{k_cq_c,k_dq_d}(t_{k_aq_a}^a,t_{k_bq_b}^b) &\equiv&
  \frac {\text{Tr}(\rho^{out} \tau_{k_cq_c} \tau_{k_dq_d})}
        {\text{Tr}(\rho^{out})} \cr
  &=& \frac{\sigma^0} {\sigma} \sum_{k_aq_a,k_bq_b} t_{k_aq_a} t_{k_bq_b}
	\frac {\text{Tr}(T\tau_{k_bq_b}^{\dagger}\tau_{k_aq_a}^{\dagger}T^{\dagger}
    \tau_{k_cq_c}\tau_{k_dq_d} )}
	{\text{Tr}(TT^{\dagger})} ~,
\label{eq3_a0}
\end{eqnarray}
with $\sigma^0$ ($\sigma$) being the unpolarized (polarized) cross section.

Defining tensors of the spin correlation transfer coefficients by:
\begin{eqnarray}
  t_{k_bq_b,k_aq_a}^{k_cq_c,k_dq_d}(\vec b(\vec a, \vec c)\vec d)  &\equiv&
	\frac {\text{Tr}(T\tau_{k_bq_b}^{\dagger}\tau_{k_aq_a}^{\dagger}T^{\dagger}
	\tau_{k_cq_c}\tau_{k_dq_d})} {\text{Tr}(TT^{\dagger})} ~,    
\label{eq4_a0}
\end{eqnarray}
one has:
\begin{eqnarray}
  \sigma t_{k_cq_c,k_dq_d}(t_{k_aq_a}^a,t_{k_bq_b}^b) &=&
  \sigma^0 \sum_{k_aq_a,k_bq_b} t_{k_aq_a} t_{k_bq_b}
 t_{k_bq_b,k_aq_a}^{k_cq_c,k_dq_d}(\vec b(\vec a, \vec c)\vec d) ~.
\label{eq5_a0}
\end{eqnarray}

If the terms with $k_a=0$ and/or $k_b=0$ in the sum of Eq.~(\ref{eq5_a0}) are 
separated, one gets:
\begin{eqnarray}
  \sigma t_{k_cq_c,k_dq_d}(t_{k_aq_a}^a,t_{k_bq_b}^b) &=&
  \sigma^0 \lbrack
t_{00,00}^{k_cq_c,k_dq_d}(b(a, \vec c)\vec d) 
+  \sum_{k_b\ne 0 q_b}  t_{k_bq_b}
 t_{k_bq_b,00}^{k_cq_c,k_dq_d}(\vec b(a, \vec c)\vec d) \cr
&+&  \sum_{k_a\ne 0 q_a}  t_{k_aq_a}
 t_{00,k_aq_a}^{k_cq_c,k_dq_d}(b(\vec a, \vec c)\vec d) \cr
&+&  \sum_{\substack{k_a\ne 0 q_a\\k_b\ne 0 q_b}} t_{k_aq_a} t_{k_bq_b}
t_{k_bq_b,k_aq_a}^{k_cq_c,k_dq_d}(\vec b(\vec a, \vec c)\vec d)
\rbrack  ~.
\label{eq6_a0}
\end{eqnarray}

It should be emphasized that in (\ref{eq6_a0}) the cross section $\sigma$
depends on the polarization of the initial state. 
The first term is the induced spin correlation of particles $c$ and $d$
in a reaction with
unpolarized initial state    
\begin{eqnarray}
  t_{k_cq_c,k_dq_d}^{(0)}(b(a, \vec c)\vec d) &\equiv&
  t_{00,00}^{k_cq_c,k_dq_d}(b(a, \vec c)\vec d)
  ~.
\label{eq7_a0}
\end{eqnarray}
The second, third, and fourth term are contributions to the spin correlation
of the outgoing particles $c$ and $d$ due to a single correlation transfer
from a polarized
 particle $a$, due to a single correlation transfer from a polarized  
 particle $b$, and due to a double correlation transfer from doubly
 spin-polarized
 state of $a$ and $b$, respectively.
 The corresponding tensors of the spin-correlation  transfer coefficients,
 single: $t_{k_aq_a}^{k_cq_c,k_dq_d}(b(\vec a, \vec c)\vec d)$ or
 $t_{k_bq_b}^{k_cq_c,k_dq_d}(\vec b(a, \vec c)\vec d)$, and  double:
 $t_{k_bq_b,k_aq_a}^{k_cq_c,k_dq_d}(\vec b(\vec a, \vec c)\vec d)$,  are:
\begin{eqnarray}
 t_{k_aq_a}^{k_cq_c,k_dq_d}(b(\vec a, \vec c)\vec d)     &\equiv&
 t_{00,k_aq_a}^{k_cq_c,k_dq_d}(\vec b(\vec a, \vec c)\vec d)    \cr
 t_{k_bq_b}^{k_cq_c,k_dq_d}(\vec b(a, \vec c)\vec d)    &\equiv&
 t_{k_bq_b,00}^{k_cq_c,k_dq_d}(\vec b(\vec a, \vec c)\vec d)   \cr
 t_{k_bq_b,k_aq_a}^{k_cq_c,k_dq_d}(\vec b(\vec a, \vec c)\vec d)     &\equiv&
 t_{k_bq_b,k_aq_a}^{k_cq_c,k_dq_d}(\vec b(\vec a, \vec c)\vec d)~
 (for~ k_a \neq 0 ~ and ~ k_b \neq 0) ~.
\label{eq8_a0}
\end{eqnarray} 
Thus the spin correlation  of the outgoing particles $c$ and $d$ is given by:
\begin{eqnarray}
  t_{k_cq_c,k_dq_d}(t_{k_aq_a}^a,t_{k_bq_b}^b) &=&
  \frac {\sigma^0} {\sigma} \lbrack
  t_{k_cq_c,k_dq_d}^{(0)}(b(a, \vec c)\vec d)
 + \sum_{k_a\ne 0 q_a}  t_{k_aq_a}
 t_{k_aq_a}^{k_cq_c,k_dq_d}(b(\vec a, \vec c)\vec d) \cr
&+&  \sum_{k_b\ne 0 q_b}  t_{k_bq_b}
 t_{k_bq_b}^{k_cq_c,k_dq_d}(\vec b(a, \vec c)\vec d) \cr
&+&  \sum_{\substack{k_a\ne 0 q_a\\k_b\ne 0 q_b}} t_{k_aq_a} t_{k_bq_b}
t_{k_bq_b,k_aq_a}^{k_cq_c,k_dq_d}(\vec b(\vec a, \vec c)\vec d)
\rbrack  ~.
\label{eq9_a0}
\end{eqnarray}

A direct calculation of the spin-correlation transfer tensors leads to the
double  spin-correlation transfer coefficients:
\begin{eqnarray}
  t_{k_bq_b,k_aq_a}^{k_cq_c,k_dq_d}(\vec b(\vec a, \vec c)\vec d)&=&
  \frac {1} { \sum_{m} |T_{m_cm_d}^{m_am_b}|^2 }
  \lbrack
  \sum_{\substack{m_c m_{c'}m_dm_{d'}\\m_a m_b m_{a'} m_{b'}}}
  T_{m_cm_d}^{m_am_b} (-1)^{q_b+q_a} \cr
 && \hat {s_b} (-1)^{s_b-m_{b'}}
  \langle s_b m_b s_b -m_{b'}|k_b-q_b \rangle  \cr
&&  \hat {s_a} (-1)^{s_a-m_{a'}}
  \langle s_a m_a s_a -m_{a'}|k_a-q_a \rangle   T_{m_{c'}m_d'}^{*m_{a'}m_{b'}} \cr
&& \hat {s_c} (-1)^{s_c-m_{c}}
  \langle s_c m_{c'} s_c -m_{c}|k_cq_c \rangle  \cr
&& \hat {s_d} (-1)^{s_d-m_{d}}
  \langle s_d m_{d'} s_d -m_{d}|k_dq_d \rangle   
  \rbrack  ~,
\label{eq10_a0}
\end{eqnarray}
and to the single spin-correlation transfer coefficients: 
\begin{eqnarray}
  t_{k_aq_a}^{k_cq_c,k_dq_d}(b(\vec a, \vec c)\vec d)&=&
  \frac {1} { \sum_{m} |T_{m_cm_d}^{m_am_b}|^2 }
  \lbrack
  \sum_{\substack{m_c m_{c'}m_dm_{d'}\\m_a m_b m_{a'} }}
  T_{m_cm_d}^{m_am_b} (-1)^{q_a} \cr
&&  \hat {s_a} (-1)^{s_a-m_{a'}}
  \langle s_a m_a s_a -m_{a'}|k_a-q_a \rangle   T_{m_{c'}m_{d'}}^{*m_{a'}m_{b}} \cr
&& \hat {s_c} (-1)^{s_c-m_{c}}
  \langle s_c m_{c'} s_c -m_{c}|k_cq_c \rangle  \cr
&& \hat {s_d} (-1)^{s_d-m_{d}}
  \langle s_d m_{d'} s_d -m_{d}|k_dq_d \rangle   
  \rbrack  ~,
\label{eq11_a0}
\end{eqnarray}
where $s_i$ and $m_i$ are particles' spins and their projections,  
$\langle s_1 m_1 s_2 m_2|k q \rangle $ is the Clebsch-Gordan coefficient, 
$\hat s \equiv \sqrt{2s+1}$, and $m \equiv {m_a, m_b, m_c, m_d}$.

It should be emphasized that the spin-correlation transfer tensors of
Eqs.~(\ref{eq10_a0}) and (\ref{eq11_a0})  are defined in a coordinate system
with the z-axis along the incoming particle's $a$ momentum.
When calculating cartesian spin-correlations of particle $c$ and $d$ 
 they have to be transformed in indices $k_c$ and $q_c$ to
 the coordinate system with the z'-axis along the outgoing particle's
 $c$ laboratory 
momentum and correspondingly for particle $d$,
by performing  rotations around the y-axis through the outgoing
particle's $c$ and $d$ laboratory angles $\theta_c^{lab}$ and $\theta_d^{lab}$,
respectively:
\begin{eqnarray}
  t_{k_bq_b,k_aq_a}^{k_cq_c,k_dq_d}(\vec b(\vec a, \vec c)\vec d) &=& \sum_{q'_cq'_d}
  D^{k_c}_{q'_c,q_c} (0 \theta_c^{lab} 0)
  D^{k_d}_{q'_d,q_d} (0 \theta_d^{lab} 0)
  t_{k_bq_b,k_aq_a}^{k_cq'_c,k_dq'_d,}(\vec b(\vec a, \vec c)\vec d) ~,
\label{eq12_a0}
\end{eqnarray}
where $D^k_{q'q}(\alpha \beta \gamma)$ is the Wigner D-matrix for 
 a rotation specified by the Euler angles $(\alpha \beta \gamma)$
~\cite{brinksatch}. 

 After performing these rotations, the cartesian spin-correlations
 of particles $c$
 and $d$ can be calculated 
 for spin $s=\frac {1} {2}$ particles using the following relations between the
  cartesian and spherical tensors  ~\cite{ohlsen1972}:
\begin{eqnarray}
  \langle \sigma_x\rangle  &=& -\frac {\sqrt{2}} {2} ( t_{1+1}-t_{1-1} ) \cr
  \langle \sigma_y\rangle  &=& i \frac {\sqrt{2}} {2} ( t_{1+1}+t_{1-1} ) \cr 
  \langle \sigma_z\rangle  &=&  t_{10} ~.
\label{eq1_aa0}
\end{eqnarray}

The definitions of the polarization tensors for the outgoing particle $c$, 
$t_{k_cq_c}(t_{k_aq_a}^a,t_{k_bq_b}^b)$, the tensors of polarization transfer
coefficients, 
$t_{k_bq_b,k_aq_a}^{k_cq_c}(\vec b(\vec a, \vec c)d) $, the induced polarization
of particle $c$, $t_{k_cq_c}^{(0)}(b(a, \vec c)d)$, the tensors of the single
polarization
transfer coefficients, $t_{k_aq_a}^{k_cq_c}(b(\vec a, \vec c)d)$ or
$t_{k_bq_b}^{k_cq_c}(\vec b(a, \vec c)d)$,
and the double polarization transfers,
$t_{k_bq_b,k_aq_a}^{k_cq_c}(\vec b(\vec a, \vec c)d)$,
 follow via substituting $k_d=0$ and $q_d=0$ in defining equations
(\ref{eq3_a0})-(\ref{eq9_a0}). One then obtains the expression for 
 $t_{k_bq_b,k_aq_a}^{k_cq_c}(\vec b(\vec a, \vec c)d)$ by omitting the last line
 in (\ref{eq10_a0})
and by setting $m_{d'}=m_d$ with the same applied for
$t_{k_aq_a}^{k_cq_c}(b(\vec a, \vec c)d)$ in (\ref{eq11_a0})
(see also ~\cite{doubl_poltrans}).

In Appendix \ref{a1} we list all possible cartesian single and double
 spin-polarization transfer
 coefficients to the outgoing neutron as well as single and double
 spin correlation transfer coefficients, and express
 them by the corresponding  spherical
 transfer tensors. Many of them (underlined)
vanish due to parity conservation ~\cite{ohlsen1972}.

 Since we look for signals of entanglement in
 the final polarization states in $np$
 elastic scattering, we consider the most general pure spin
state for the outgoing $np$ pair, which is given by:
\begin{eqnarray}
  |\psi_{np}^{spin} \rangle &=& \alpha_{+\frac {1} {2} +\frac {1} {2}}
  | {+\frac {1} {2}} {+\frac {1} {2}}\rangle 
  + \alpha_{+\frac {1} {2} -\frac {1} {2}} |{+\frac {1} {2}}
  {-\frac {1} {2}}\rangle  \cr 
  &+& \alpha_{-\frac {1} {2} +\frac {1} {2}} |{-\frac {1} {2}}
  {+\frac {1} {2}}\rangle  
  + \alpha_{-\frac {1} {2} -\frac {1} {2}} |{-\frac {1} {2}}
  {-\frac {1} {2}}\rangle  ~,
\label{eq13_a0}
\end{eqnarray}
with complex expansion coefficients $\alpha_{+\frac {1} {2} +\frac {1} {2}}$,
$\alpha_{+\frac {1} {2} -\frac {1} {2}}$, $\alpha_{-\frac {1} {2} +\frac {1} {2}}$, and
$\alpha_{-\frac {1} {2} -\frac {1} {2}}$ satisfying:
$|\alpha_{+\frac {1} {2} +\frac {1} {2}}|^2 + |\alpha_{+\frac {1} {2} -\frac {1} {2}}|^2 +
|\alpha_{-\frac {1} {2} +\frac {1} {2}}|^2 + |\alpha_{-\frac {1} {2} -\frac {1} {2}}|^2=1$, 
and the first and second spin projection referring to the neutron and proton,
respectively. 
The corresponding density matrix $\rho^{np}$ is given by:
\begin{eqnarray}
 \rho^{np} &=& |\psi_{np}^{spin} \rangle  \langle \psi_{np}^{spin} | ~,
\label{eq14_a0}
\end{eqnarray}
and has a unit trace: $\text{Tr}(\rho^{np}) = 1$. It also satisfies
the pure state
idempotent condition:
\begin{eqnarray}
  (\rho^{np})^2 &=& \rho^{np} ~.
\label{eq15_a0}
\end{eqnarray}

The neutron polarizations $\langle \sigma_i^n\rangle $, the proton
polarizations 
$\langle \sigma_i^p\rangle $, and the neutron-proton spin-correlations
 $\langle \sigma_i^n \sigma_j^p\rangle $, which according to
(\ref{eq1_a0_0}) completely determine the density matrix of
the spin state (\ref{eq13_a0}), are given in terms of the expansion coefficients
$\alpha$ by:
\begin{eqnarray}
 \langle \sigma_x^n\rangle  &\equiv& \langle \psi_{np}^{spin}|\sigma_x^n |
	\psi_{np}^{spin} \rangle  = \text{Tr} (\rho^{np}\sigma_x^n) 
  =
  2\mathfrak{Re}(\alpha_{+\frac {1} {2} +\frac {1} {2}}
  \alpha^*_{-\frac {1} {2} +\frac {1} {2}}
    + \alpha_{-\frac {1} {2} -\frac {1} {2}}
    \alpha^*_{+\frac {1} {2} -\frac {1} {2}})   ~, \cr
  \langle \sigma_y^n\rangle  &=& 
  -2\mathfrak{Im}(\alpha_{+\frac {1} {2} +\frac {1} {2}}
  \alpha^*_{-\frac {1} {2} +\frac {1} {2}}
    - \alpha_{-\frac {1} {2} -\frac {1} {2}}
    \alpha^*_{+\frac {1} {2} -\frac {1} {2}})   ~, \cr
  \langle \sigma_z^n\rangle    &=&
  | \alpha_{+\frac {1} {2} +\frac {1} {2}} |^2 + |\alpha_{+\frac {1} {2} -\frac {1} {2}}|^2
 -| \alpha_{-\frac {1} {2} +\frac {1} {2}} |^2 - |\alpha_{-\frac {1} {2} -\frac {1} {2}}|^2 
  ~. \cr
  \langle \sigma_x^p\rangle   &=&
  2\mathfrak{Re}(\alpha_{+\frac {1} {2} +\frac {1} {2}}
  \alpha^*_{+\frac {1} {2} -\frac {1} {2}}
    + \alpha_{-\frac {1} {2} -\frac {1} {2}}
    \alpha^*_{-\frac {1} {2} +\frac {1} {2}})   ~, \cr
  \langle \sigma_y^p\rangle  &=& 
  -2\mathfrak{Im}(\alpha_{+\frac {1} {2} +\frac {1} {2}}
  \alpha^*_{+\frac {1} {2} -\frac {1} {2}}
    - \alpha_{-\frac {1} {2} -\frac {1} {2}}
    \alpha^*_{-\frac {1} {2} +\frac {1} {2}})   ~, \cr
  \langle \sigma_z^p\rangle    &=&
  | \alpha_{+\frac {1} {2} +\frac {1} {2}} |^2 + |\alpha_{-\frac {1} {2} +\frac {1} {2}}|^2
 -| \alpha_{+\frac {1} {2} -\frac {1} {2}} |^2 - |\alpha_{-\frac {1} {2} -\frac {1} {2}}|^2 
 ~. \cr
  \langle \sigma_x^n \sigma_x^p\rangle  &=&
  2\mathfrak{Re}(\alpha_{+\frac {1} {2} +\frac {1} {2}}
  \alpha^*_{-\frac {1} {2} -\frac {1} {2}} 
    +\alpha_{+\frac {1} {2} -\frac {1} {2}}
    \alpha^*_{-\frac {1} {2} +\frac {1} {2}})   ~, \cr
  \langle \sigma_y^n \sigma_y^p\rangle  
  &=&
  -2\mathfrak{Re} (\alpha_{+\frac {1} {2} +\frac {1} {2}}
  \alpha^*_{-\frac {1} {2} -\frac {1} {2}} 
    - \alpha_{+\frac {1} {2} -\frac {1} {2}}
    \alpha^*_{-\frac {1} {2} +\frac {1} {2}})   ~, \cr
  \langle \sigma_z^n \sigma_z^p\rangle  &=&
  | \alpha_{+\frac {1} {2} +\frac {1} {2}} |^2 
  - | \alpha_{+\frac {1} {2} -\frac {1} {2}} |^2 
  - | \alpha_{-\frac {1} {2} +\frac {1} {2}} |^2
  + | \alpha_{-\frac {1} {2} -\frac {1} {2}} |^2  ~,  \cr
  \langle \sigma_x^n \sigma_y^p\rangle  &=&
  -2\mathfrak{Im} (\alpha_{+\frac {1} {2} +\frac {1} {2}}
  \alpha^*_{-\frac {1} {2} -\frac {1} {2}}
    + \alpha_{-\frac {1} {2} +\frac {1} {2}}
    \alpha^*_{+\frac {1} {2} -\frac {1} {2}})  ~, \cr
  \langle \sigma_y^n \sigma_x^p\rangle  &=&
  -2\mathfrak{Im} ( \alpha_{+\frac {1} {2} +\frac {1} {2}}
  \alpha^*_{-\frac {1} {2} -\frac {1} {2}}
    + \alpha_{+\frac {1} {2} -\frac {1} {2}}
    \alpha^*_{-\frac {1} {2} +\frac {1} {2}} )  ~, \cr
  \langle \sigma_x^n \sigma_z^p\rangle  &=&
  2\mathfrak{Re} (\alpha_{+\frac {1} {2} +\frac {1} {2}}
  \alpha^*_{-\frac {1} {2} +\frac {1} {2}}
    - \alpha_{-\frac {1} {2} -\frac {1} {2}}
    \alpha^*_{+\frac {1} {2} -\frac {1} {2}})  ~, \cr
  \langle \sigma_z^n \sigma_x^p\rangle  &=&
  2\mathfrak{Re} ( \alpha_{+\frac {1} {2} +\frac {1} {2}}
  \alpha^*_{+\frac {1} {2} -\frac {1} {2}}
    - \alpha_{-\frac {1} {2} -\frac {1} {2}}
    \alpha^*_{-\frac {1} {2} +\frac {1} {2}} )  ~, \cr
  \langle \sigma_y^n \sigma_z^p\rangle  &=&
  -2\mathfrak{Im} (\alpha_{+\frac {1} {2} +\frac {1} {2}}
  \alpha^*_{-\frac {1} {2} +\frac {1} {2}}
    + \alpha_{-\frac {1} {2} -\frac {1} {2}}
    \alpha^*_{+\frac {1} {2} -\frac {1} {2}})  ~, \cr
  \langle \sigma_z^n \sigma_y^p\rangle  &=&
  -2\mathfrak{Im} ( \alpha_{+\frac {1} {2} +\frac {1} {2}}
  \alpha^*_{+\frac {1} {2} -\frac {1} {2}}
    + \alpha_{-\frac {1} {2} -\frac {1} {2}}
    \alpha^*_{-\frac {1} {2} +\frac {1} {2}} )  
    \label{eq16_a0}    ~.
\end{eqnarray}

The pure spin state of Eq.~(\ref{eq13_a0}) becomes strongly entangled when
either coefficients
$\alpha_{+\frac {1} {2} -\frac {1} {2}}$ and $\alpha_{-\frac {1} {2} +\frac {1} {2}}$
or $\alpha_{+\frac {1} {2} +\frac {1} {2}}$ and $\alpha_{-\frac {1} {2} -\frac {1} {2}}$
vanish. By entanglement we mean that the results of the neutron and proton
spin projection measurements in such  states are strongly correlated.
Namely, the result  of the proton
spin projection measurement in the state with 
$\alpha_{+\frac {1} {2} -\frac {1} {2}}$ and $\alpha_{-\frac {1} {2} +\frac {1} {2}}$
equal zero must be identical to the outcome  of the  neutron spin
projection measurement. In contrast, they must have opposite signs when
$\alpha_{+\frac {1} {2} +\frac {1} {2}}$ and $\alpha_{-\frac {1} {2} -\frac {1} {2}}$
vanish.

A representative case for maximal entanglement is provided by  the  basis of
Bell states~\cite{bookqinf}:
\begin{eqnarray}
  | \psi_I\rangle  &=& \frac {1} {\sqrt{2}}
  (| +\frac {1} {2} +\frac {1} {2}\rangle  +
    |-\frac {1} {2} -\frac {1} {2} \rangle ) \cr
  | \psi_{II}\rangle  &=& \frac {1} {\sqrt{2}}
  ( | +\frac {1} {2} +\frac {1} {2}\rangle  -
    |-\frac {1} {2} -\frac {1} {2} \rangle  ) \cr
  | \psi_{III}\rangle  &=& \frac {1} {\sqrt{2}}
  ( | -\frac {1} {2} +\frac {1} {2}\rangle  +
    |+\frac {1} {2} -\frac {1} {2} \rangle  ) \cr
  | \psi_{IV}\rangle  &=& \frac {1} {\sqrt{2}} 
  ( | +\frac {1} {2} -\frac {1} {2}\rangle  -
    |-\frac {1} {2} +\frac {1} {2} \rangle  ) ~.
\label{eq17_a0}
\end{eqnarray}
Since in these states the probabilities of positive and negative spin
 projections
 either for the neutron or proton are equal, all 
 components  of the polarization $\langle \sigma_i^{n(p)}\rangle $ 
 vanish. 
 In Table~\ref{tab1} we show values of all polarizations and
spin correlations for  the Bell states of (\ref{eq17_a0}), which are
considered the most strongly entangled spin states.
We also show the corresponding values of these observables for two general
entangled states with
$\alpha_{+\frac {1} {2} -\frac {1} {2}}=\alpha_{-\frac {1} {2} +\frac {1} {2}}=0$ in the
second, and  with 
$\alpha_{+\frac {1} {2} +\frac {1} {2}}=\alpha_{-\frac {1} {2} -\frac {1} {2}}=0$ in the
third column of the Table~\ref{tab1}. At this point we notice
that any admixture of a component from one of these states into the other
will cause a decrease of entanglement of that state. It is also interesting
to observe
that for a general entangled state, the spin-correlation
$\langle \sigma_y^n \sigma_y^p\rangle $
can take any value between $-1$ and $+1$, which suggests that to ascertain 
 entanglement one should check values of all the observables determining the
 density matrix and not only the value of the spin-correlation
 in the $y$-direction.

Since in $np$ elastic scattering parity conservation causes 
the following polarizations and spin correlations to vanish independently: 
$ \langle \sigma_x^{n(p)}\rangle $, $ \langle \sigma_z^{n(p)}\rangle $,
$ \langle \sigma_x^n \sigma_y^p\rangle $,
$ \langle \sigma_y^n \sigma_x^p\rangle $,
$ \langle \sigma_y^n \sigma_z^p\rangle $, and
$ \langle \sigma_z^n \sigma_y^p\rangle $ ~\cite{ohlsen1972}, values of
polarizations 
$ \langle \sigma_y^{n(p)}\rangle $ and spin correlations
$ \langle \sigma_y^n \sigma_y^p\rangle $,
$ \langle \sigma_x^n \sigma_x^p\rangle $,
$ \langle \sigma_z^n \sigma_z^p\rangle $,
$ \langle \sigma_x^n \sigma_z^p\rangle $, and
$ \langle \sigma_z^n \sigma_x^p\rangle $ will be crucial
 for realization of entanglement. 
 Thus three features of the entangled (Bell) states jointly,
 namely, values of
 spin-correlations approaching $\pm 1$ or $0$,  
values of the neutron or proton $y$ component of the polarization 
nearing zero,
and purity of the state, which is equivalent to the corresponding
density matrix fulfilling the idempotent condition (\ref{eq15_a0}),  
can be considered as strong indicators of entanglement in the final $NN$
scattering spin state of the $np$ pair.

%\begin{center}
\begin{table}[ht!]
\centering  
\begin{tabular}{ |c|c|c|c|c|c|c| }
\hline
~~ & $\alpha_{+\frac {1} {2} -\frac {1} {2}}=\alpha_{-\frac {1} {2} +\frac {1} {2}}=0$
   & $\alpha_{+\frac {1} {2} +\frac {1} {2}}=\alpha_{-\frac {1} {2} -\frac {1} {2}}=0$ 
 & $ | \psi_I\rangle $ & $| \psi_{II}\rangle $ &  $ |\psi_{III}\rangle $ & $ |\psi_{IV}\rangle $ \\
\hline
$ \langle \sigma_y^{n(p)}\rangle $  & 0  & 0 & 0 & 0  &0 & 0 \\
\hline
$ \langle \sigma_x^{n(p)}\rangle $  & 0  & 0 & 0 & 0  &0 & 0 \\
\hline
$ \langle \sigma_z^{n(p)}\rangle $  & $1-2|\alpha_{-\frac {1} {2} -\frac {1} {2}} |^2$
& $1-2|\alpha_{-\frac {1} {2} +\frac {1} {2}} |^2$
& 0  & 0  &0 & 0 \\
\hline
$ \langle \sigma_y^n \sigma_y^p\rangle $  & $-2\mathfrak{Re} (\alpha_{+\frac {1} {2} +\frac {1} {2}}
  \alpha^*_{-\frac {1} {2} -\frac {1} {2}} )$
& $2\mathfrak{Re} ( \alpha_{+\frac {1} {2} -\frac {1} {2}}
    \alpha^*_{-\frac {1} {2} +\frac {1} {2}})$
    & -1   & +1  & +1  & -1 \\
\hline
$ \langle \sigma_x^n \sigma_x^p\rangle $  & $ 2\mathfrak{Re} (\alpha_{+\frac {1} {2} +\frac {1} {2}}
  \alpha^*_{-\frac {1} {2} -\frac {1} {2}} )$
& $2\mathfrak{Re} ( \alpha_{+\frac {1} {2} -\frac {1} {2}}
    \alpha^*_{-\frac {1} {2} +\frac {1} {2}})$
& +1  & -1  & +1 & -1 \\
\hline
$ \langle \sigma_z^n \sigma_z^p\rangle $  & +1  & -1 & +1  & +1  & -1 & -1 \\
\hline
$ \langle \sigma_x^n \sigma_y^p\rangle $  & $ -2\mathfrak{Im} (\alpha_{+\frac {1} {2} +\frac {1} {2}}
  \alpha^*_{-\frac {1} {2} -\frac {1} {2}} ) $
& $ -2\mathfrak{Im} ( \alpha_{-\frac {1} {2} +\frac {1} {2}}
    \alpha^*_{+\frac {1} {2} -\frac {1} {2}}) $
& 0 & 0  & 0 & 0  \\
\hline
$ \langle \sigma_y^n \sigma_x^p\rangle $  & $ -2\mathfrak{Im} (\alpha_{+\frac {1} {2} +\frac {1} {2}}
  \alpha^*_{-\frac {1} {2} -\frac {1} {2}} ) $
& $ -2\mathfrak{Im} ( \alpha_{+\frac {1} {2} -\frac {1} {2}}
    \alpha^*_{-\frac {1} {2} +\frac {1} {2}}) $
& 0 & 0  & 0 & 0  \\
\hline
$ \langle \sigma_x^n \sigma_z^p\rangle $  & 0 & 0 & 0 & 0  & 0 & 0  \\
\hline
$ \langle \sigma_z^n \sigma_x^p\rangle $  & 0 & 0 & 0 & 0  & 0 & 0  \\
\hline
$ \langle \sigma_y^n \sigma_z^p\rangle $  & 0 & 0 & 0 & 0  & 0 & 0  \\
\hline
$ \langle \sigma_z^n \sigma_y^p\rangle $  & 0 & 0 & 0 & 0  & 0 & 0  \\
\hline
\end{tabular}
\caption{Values of the neutron (proton) polarizations and
  spin correlations in  the general entangled state of (\ref{eq13_a0}) with
  $\alpha_{+\frac {1} {2} -\frac {1} {2}}=\alpha_{-\frac {1} {2} +\frac {1} {2}}=0$ or
  $\alpha_{+\frac {1} {2} +\frac {1} {2}}=\alpha_{-\frac {1} {2} -\frac {1} {2}}=0$   
  and for the 
  set of Bell states (\ref{eq17_a0}).}
\label{tab1}
\end{table}
%\end{center}

 It is obvious that the properties of the final spin states in $np$ scattering
 are, to some extent, determined by the incoming polarization state,
 which in nuclear physics experiments is typically taken in the form
 of (\ref{eq1_a0}), with the polarizations of the beam and target
 prepared entirely independently.
 Such a state is clearly not entangled and, for partially polarized
 particles, also impure, leading to final spin states that are
 predominantly statistical mixtures.
Despite not being pure, they can still be entangled ~\cite{vinc2024}.

In the following, we aim to quantify the degree of spin entanglement
of the outgoing $np$ pair using two measures proposed in the literature:
the entanglement power $\epsilon$, defined
in ~\cite{beane2019,bai2022,bai2024,kirch2,liu2023}, and
the concurrence $C$, introduced in \cite{liu2023,bai2024}.
These measures can quantify the degree of entanglement for
states that are neither maximally entangled nor simple product states.

For the reader’s convenience, we briefly recall their definitions.
At the same time, we emphasize that, since our interest lies directly
in polarization and spin-correlation observables in $np$ elastic
scattering, which determine the final spin density matrix, our basic
operator is the transition operator $T$, in contrast to previous
studies ~\cite{kirch2,liu2023,bai2024,bai2022,bai2023,beane2019,beane2021},
where the unitary $S$ matrix was used instead.

The spin state of a two-particle system, such as the $np$ pair in
NN scattering, is generally described by the density matrix $\rho_{np}$
with unit trace.
A quantity that measures the degree of entanglement of this state is
the entanglement power $\epsilon(\rho_{np})$, defined as:
\begin{eqnarray}
\epsilon(\rho_{np}) &=& 1 - Tr(\rho_n^2) ~,
\label{a_eq1}
\end{eqnarray}
where $\rho_{n(p)} \equiv Tr_{p(n)}(\rho_{np})$ is the reduced density
matrix obtained by tracing $\rho_{np}$ over the subsystem $p(n)$.

Another measure used to quantify strength of entanglement
is the concurrence $C(\rho_{np})$, defined through average value of
spin correlation in $y$-direction  ~\cite{liu2023,bai2024}:
\begin{eqnarray}
  C(\rho_{np}) &\equiv& \frac {1} {2}~ | < \hat{\sigma}_y^n
  \otimes \hat{\sigma}_y^p > |
  =  \frac {1} {2}~ | Tr(\rho_{np}~ \hat{\sigma}_y^n
  \otimes \hat{\sigma}_y^p) |    ~.
  \label{a_eq2}
\end{eqnarray}
We also applied another form of concurrence ${\bar C(}\rho_{np})$
proposed in ~\cite{bai2024}:
\begin{eqnarray}
  {\bar C}(\rho_{np}) &\equiv& \frac {1} {2}~
  \sqrt {1 - \langle \hat{\sigma}_x^n \rangle^2
    - \langle \hat{\sigma}_y^n \rangle^2
    - \langle \hat{\sigma}_y^n \rangle^2 }
  =   \frac {1} {2}~
  \sqrt {1  - \langle \hat{\sigma}_y^n \rangle^2 }   ~.
  \label{a_eq2a}
\end{eqnarray}

These measures reach their maximum value of $0.5$ for maximally
entangled Bell states. In addition, the entanglement power vanishes
for a product spin state.
If the state corresponds to one of the entangled states from columns
2 or 3 of Table~\ref{tab1}, the entanglement power becomes
$\epsilon = 2|\alpha_0|^2(1 - |\alpha_0|^2)$,
where $\alpha_0 = \alpha_{-\frac{1}{2} -\frac{1}{2}}$ (column 2)
or $\alpha_0 = \alpha_{-\frac{1}{2} +\frac{1}{2}}$ (column 3).
It vanishes for $|\alpha_0|^2 = 0$ or $1$ (in which case the state
becomes a pure product spin state), and reaches its maximum
value $\epsilon = 0.5$ at $\alpha_0 = \frac{1}{\sqrt{2}}$ (Bell states).
The same holds true for concurrence.

Since the density matrix $\rho_{np}$ depends on the c.m. angle $\Theta_{c.m.}$,
the quantities $\epsilon(\rho_{np})$, $C(\rho_{np})$, and ${\bar C}(\rho_{np})$
are also angle-dependent.
By averaging over the angle, one obtains the integrated
measures $\epsilon(\rho_{np})_{\Theta_{int}}$:
\begin{eqnarray}
\epsilon(\rho_{np})_{\Theta_{int}} &\equiv& \frac {1} {2}
\int_0^{\pi} \sin\Theta ~ d\Theta ~ \epsilon(\rho_{np}) ~,
\label{a_eq3}
\end{eqnarray}
and similarly for the angle-integrated
concurrences $C(\rho_{np})_{\Theta_{int}}$ and ${\bar C}(\rho_{np})_{\Theta_{int}}$.

\section{Results and discussion}
\label{results}

Neutron-proton scattering caused by their interaction 
via the $v_{np}$ potential is
described in terms of the transition operator $T$ satisfying the
Lippmann-Schwinger (LS)  equation~\cite{book}
\begin{eqnarray}
T  &=& v_{np}  + v_{np} G_0 T \, ,
\label{eq1a}
\end{eqnarray}
where $G_0$ is the free two-body resolvent.

To investigate effects induced on the final $np$ polarization states by
varying polarizations of the incoming neutron and proton, 
we solved the LS equation (\ref{eq1a}) using the AV18 neutron-proton
 potential ~\cite{AV18} 
in the momentum-space partial-wave basis,
taking into account all states with the 
 $np$ total angular momenta $j$ up to $j_{max}=5$, and choosing
 three incoming neutron energies
 $E_{lab}=10, 50$ and $100$~MeV.
 In standard scattering experiments the spin states of the incoming particles
  are prepared independently, reducing the possible form 
of the initial spin-states  
 to the tensor product of the $n$ and $p$ $\rho$-matrices:   
$\rho^{in}=\rho^{n} \otimes \rho^{p}$, as expressed in (\ref{eq1_a0}).
 At each energy and c.m. angle of the outgoing neutron $\Theta_{c.m.}$
 we computed the induced neutron-proton spin correlations (\ref{eq7_a0})
 and the induced neutron and proton polarizations,    
 together with the corresponding neutron-proton single and double
 spin-correlation transfers (\ref{eq8_a0}), as well as
 neutron and proton single and double polarization transfers.
 This enabled us to 
 specify the final $np$ polarization state by calculating
 all contributing terms to the final $np$ spin-correlations (\ref{eq9_a0}) and
 the final neutron and proton polarizations, for all allowed  values
 of the incoming neutron and proton polarizations, which we 
 restricted to be nonvanishing only in their $y$-components:
 $p_y^n,p_y^p \in [-1.0,1.0]$.
 Parity conservation implies that $x$ and $z$ 
  components of the outgoing  neutron and proton polarizations vanish and 
 the only nonvanishing is $y$ constituent.    
  The four contributions to the final neutron polarization $P_{y'}^{n}$: one from
  the induced polarization $P_{y'}^{(0)}$, one from the single neutron
  $K_{0,y}^{y'}$, one from the 
  single proton $K_{y,0}^{y'}$, and one from the  double
  spin-polarization $K_{y,y}^{y'}$ transfer, make:
\begin{eqnarray}
  \langle \sigma_{y'}^{n}\rangle  \equiv P_{y'}^{n} &=& \frac {P_{y'}^{(0)} +
    p_y^{n} K_{0,y}^{y',0}
      +  p_y^p K_{y,0}^{y',0}
      +  p_y^n p_y^p K_{y,y}^{y',0} } {1+p_y^n A_y^n
    +  p_y^p A_y^p
    +  p_y^n p_y^p C_{y,y}  } ~,
\label{eqqq_1}
\end{eqnarray}    
where $A_y^n$ and $A_y^p$ are the neutron and proton analyzing powers and
$C_{y,y}$ is the spin correlation coefficient. Similarly, the only
nonvanishing $np$ spin correlations are: $ \langle \sigma_{y'}^n
\sigma_{y'}^p\rangle $,
$ \langle \sigma_{x'}^n \sigma_{x'}^p\rangle $, $ \langle \sigma_{z'}^n
\sigma_{z'}^p\rangle $,
$ \langle \sigma_{x'}^n \sigma_{z'}^p\rangle $, $ \langle \sigma_{z'}^n
\sigma_{x'}^p\rangle $, which 
are given by:
\begin{eqnarray}
  \langle \sigma_{y'}^n \sigma_{y'}^p\rangle  &=& \frac
          { \langle \sigma_{y'}^n \sigma_{y'}^p\rangle ^{(0)}
    + p_y^n K_{0,y}^{y',y'}
      +  p_y^p K_{y,0}^{y',y'}
      +  p_y^n p_y^p K_{y,y}^{y',y'} } {1+p_y^n A_y^n
    +  p_y^p A_y^p
    +  p_y^n p_y^p C_{y,y}  } ~,    \label{eqqq_2}  \\
  \langle \sigma_{z'}^n \sigma_{z'}^p\rangle  &=& \frac
          { \langle \sigma_{z'}^n \sigma_{z'}^p\rangle ^{(0)}
    + p_y^n K_{0,y}^{z',z'}
      +  p_y^p K_{y,0}^{z',z'}
      +  p_y^n p_y^p K_{y,y}^{z',z'} } {1+p_y^n A_y^n
    +  p_y^p A_y^p
    +  p_y^n p_y^p C_{y,y}  } ~,\\   \label{eqqq_3} 
          \langle \sigma_{x'}^n \sigma_{x'}^p\rangle  &=& \frac
                  { \langle \sigma_{x'}^n \sigma_{x'}^p\rangle ^{(0)}
    + p_y^n K_{0,y}^{x',x'}
      +  p_y^p K_{y,0}^{x',x'}
      +  p_y^n p_y^p K_{y,y}^{x',x'} } {1+p_y^n A_y^n
    +  p_y^p A_y^p
    +  p_y^n p_y^p C_{y,y}  } ~,   \label{eqqq_4} \\
                  \langle \sigma_{x'}^n \sigma_{z'}^p\rangle  &=& \frac
                          { \langle \sigma_{x'}^n \sigma_{z'}^p\rangle ^{(0)}
    + p_y^n K_{0,y}^{x',z'}
      +  p_y^p K_{y,0}^{x',z'}
      +  p_y^n p_y^p K_{y,y}^{x',z'} } {1+p_y^n A_y^n
    +  p_y^p A_y^p
    +  p_y^n p_y^p C_{y,y}  } ~,   \label{eqqq_5} \\
                          \langle \sigma_{z'}^n \sigma_{x'}^p\rangle  &=& \frac
                         { \langle \sigma_{z'}^n \sigma_{x'}^p\rangle ^{(0)}
    + p_y^n K_{0,y}^{z',x'}
      +  p_y^p K_{y,0}^{z',x'}
      +  p_y^n p_y^p K_{y,y}^{z',x'} } {1+p_y^n A_y^n
    +  p_y^p A_y^p
    +  p_y^n p_y^p C_{y,y}  } ~.   \label{eqqq_6} 
\end{eqnarray}  

\begin{figure}
\includegraphics[scale=1.0]{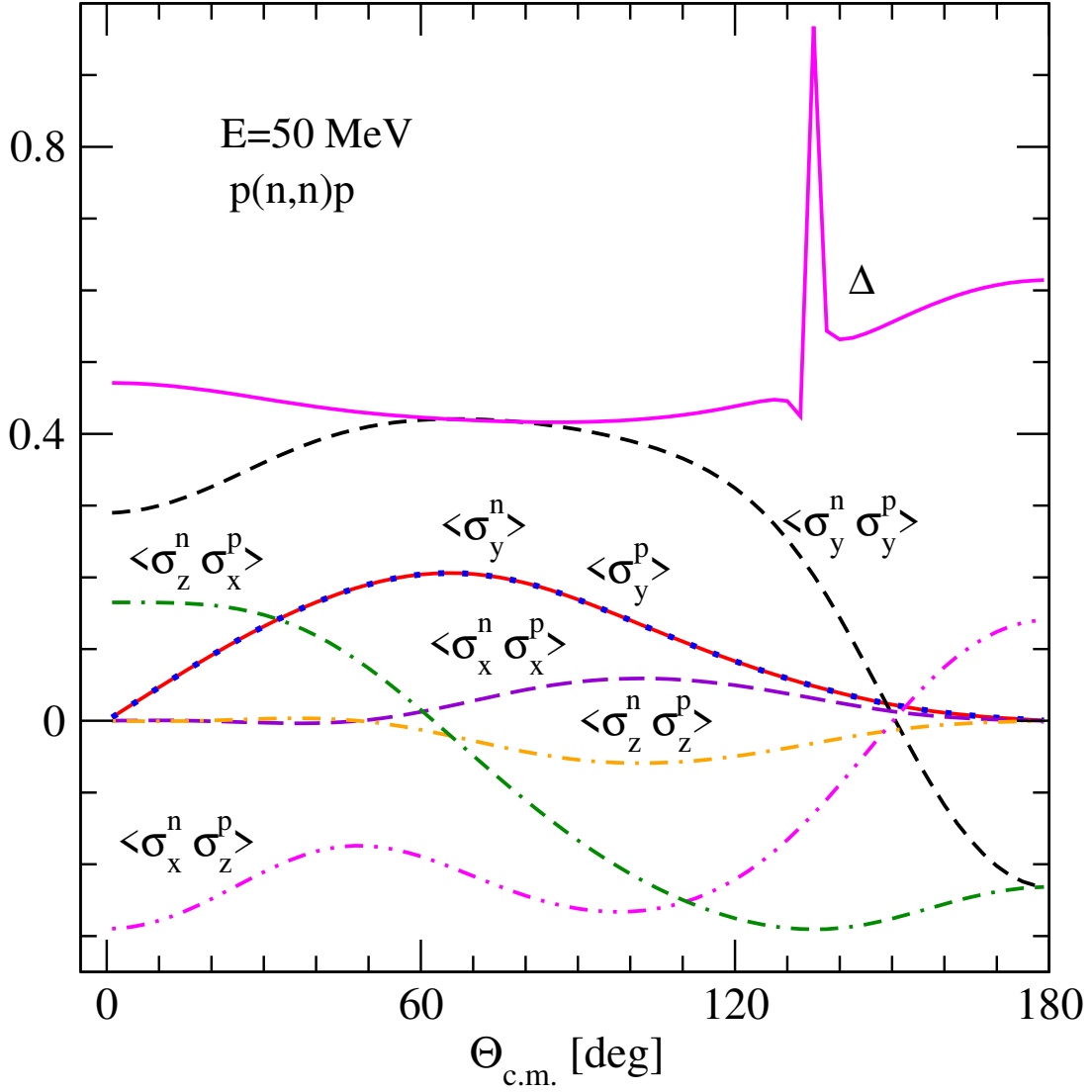}  
\caption{
(color online) 
  The induced outgoing neutron $\langle \sigma_y^n\rangle $ and proton
  $\langle \sigma_y^p\rangle $ polarizations
  together with all nonvanishing induced spin correlations
  $\langle \sigma_i^n \sigma_i^p\rangle $ 
  in  unpolarized $p(n,n)p$ scattering at  $E_{lab}^n=50$~MeV,
  as a function of the  c.m.
  scattering angle $\Theta_{c.m.}$, as predicted by the AV18 potential.
  The magenta solid line shows that the final polarization state
  at each angle is far
  from being pure and is described by a
  spin density matrix which does not fulfill the purity
  condition (\ref{eq15_a0}) 
  (see (\ref{eqqq_7}) for the definition of $\bigtriangleup$).}
\label{fig1}
\end{figure}

In Fig.~\ref{fig1} we show at $E_{lab}=50$~MeV and for each c.m. angle
 $\Theta_{c.m.}$ all nonvanishing induced neutron and proton polarizations
and  spin correlations, which arise in unpolarized $np$
scattering. The induced neutron and proton polarizations
$\langle \sigma_y\rangle $
as well as spin correlation $\langle \sigma_y^n \sigma_y^p\rangle $ reach
their maximal
values of about $0.2$ and $0.4$, respectively,  at $\Theta_{c.m.} \approx 70^o$.
The induced spin correlations $\langle \sigma_x^n \sigma_x^p\rangle $
and $\langle \sigma_z^n \sigma_z^p\rangle $
are much smaller and have the same magnitudes but opposite signs.

The initial state of
the  $np$ scattering with the only nonvanishing $y$-component of the neutron
and proton polarizations, is described by the initial spin density matrix
$\rho^{in}=\frac {1} {2} ( I^n + p_y^n \sigma_y^n) 
\otimes \frac {1} {2} ( I^p + p_y^p \sigma_y^p)$. Such a state is usually 
impure and does not fulfill  the idempotent condition
(\ref{eq15_a0}). 
To characterize quantitatively the fulfillment of this purity condition
 we introduce parameter $\bigtriangleup$, defined as the  relative
deviation between matrix elements of $\rho$ and $\rho^2$ averaged
over sixteen elements of the $4\times4$ $\rho$-matrix:
\begin{eqnarray}
  \bigtriangleup &=& \frac {1} {16} \sum_{i,j=1}^4 \frac{| \rho_{i,j} -
    (\rho^2)_{i,j} | } {| \rho_{i,j} |} ~,
\label{eqqq_7}
\end{eqnarray}
with the summation over nonzero elements of $\rho$. The direct calculation
 of $\bigtriangleup$ for $\rho^{in}$ gives:
\begin{eqnarray}
  \bigtriangleup^{in} &=& \frac {1} {16} [ 4-(1+(p_y^n)^2)(1+(p_y^p)^2)
    + 2p_y^p(1-(p_y^n)^2) + 2p_y^n(1-(p_y^p)^2) ] ~.
\label{eqqq_7a}
\end{eqnarray}
For unpolarized $np$ scattering ($p_y^n=p_y^p=0$) one gets
$\bigtriangleup = 0.1875$ and
the initial density matrix  evidently
does not satisfy the purity condition (\ref{eq15_a0}), being thus  a statistical
mixture of spin states. It  immediately raises the suspicion that
the spin state of the outgoing $np$ pair in unpolarized $np$ scattering
 would be also such a mixture of spin states.
Indeed,  in Fig.~\ref{fig1} the magenta solid line shows
$\bigtriangleup$ for the final induced polarization state, making it evident
that this state at no c.m. angle is pure.
%Therefore, 
%one should not probably expect to find any indication of entanglement in
%the final polarization states -- not only for unpolarized $np$ scattering but
%also when the incoming neutron and proton are both polarized.  

To study in more detail the properties of the final states
 we investigated at the
three chosen energies all nonvanishing
final polarizations and spin-correlations as functions of the
allowed polarizations of the incoming neutron $p_y^n$ and proton $p_y^p$.
%
% Since conclusions do not depend on the energy or the
%scattering angle, in the following
%
 We illustrate the results by showing for each
quantity a surface of its magnitude variations 
 as well as a map of their projections on the plane $p_y^n - p_y^p$, 
 for two exemplary c.m. angles of the outgoing
 neutron at $E_{lab}=50$~MeV.

\begin{figure}
\begin{center}
\begin{tabular}{c}
%\resizebox{138mm}{!}{\includegraphics[angle=270]{gnu1_3d_p_fig1a_e50p0.eps}} \\
%\resizebox{138mm}{!}{\includegraphics[angle=270]{gnu1_3d_p_fig1b_e50p0.eps}} \\
\resizebox{138mm}{!}{\includegraphics[angle=270]{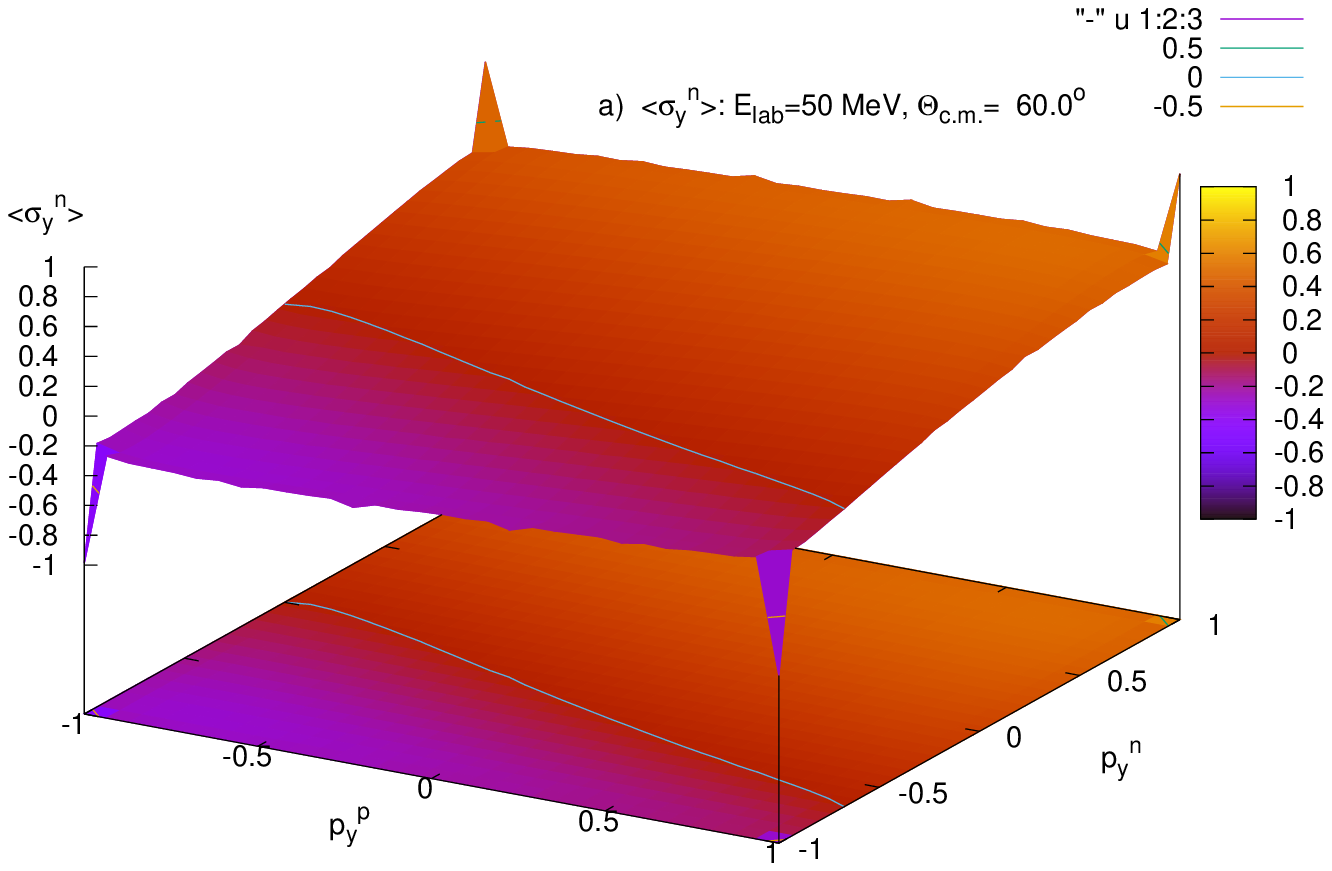}} \\
\resizebox{138mm}{!}{\includegraphics[angle=270]{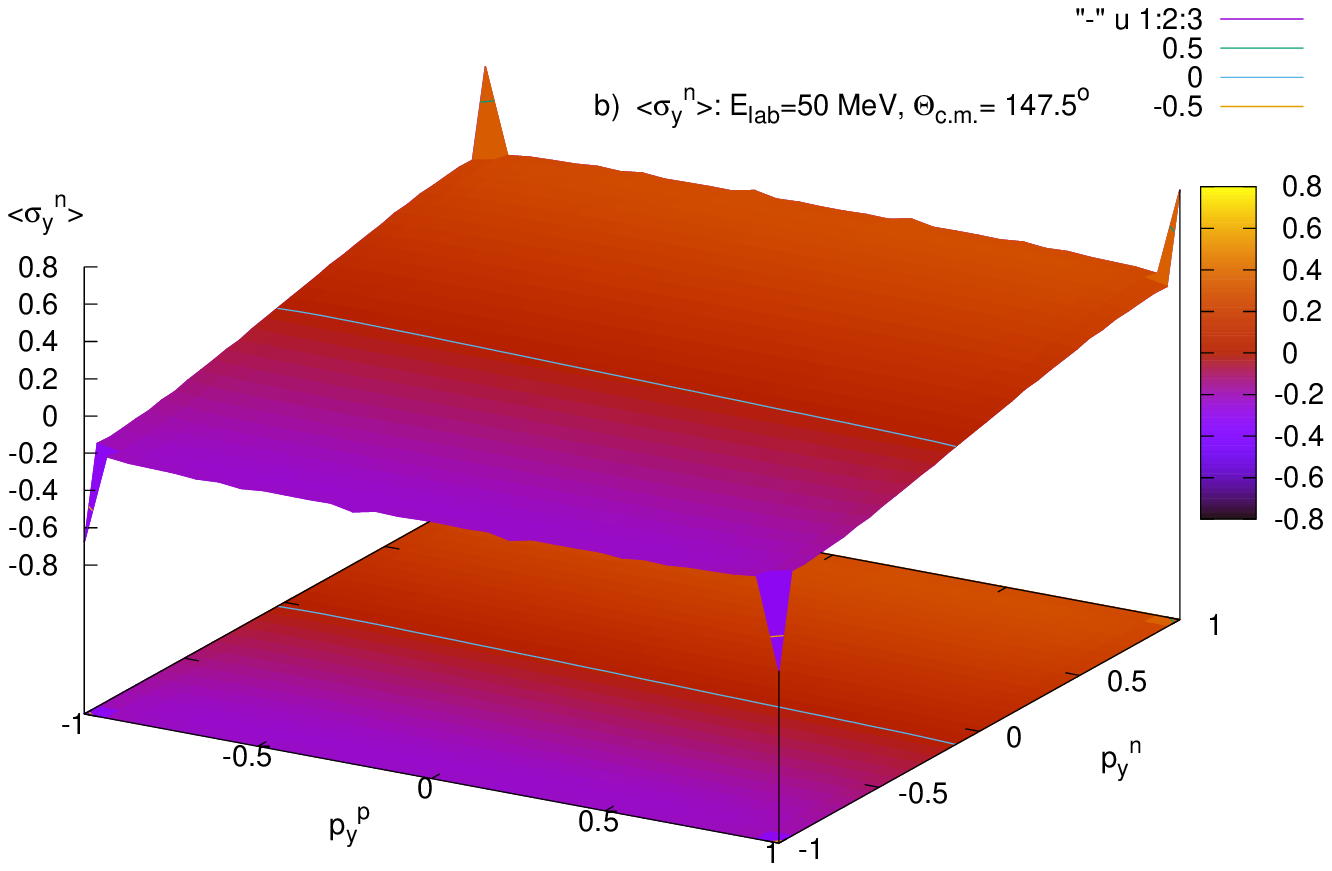}} \\
\end{tabular}%
\caption{
(color online) 
  The polarization $\langle \sigma_y^n\rangle $ of the outgoing neutron
  in $\vec p(\vec n, {\vec n})p$ scattering  at $E_{lab}=50$~MeV
  at two c.m. angles
  $\Theta_{c.m.}=60.0^o$ a) and $147.5^o$ b),
  as a function of polarizations of the incoming neutron $p_y^n$ and
  proton $p_y^p$, predicted with the AV18 potential.}
\label{fig2}
\end{center}
\end{figure}

\begin{figure}
\begin{center}
\begin{tabular}{c}
%\resizebox{138mm}{!}{\includegraphics[angle=270]{gnu1_3d_p_proton_a_e50p0.eps}} \\
  %\resizebox{138mm}{!}{\includegraphics[angle=270]{gnu1_3d_p_proton_b_e50p0.eps}} \\
\resizebox{138mm}{!}{\includegraphics[angle=270]{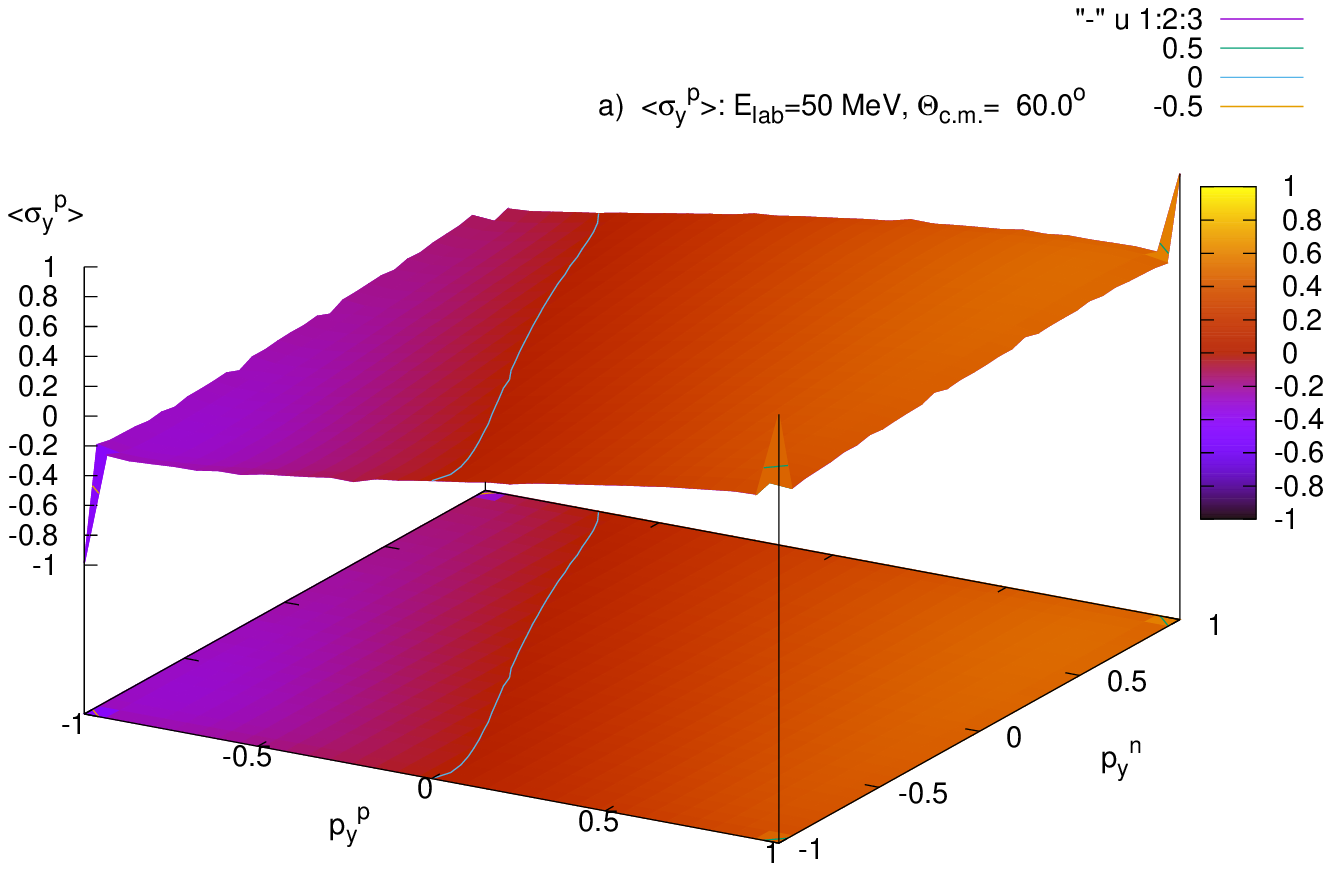}} \\
\resizebox{138mm}{!}{\includegraphics[angle=270]{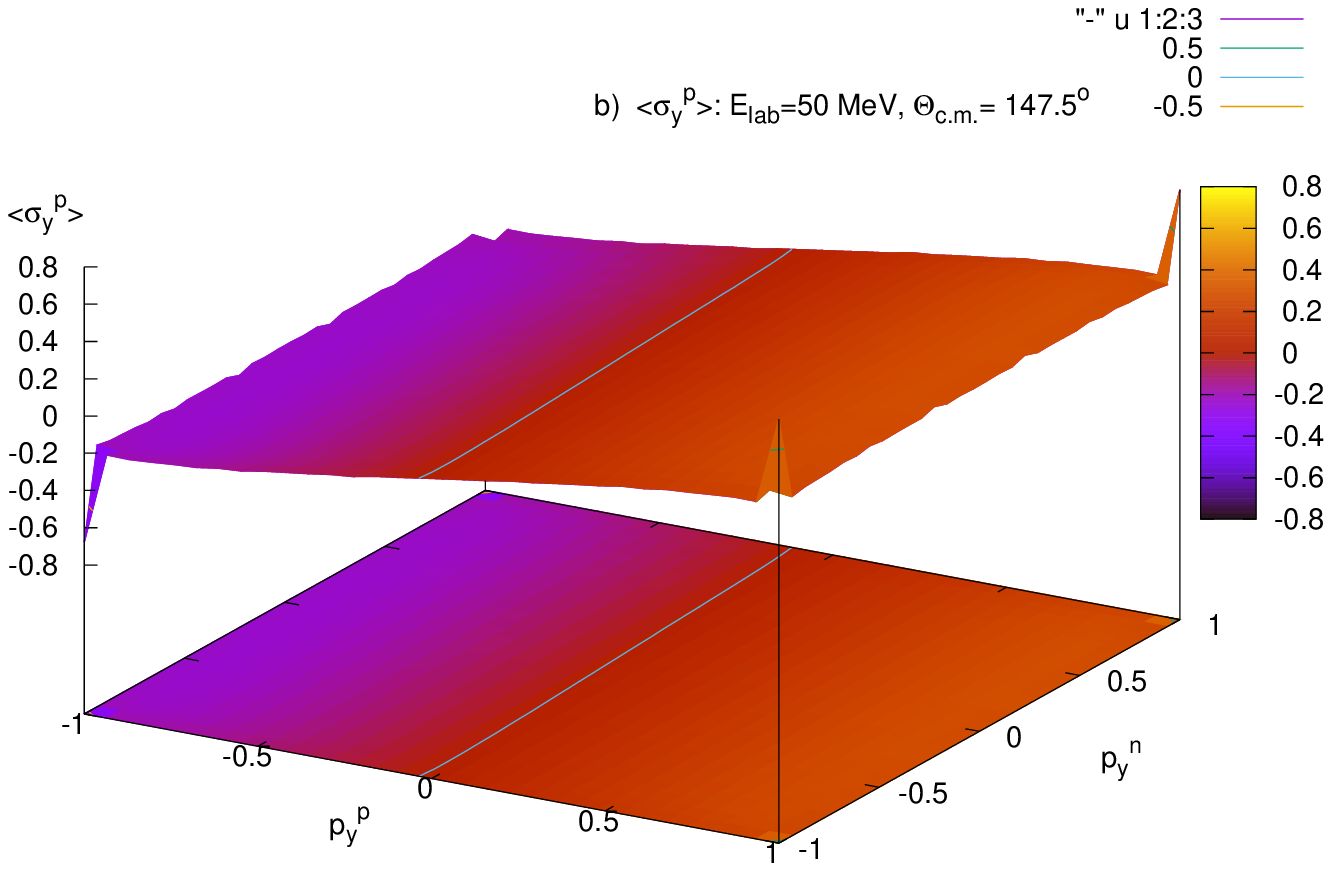}} \\  
\end{tabular}%
\caption{
(color online) 
  The polarization $\langle \sigma_y^p\rangle $ of the outgoing proton 
  in $\vec p(\vec n,n){\vec p}$ scattering  at $E_{lab}=50$~MeV
  at two c.m. angles
  $\Theta_{c.m.}=60.0^o$ a) and $147.5^o$ b),
  as a function of polarizations of the incoming neutron $p_y^n$ and
  proton $p_y^p$, predicted with the AV18 potential.}
\label{fig3}
\end{center}
\end{figure}

In Figs.~\ref{fig2}a and \ref{fig2}b we show such a surface and a map 
for the outgoing neutron polarization $\langle \sigma_y^n\rangle $
at two scattering angles: $\Theta_{c.m.}=60^o$ and $\Theta_{c.m.}=147.5^o$,
respectively.   
 Different colors of the projected map reflect values of the neutron
final polarization  as shown in the legends on the right side of these plots.
The outgoing neutron  polarization in Fig.~\ref{fig2}a takes on values 
in the interval [-1.0,1.0] and in Fig.~\ref{fig2}b in  [-0.8,0.8].
The blue solid
lines indicate where it vanishes. In large regions of the
$p_y^n - p_y^p$ plane values of the outgoing neutron polarization are far away
from vanishing, what would happen for entangled states
(see Table.~\ref{tab1}).
 It is interesting
to note that maximal magnitudes of the outgoing  neutron polarizations are
reached at initial maximal polarizations $p_y^n=\pm1$ and $p_y^p=\pm1$. 
  The corresponding surfaces and maps for  the outgoing
  proton polarization are shown in Figs.~\ref{fig3}a and \ref{fig3}b.
  This observable behaves just as
the outgoing neutron polarization.   

\begin{figure}
\begin{center}
\begin{tabular}{c}
%\resizebox{138mm}{!}{\includegraphics[angle=270]{gnu1_3d_sig_fig2a_e50p0.eps}} \\
%\resizebox{138mm}{!}{\includegraphics[angle=270]{gnu1_3d_sig_fig2b_e50p0.eps}} \\
\resizebox{138mm}{!}{\includegraphics[angle=270]{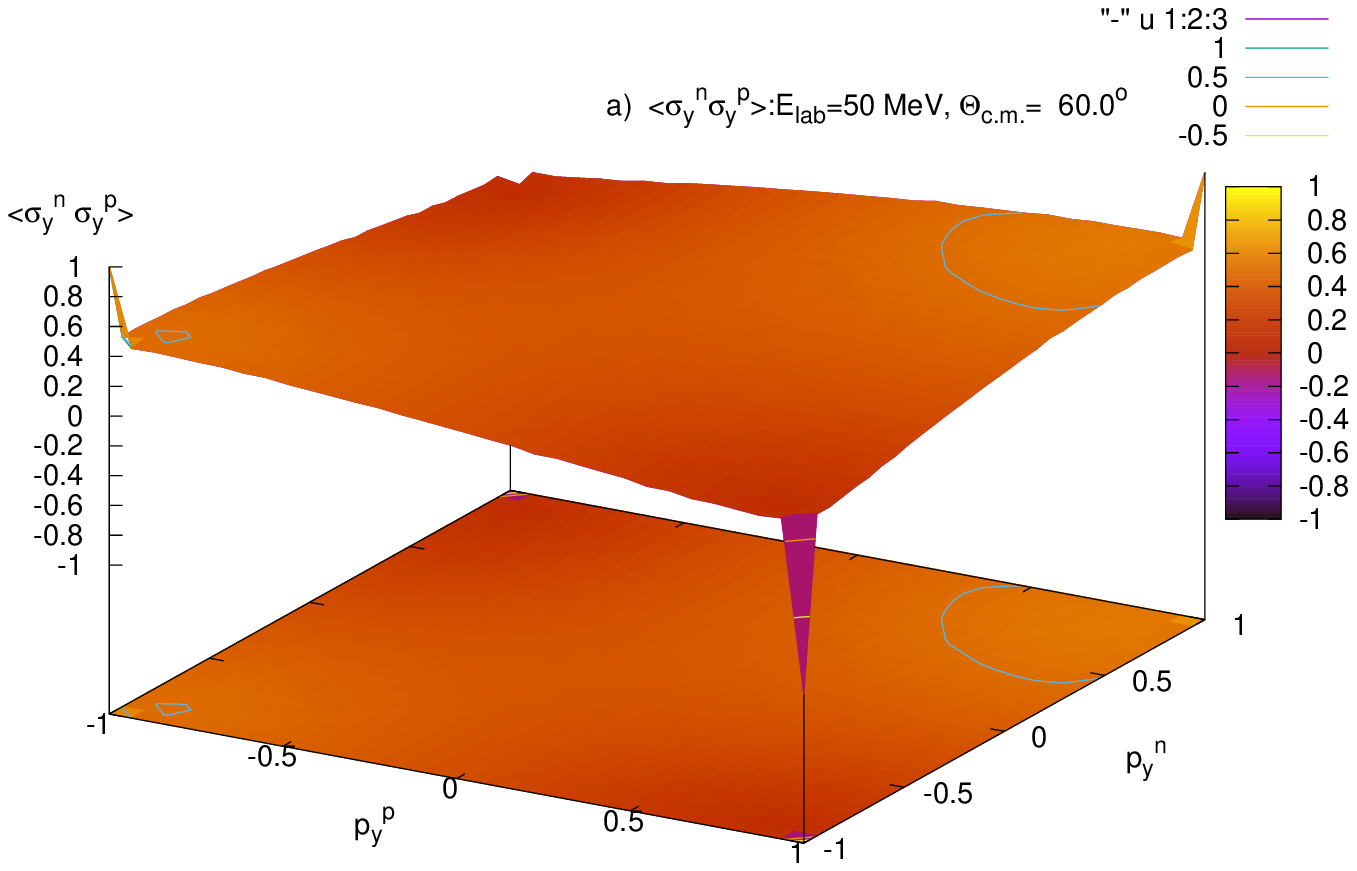}} \\
\resizebox{138mm}{!}{\includegraphics[angle=270]{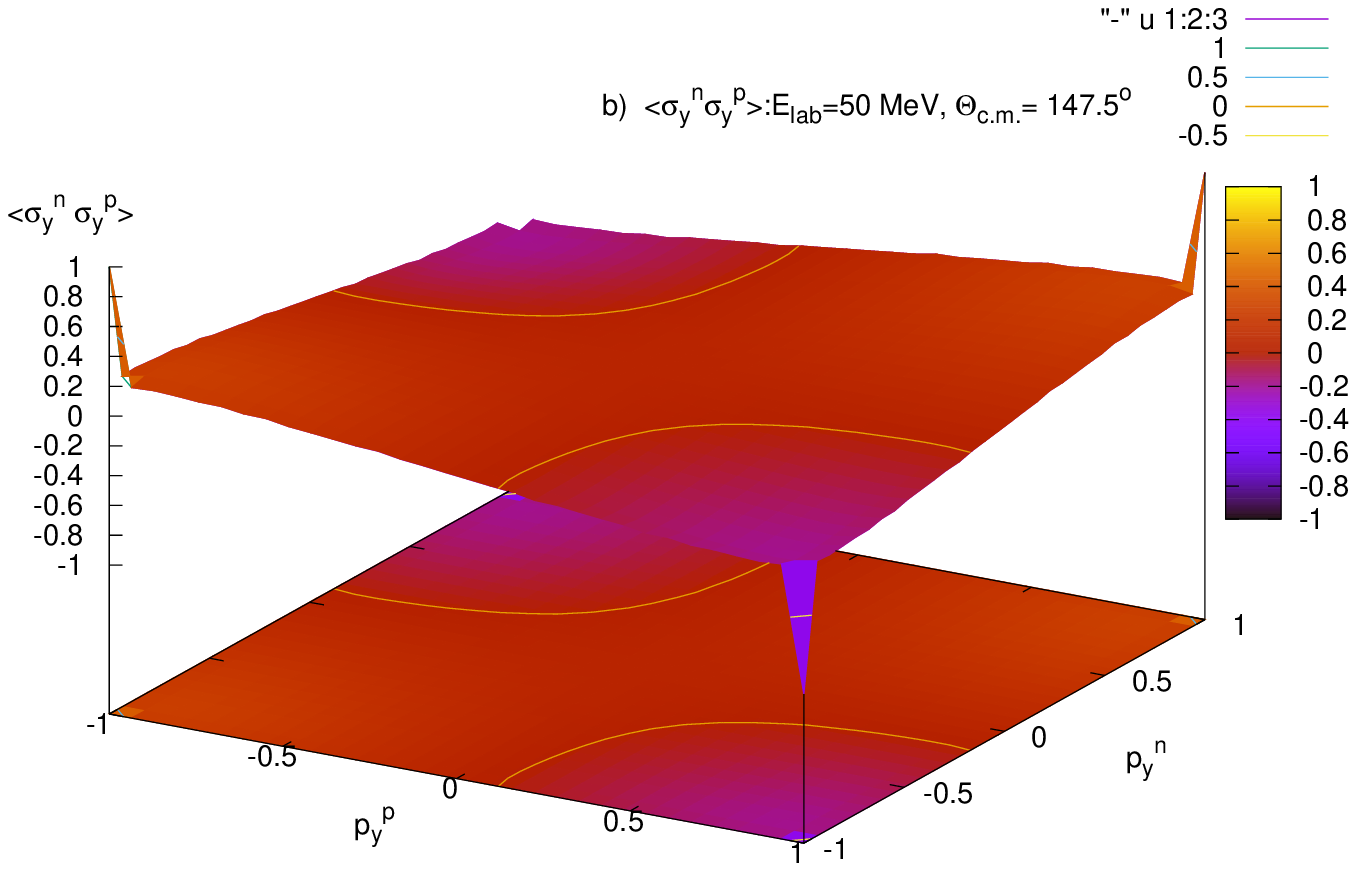}} \\
\end{tabular}%
\caption{
  (color online)
  The spin correlation $\langle \sigma_y^n\sigma_y^p\rangle $ of
  the outgoing neutron and
  proton  in $\vec p(\vec n,\vec n)\vec p$ scattering  at $E_{lab}=50$~MeV at
  two c.m. angles
  $\Theta_{c.m.}=60.0^o$ a) and $147.5^o$ b),
  as a function of polarizations of the incoming neutron $p_y^n$ and
  proton $p_y^p$, predicted with the AV18 potential.}
\label{fig4}
\end{center}
\end{figure}

The corresponding picture for the spin correlation
$\langle \sigma_y^n \sigma_y^p\rangle $ is
shown in  Figs.~\ref{fig4}a and \ref{fig4}b. Here the surface is flatter
than for
the neutron $\langle \sigma_y^n\rangle $ or proton
$\langle \sigma_y^p\rangle $ polarizations,
and predominantly one finds values
of $\langle \sigma_y^n \sigma_y^p\rangle $ in the range $(-0.2,0.2)$,
which are far away from
$\langle \sigma_y^n \sigma_y^p\rangle =\pm 1$ expected for the Bell states
 (see Table.~\ref{tab1}). The value
$\langle \sigma_y^n \sigma_y^p\rangle =+1$ or $-1$ is approached only for
combinations of the initial neutron and proton polarizations
$p_y^n, p_y^p=\pm 1$. 
As we will see later, the final polarization state in this case becomes pure.

\begin{figure}
\begin{center}
\begin{tabular}{c}
%\resizebox{138mm}{!}{\includegraphics[angle=270]{gnu1_3d_sig_xx_a_e50p0.eps}} \\
%\resizebox{138mm}{!}{\includegraphics[angle=270]{gnu1_3d_sig_xx_b_e50p0.eps}} \\
\resizebox{138mm}{!}{\includegraphics[angle=270]{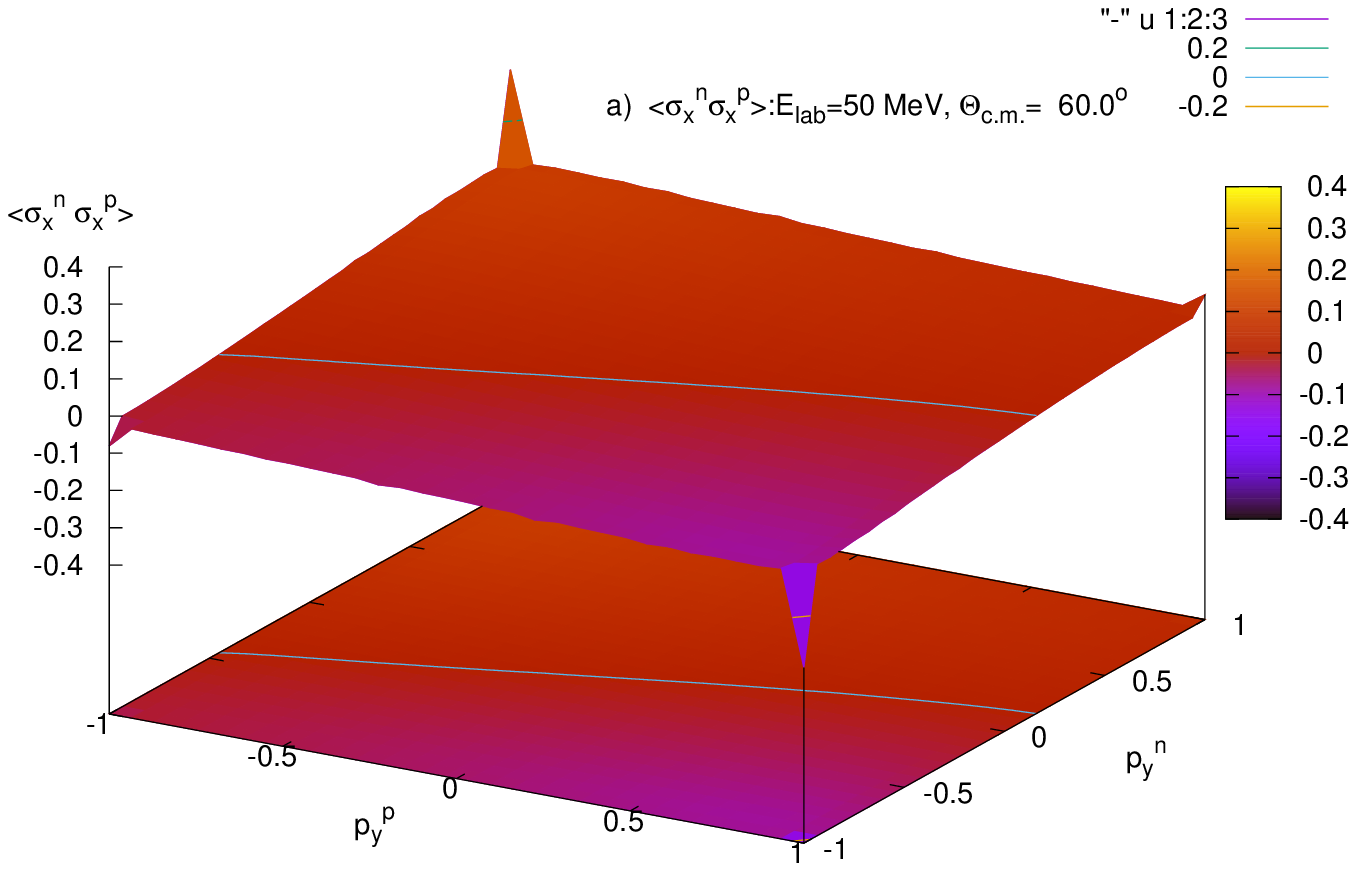}} \\
\resizebox{138mm}{!}{\includegraphics[angle=270]{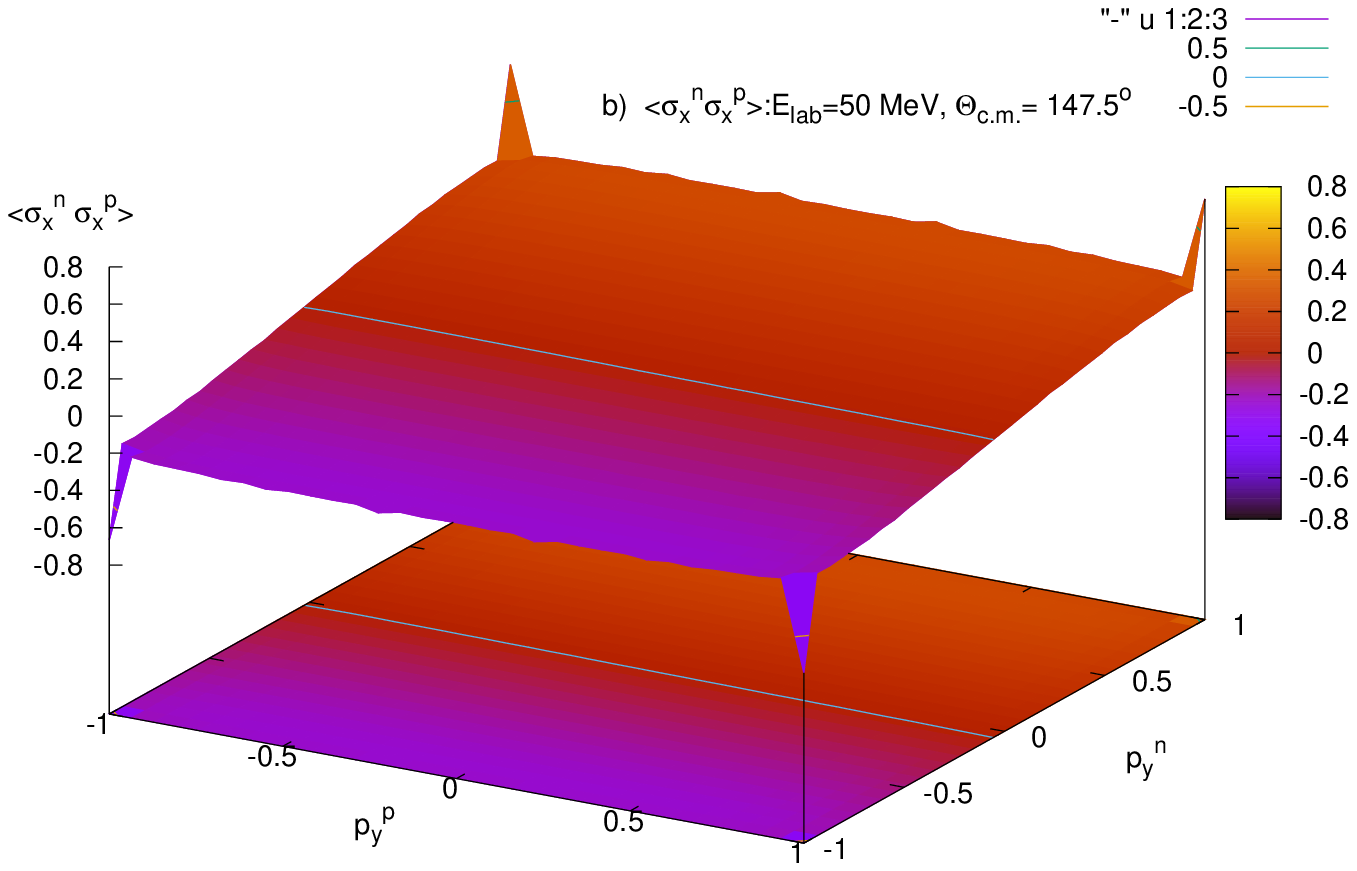}} \\
\end{tabular}%
\caption{
  (color online)
  The same as in Fig.\ref{fig4} but for
  spin correlation $\langle \sigma_x^n\sigma_x^p\rangle $.}
\label{fig5}
\end{center}
\end{figure}

\begin{figure}
\begin{center}
\begin{tabular}{c}
%\resizebox{138mm}{!}{\includegraphics[angle=270]{gnu1_3d_sig_zz_a_e50p0.eps}} \\
%\resizebox{138mm}{!}{\includegraphics[angle=270]{gnu1_3d_sig_zz_b_e50p0.eps}} \\
\resizebox{138mm}{!}{\includegraphics[angle=270]{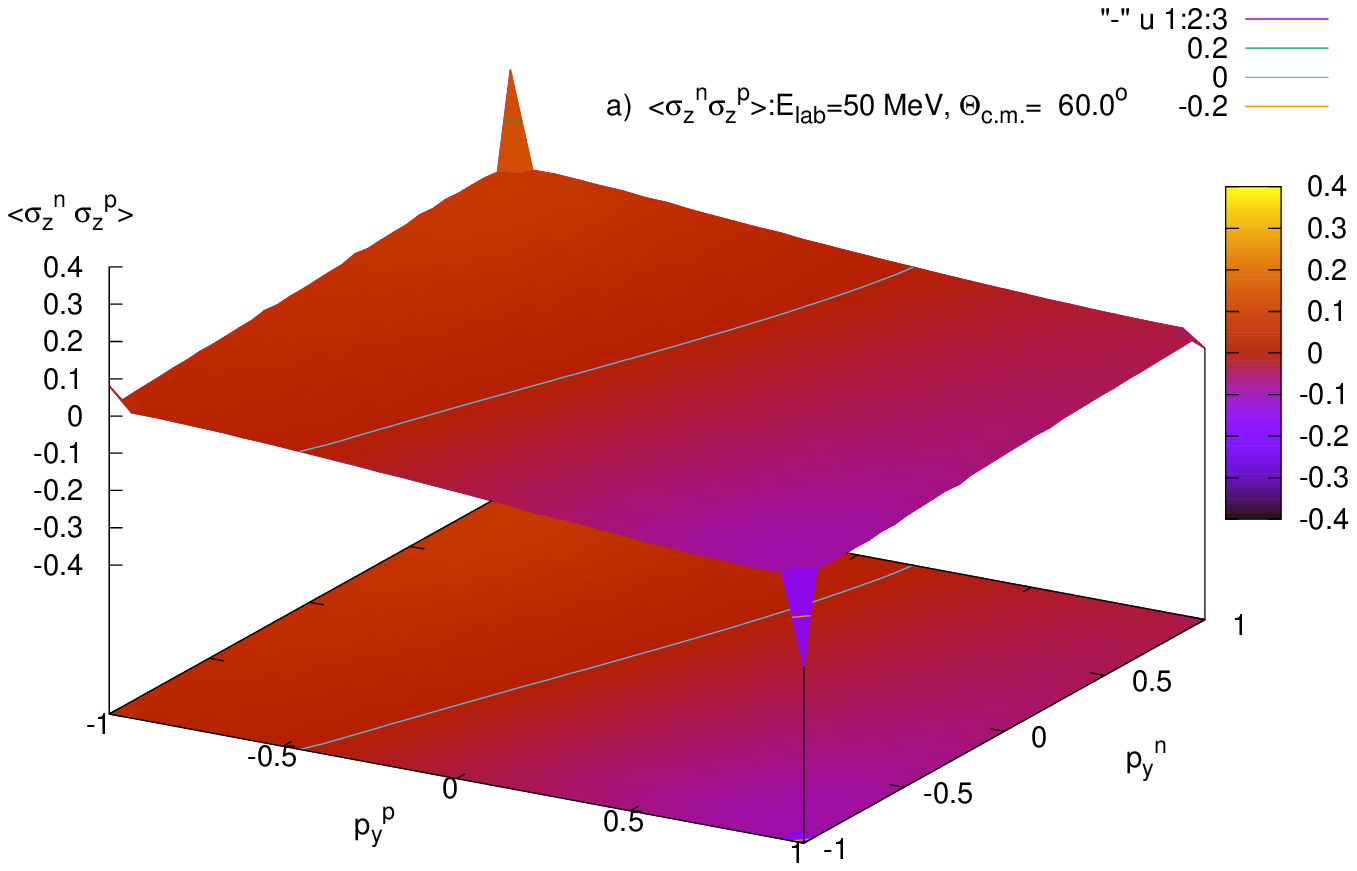}} \\
\resizebox{138mm}{!}{\includegraphics[angle=270]{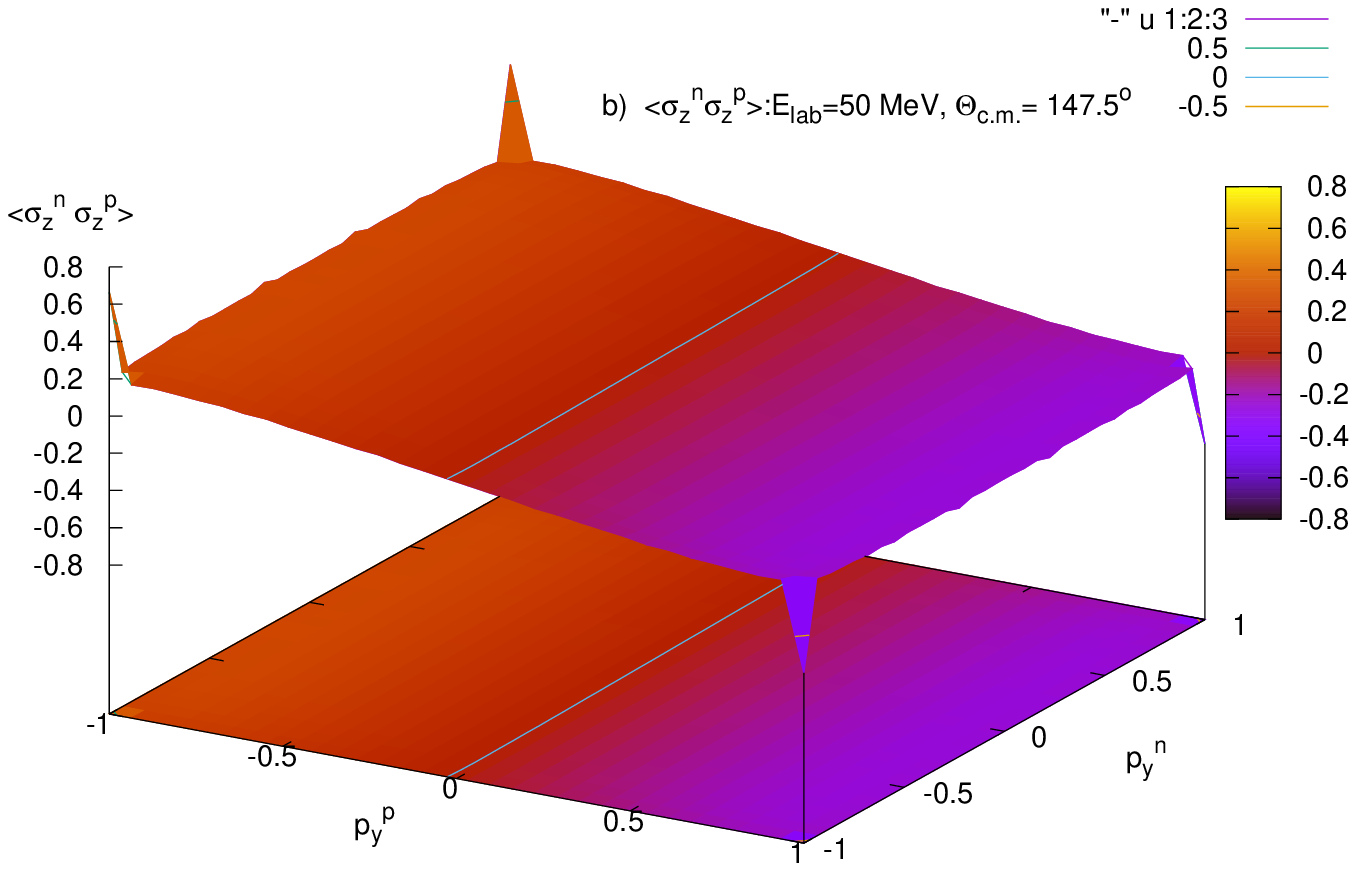}} \\
\end{tabular}%
\caption{
  (color online)
  The same as in Fig.\ref{fig4} but for
  the spin correlation $\langle \sigma_z^n\sigma_z^p\rangle $.}
\label{fig6}
\end{center}
\end{figure}

\begin{figure}
\begin{center}
\begin{tabular}{c}
%\resizebox{138mm}{!}{\includegraphics[angle=270]{gnu1_3d_sig_xz_a_e50p0.eps}} \\
%\resizebox{138mm}{!}{\includegraphics[angle=270]{gnu1_3d_sig_xz_b_e50p0.eps}} \\
\resizebox{138mm}{!}{\includegraphics[angle=270]{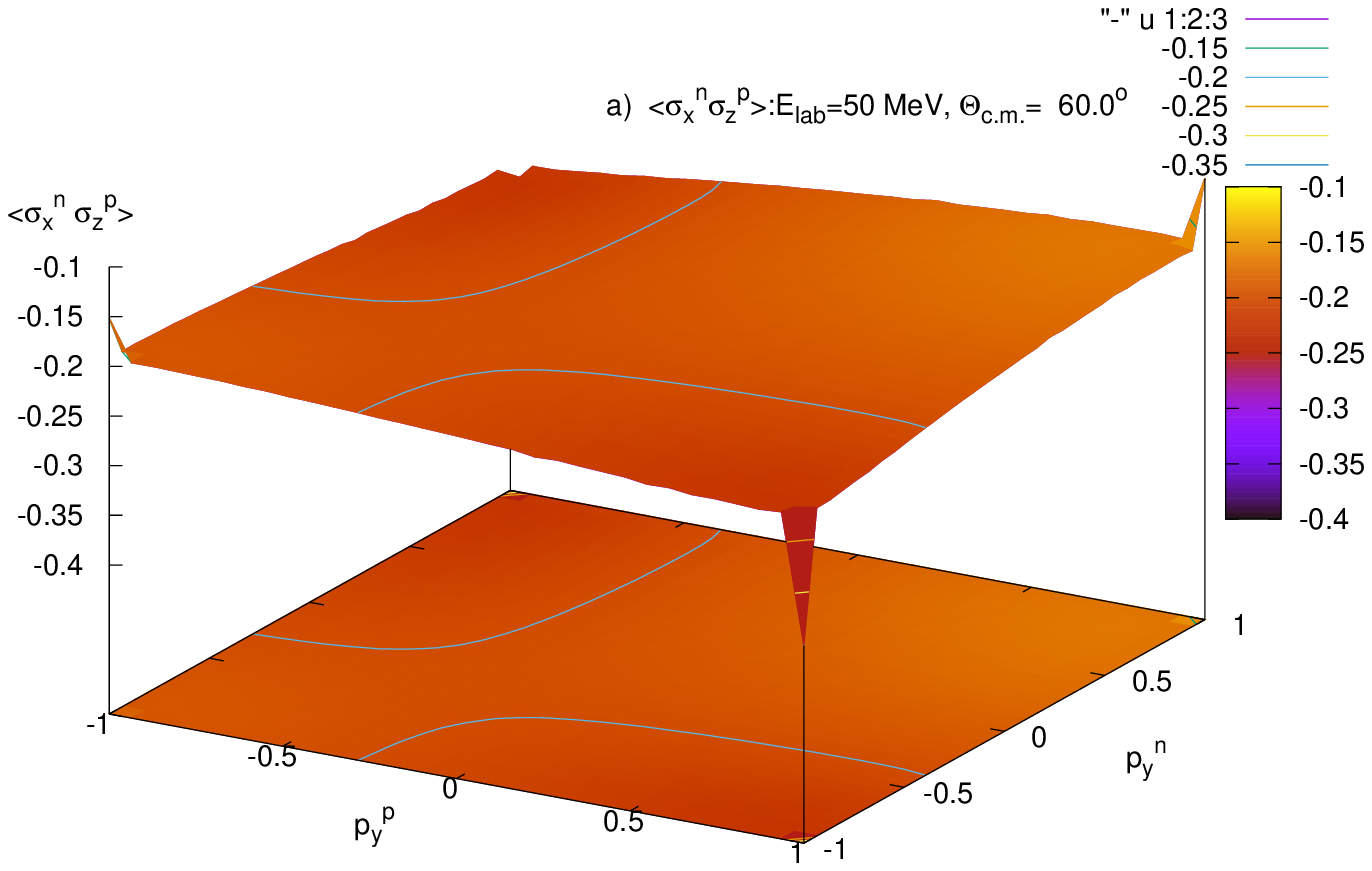}} \\
\resizebox{138mm}{!}{\includegraphics[angle=270]{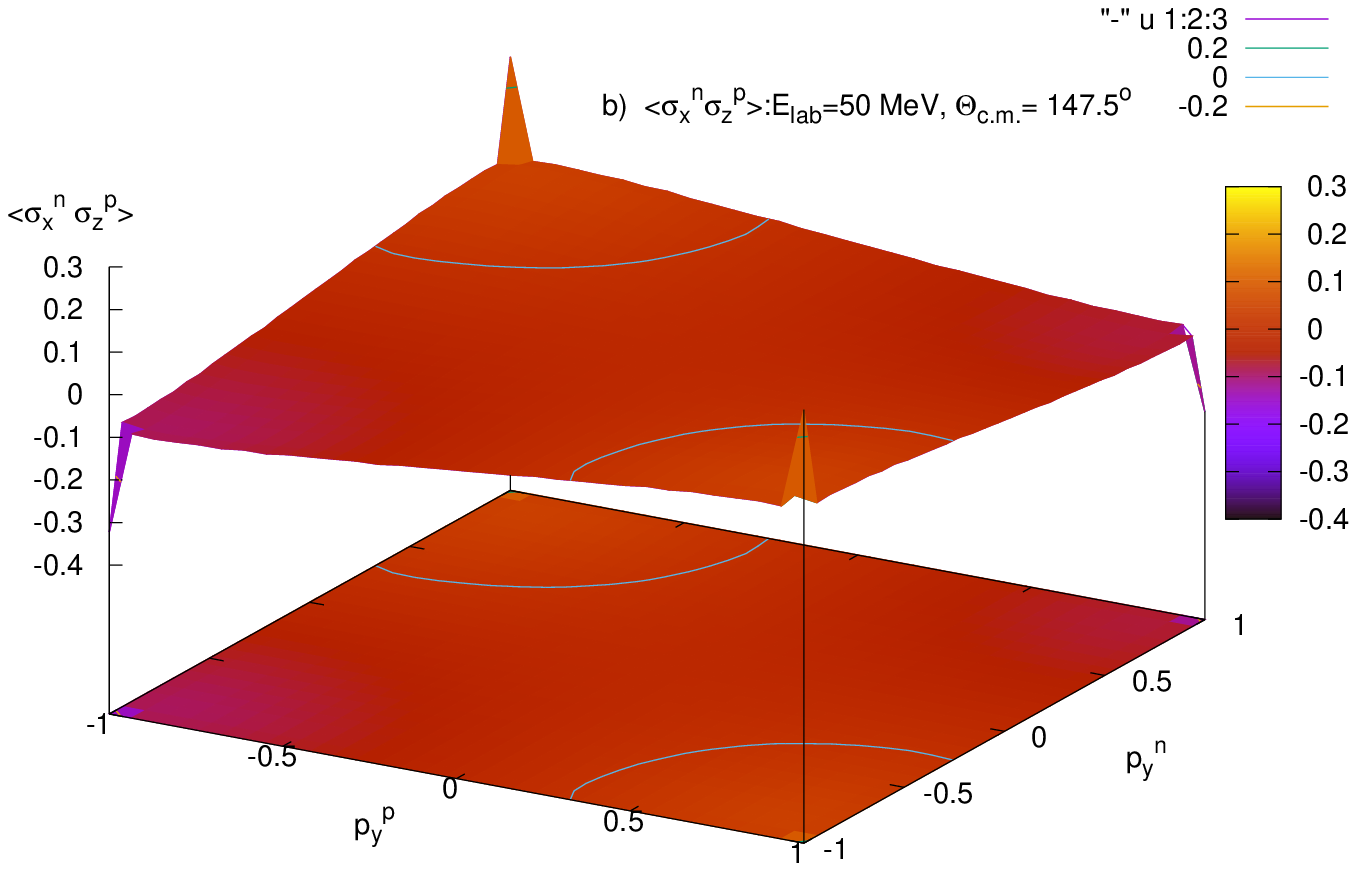}} \\
\end{tabular}%
\caption{
  (color online)
  The same as in Fig.\ref{fig4} but for
  the spin correlation $\langle \sigma_x^n\sigma_z^p\rangle $.}
\label{fig7}
\end{center}
\end{figure}

\begin{figure}
\begin{center}
\begin{tabular}{c}
%\resizebox{138mm}{!}{\includegraphics[angle=270]{gnu1_3d_sig_zx_a_e50p0.eps}} \\
%\resizebox{138mm}{!}{\includegraphics[angle=270]{gnu1_3d_sig_zx_b_e50p0.eps}} \\
\resizebox{138mm}{!}{\includegraphics[angle=270]{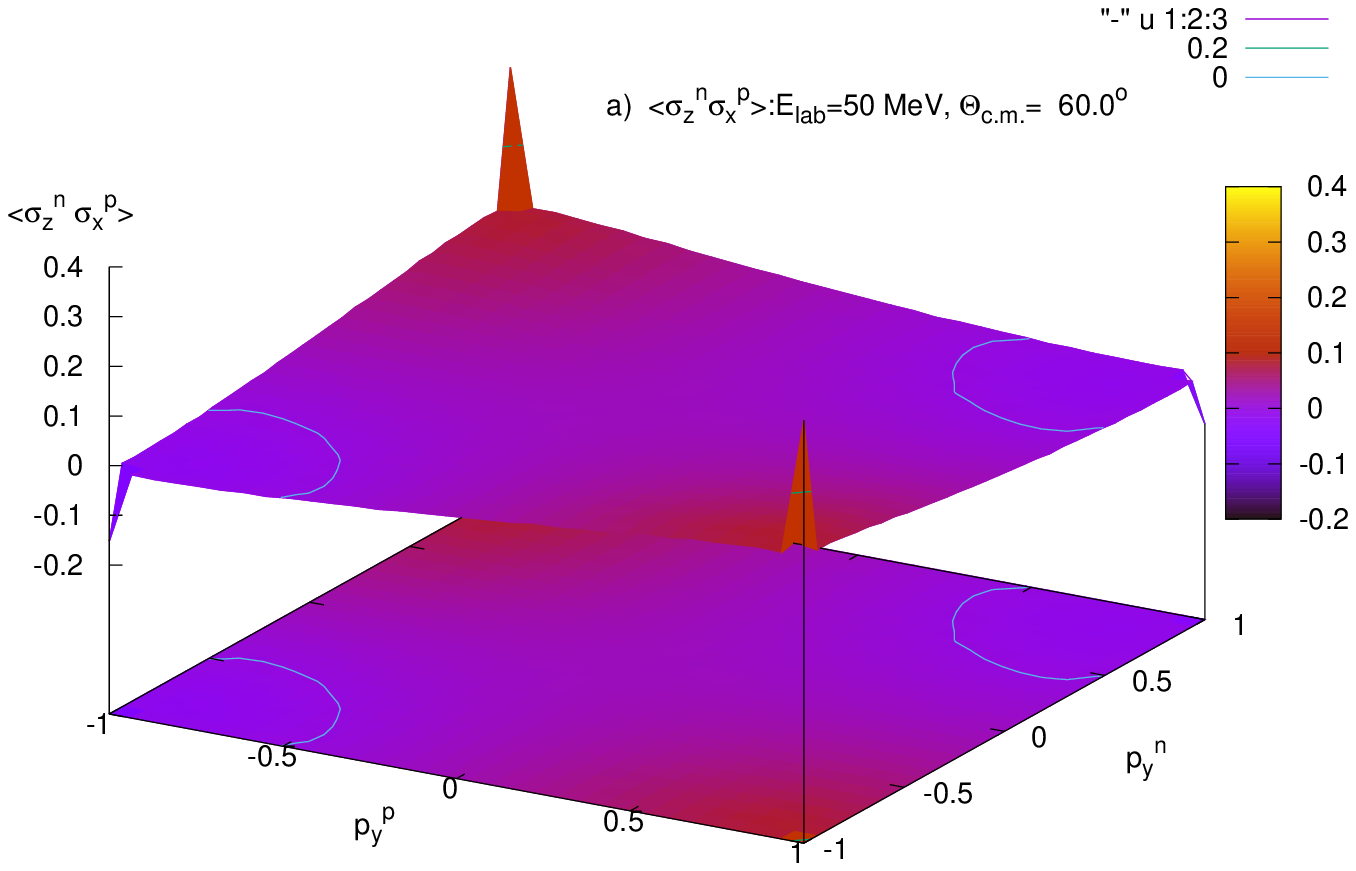}} \\
\resizebox{138mm}{!}{\includegraphics[angle=270]{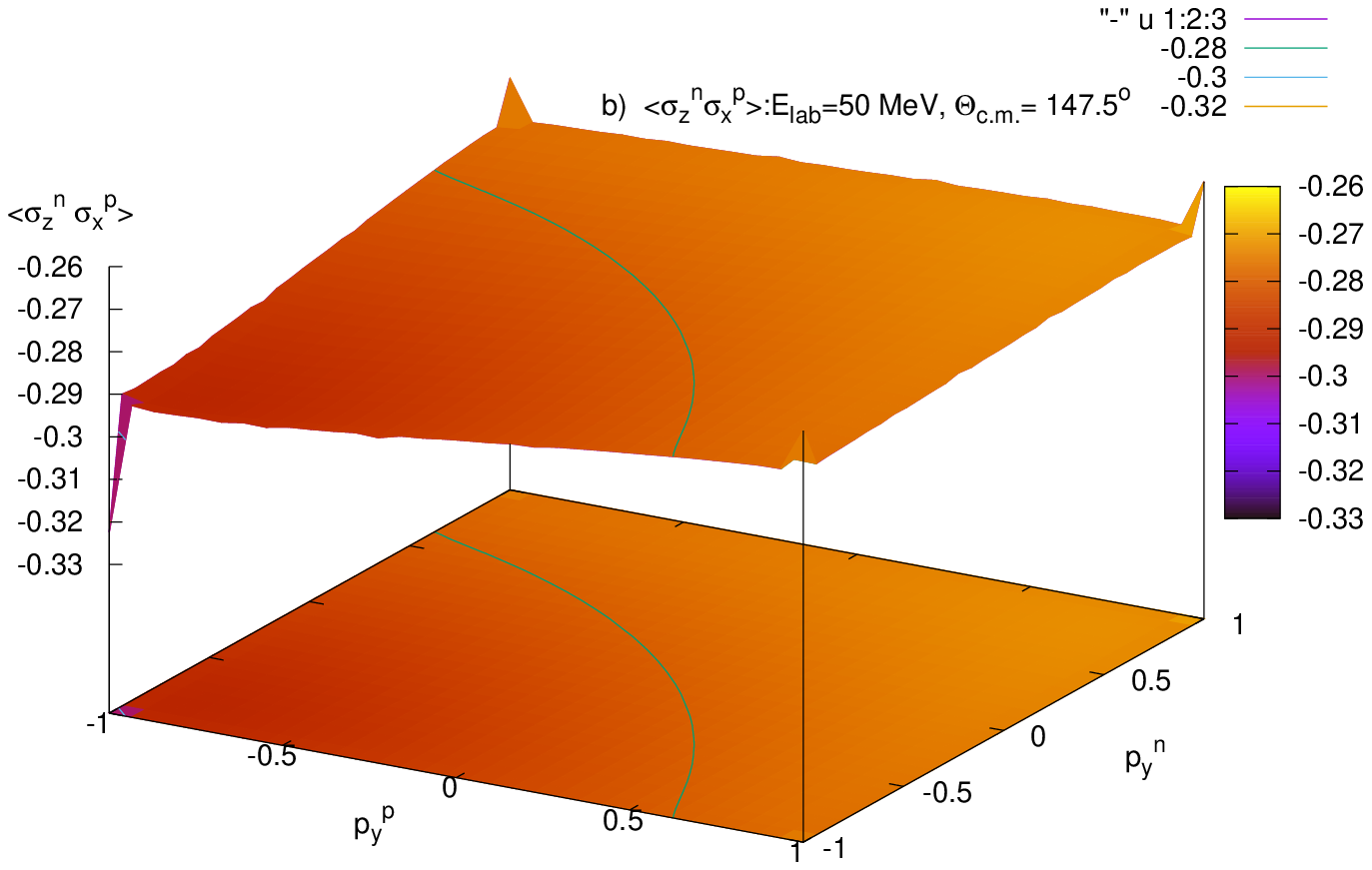}} \\
\end{tabular}%
\caption{
  (color online)
  The same as in Fig.\ref{fig4} but for
  the spin correlation $\langle \sigma_z^n\sigma_x^p\rangle $.}
\label{fig8}
\end{center}
\end{figure}

The remaining nonvanishing spin correlations are shown in
Figs.~\ref{fig5}a-\ref{fig8}b.  Spin correlations
$\langle \sigma_x^n \sigma_x^p\rangle $
(Figs.~\ref{fig5}a and \ref{fig5}b) and $\langle \sigma_z^n \sigma_z^p\rangle $ 
(Figs.~\ref{fig6}a and \ref{fig6}b) behave identically, covering mostly
the range of values  $(-0.2,0.2)$ similarly to
$\langle \sigma_y^n \sigma_y^p\rangle $.
However, for specific combinations of $p_y^n, p_y^p=\pm 1$ they
take on smaller values than
$\langle \sigma_y^n \sigma_y^p\rangle $, differing significantly
from those for the Bell states. Also spin correlations
$\langle \sigma_x^n \sigma_z^p\rangle $ and
$\langle \sigma_z^n \sigma_x^p\rangle $ shown in
Figs.~\ref{fig7} and \ref{fig8}, respectively, assume smaller values, and
in spite of being closer to zero, they are clearly incompatible with the
vanishing values of the corresponding correlations for maximally
entangled states. In particular 
$\langle \sigma_x^n \sigma_z^p\rangle $ at $\Theta_{c.m.}=60^o$ or
$\langle \sigma_z^n \sigma_x^p\rangle $ at
$\Theta_{c.m.}=147.5^o$ are admittedly small but
clearly non-zero for all values of the neutron and
proton polarizations.

\begin{figure}
\begin{center}
\begin{tabular}{c}
%\resizebox{131mm}{!}{\includegraphics[angle=270]{gnu1_3d_rho_fig3a_e50p0_surf.eps}} \\
%\resizebox{131mm}{!}{\includegraphics[angle=270]{gnu1_3d_rho_fig3b_e50p0_surf.eps}} \\
\resizebox{131mm}{!}{\includegraphics[angle=270]{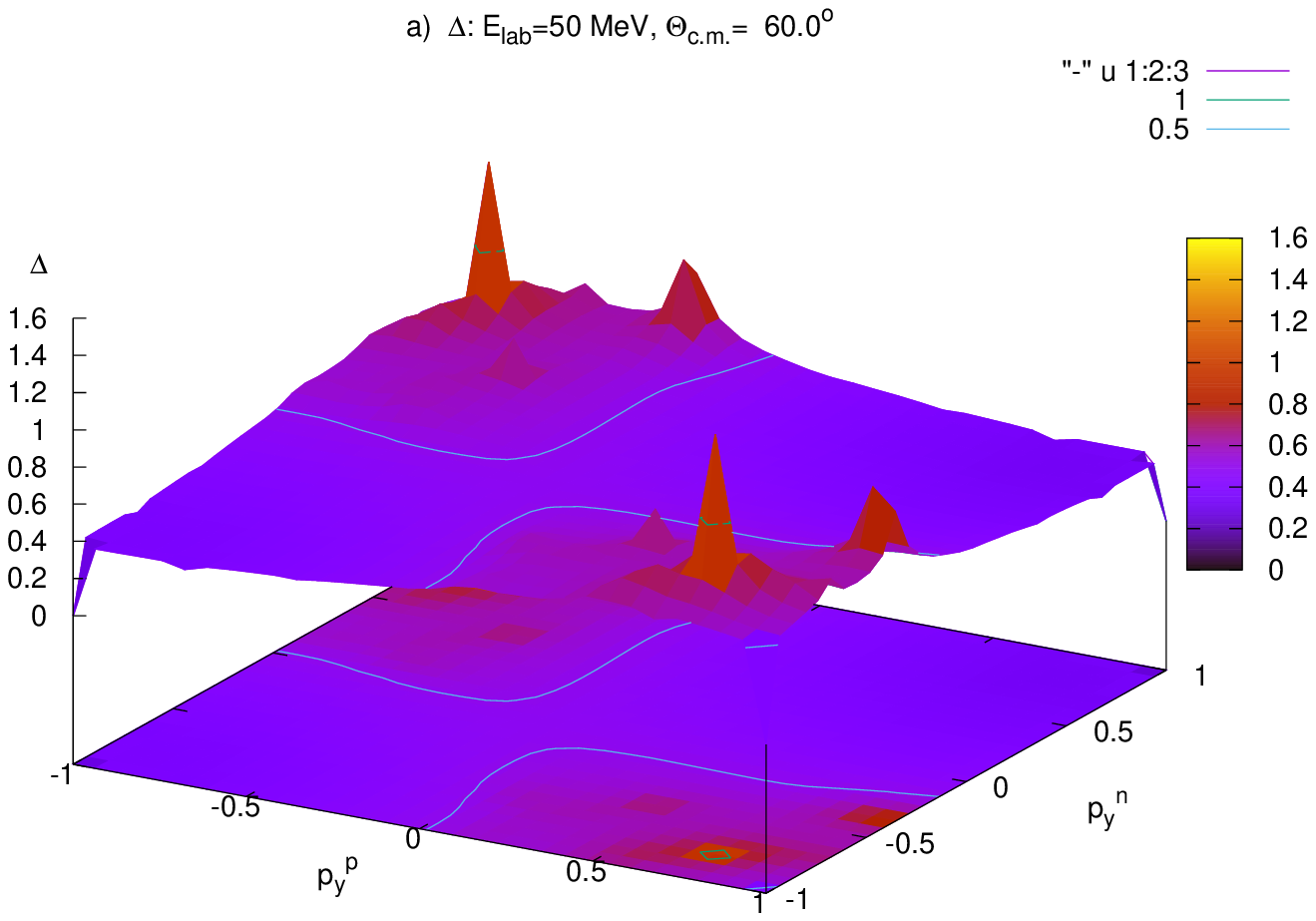}} \\
\resizebox{131mm}{!}{\includegraphics[angle=270]{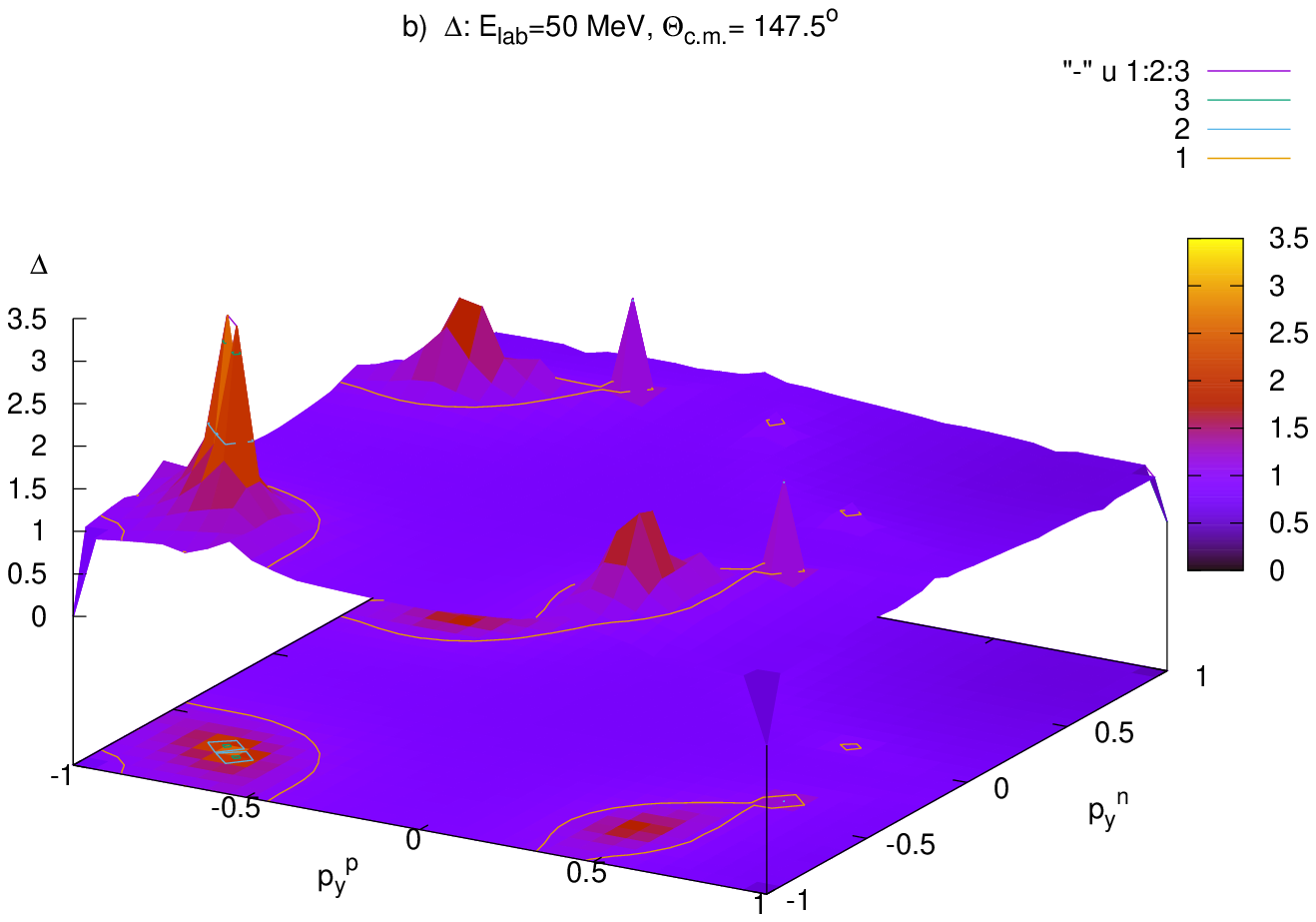}} \\
\end{tabular}%
\caption{
(color online) 
  The $\bigtriangleup$ values (see (\ref{eqqq_7}) for definition)
  for the density matrix $\rho^{out}$, describing
  the final neutron and proton  polarization state
  in $\vec p(\vec n,\vec n)\vec p$ scattering  at $E_{lab}=50$~MeV, at
  two c.m. angles
  $\Theta_{c.m.}=60.0^o$ a) and $147.5^o$ b),
  as a function of polarizations of the incoming neutron $p_y^n$ and
  proton $p_y^p$, predicted with the AV18 potential.}
\label{fig9}
\end{center}
\end{figure}

%Since entangled states, in particular the  Bell states, are pure, 
%
We checked the idempotent condition
(\ref{eq15_a0}) for final $np$ spin density matrix by calculating
the quantity  $\bigtriangleup$ at each c.m. angle as a function of the
initial neutron and proton polarizations. 
The corresponding  maps of $\bigtriangleup$
are shown in  Fig.~\ref{fig9} for the same c.m. angles as in
Figs.~\ref{fig2}-\ref{fig8}. In nearly all regions of the $p_y^n - p_y^p$
plane  $\bigtriangleup \gtrapprox 0.5$,
except at four locations
with $p_y^n=\pm 1$ and $p_y^p=\pm 1$, where $\bigtriangleup$ vanishes. This
means that practically all final polarization states are not pure
spin states but statistical mixtures of spin states, 
except for these four states at each angle, 
which correspond to combinations of polarization values $p_y^n=\pm 1$
and $p_y^p=\pm 1$, and  which are clearly pure states.
They are generated from pure initial spin states
for identical combinations of $p_y^n$ and $p_y^p$ values, with all neutrons
or protons occupying the same spin state with the spin projections
$m=+\frac {1} {2}$ or $m=-\frac {1} {2}$ along $y$-direction.
    Since the neutron or proton polarization in these four specific final pure
    states is relatively large, they appear to be fundamentally different
    from the entangled Bell states, for which this quantity vanishes.

%%%%%%%%%%%%%%%%%%%%%%%%%%%%%%%%%%

Let us now focus on these specific pure initial states for  different
combinations of the maximal neutron $p_y^n=\pm 1$ and proton $p_y^p=\pm 1$
polarizations. The state of the
neutron or proton with $p_y=+1$ can be obtained from the states with spin
projections $m=\pm \frac {1} {2}$ along the
$z$-axis by a rotation
\[
| \frac {1} {2} m \rangle ^y =
\sum\limits_{m'}
D^{\frac {1} {2} }_{m'm} (\alpha \beta \gamma)
| \frac {1} {2} m' \rangle ^z 
\]
with the Euler angles $\alpha=\frac {\pi} {2}$, 
$\beta=\frac {\pi} {2}$ and $\gamma=0$~\cite{brinksatch}.
%performing rotation of the axes $(x, y, z)$ to 
%a new $(x', y', z')$ system of axes
%with $z'$ parallel to  $y$, and  
%$x'$ and $y'$ opposite to $z$ and $x$, respectively. The corresponding 
%Wigner D-matrix $D^{\frac {1} {2} }_{m'm} (\alpha \beta \gamma)$  
% is specified by the Euler angles $\alpha=\frac {\pi} {2}$, 
%$\beta=\frac {\pi} {2}$, and $\gamma=0$ ~\cite{brinksatch} and 
% the states $| \frac {1} {2} \pm  \frac {1} {2}\rangle ^y$ can be written as:
%\begin{eqnarray}
%|\frac {1} {2},+\frac {1} {2} \rangle ^y &=&  |\frac {1} {2},+\frac {1} {2} \rangle ^{z'} =
%\frac {\sqrt{2}} {2}e^{-i\frac {\pi} {4}} (+|\frac {1} {2},+\frac {1} {2} \rangle  +
%e^{i\frac {\pi} {2}}  |\frac {1} {2},-\frac {1} {2} \rangle  ) \label{new1} \\
%|\frac {1} {2},-\frac {1} {2} \rangle ^y &=&  |\frac {1} {2},-\frac {1} {2} \rangle ^{z'} =
%\frac {\sqrt{2}} {2}e^{-i\frac {\pi} {4}} (-|\frac {1} {2},+\frac {1} {2} \rangle  +
%e^{i\frac {\pi} {2}}  |\frac {1} {2},-\frac {1} {2} \rangle  ) \label{new2} ~.
%\end{eqnarray}
This leads to
\begin{eqnarray}
|\frac {1} {2},+\frac {1} {2} \rangle ^y &=&  
\frac {\sqrt{2}} {2}e^{-i\frac {\pi} {4}} \left(+|\frac {1} {2},+\frac {1} {2} \rangle  +
%e^{i\frac {\pi} {2}}  |\frac {1} {2},-\frac {1} {2} \rangle  \right) \label{new1} \, , \\
i\, |\frac {1} {2},-\frac {1} {2} \rangle  \right) \label{new1} \, , \\
|\frac {1} {2},-\frac {1} {2} \rangle ^y &=& 
\frac {\sqrt{2}} {2}e^{-i\frac {\pi} {4}} \left(-|\frac {1} {2},+\frac {1} {2} \rangle  +
%e^{i\frac {\pi} {2}}  |\frac {1} {2},-\frac {1} {2} \rangle  \right) \label{new2} ~.
i\, |\frac {1} {2},-\frac {1} {2} \rangle  \right) \label{new2} ~.
\end{eqnarray}
Thus, neglecting the irrelevant overall % phase space
factor $e^{-i\frac {\pi} {2}}$, the
four pure initial states for  different
combinations of the maximal neutron $p_y^n=\pm 1$ and proton $p_y^p=\pm 1$
polarizations are given by (\ref{eq13_a0}), with complex coefficients
$\alpha_{mm'}$ shown in Table ~\ref{tab2}. Using these coefficients, one can
calculate all polarizations and spin correlations, given in
Eq.~(\ref{eq16_a0}) which are needed to specify this state. 
Since these states are cross product states 
 it is evident that they are not entangled. 

We have seen that a statistical mixture of initial spin states leads to impure
final state. When the initial spin state is pure, the final
state will also be pure. Namely, a pure initial state $|\psi\rangle $,
normalized to one, leads
to a idempotent density matrix $\rho^{in}=|\psi\rangle \langle \psi|$
with a unit trace.
The corresponding final state is described by the density matrix
$\rho^{out}$:
\begin{eqnarray}
  \rho^{out} &=& \frac {T |\psi\rangle  \langle \psi | T^{\dagger}}
      {\text{Tr} (T |\psi\rangle 
    \langle \psi | T^{\dagger}  )}
\label{new3} ~.
\end{eqnarray}
Its $(i,j)$-th element squared reproduces itself
(in the following $\{m_k^{out(in)}\}$
stands for the set of the outgoing (incoming) nucleons spin projections):
\begin{eqnarray}
  (\rho^{out})^2_{ij} &=& \frac { (T |\psi\rangle  \langle \psi | T^{\dagger}
    \sum_{\{ m_k^{out} \} } |\{m_k^{out}\}\rangle  \langle \{m_k^{out}\}|  
    T |\psi\rangle  \langle \psi | T^{\dagger} )_{i,j} }
  { \text{Tr} (T |\psi\rangle 
    \langle \psi | T^{\dagger}  ) \text{Tr}
    (T |\psi\rangle    \langle \psi | T^{\dagger}  ) } \cr
  &=&  \frac { (T |\psi\rangle   \sum_{\{ m_k^{out} \} }
    | \langle \{m_k^{out}\}| T |\psi\rangle  |^2
    \langle \psi | T^{\dagger} )_{i,j} }
  { \sum_{\{ m_k^{out} \} } \langle \{m_k^{out}\} | T |\psi\rangle 
    \langle \psi | T^{\dagger} |\{m_k^{out}\}\rangle 
	\text{Tr} (T |\psi\rangle    \langle \psi | T^{\dagger}  ) } \cr
  &=&  \frac { \sum_{\{ m_k^{out} \} } | \langle \{m_k^{out}\}| T |\psi\rangle  |^2
     (T |\psi\rangle  \langle \psi | T^{\dagger} )_{i,j} }
  { \sum_{\{ m_k^{out} \} } | \langle \{m_k^{out}\} | T |\psi\rangle  |^2 
    \text{Tr} (T |\psi\rangle    \langle \psi | T^{\dagger}  ) }
  = (\rho^{out})_{ij} ~.
\label{new4} 
\end{eqnarray}
The average value of an observable $O$ in such a pure final state is given by:
\begin{eqnarray}
  && \langle O\rangle  = \text{Tr} (\rho^{out}O) =  \frac
  { \text{Tr} (T |\psi\rangle  \langle \psi | T^{\dagger} O ) }
	{ \text{Tr} (T |\psi\rangle  \langle \psi | T^{\dagger}  ) }
\cr
&=&  \frac { \sum \limits_{ \{ m_{{out} }\} } \sum \limits_{ \{ m'_{ {out} }\} }
    [ \sum \limits_{ \{ m_{ {in} } \} }
      T_{ \{ m_{in} \} }^{ \{ m_{out} \} } \alpha_{ \{ m_{in} \} } ]
    [\sum \limits_{ \{ m'_{ {in} } \} }
      T_{ \{ m'_{in} \} }^{* \{ m'_{out} \} } {\alpha^*}_{ \{ m'_{in} \} } ]
      \langle  \{ m'_{out} \} | O | \{ m_{out} \}\rangle  }   
   { \sum \limits_{ \{ m_{out} \} } \vert   \sum \limits_{\{ m_{in} \} }
       T_{ \{ m_{in} \} }^{ \{ m_{out} \} }   \alpha_{ \{ m_{in} \} } \vert^2   } ~.
\label{new5} 
\end{eqnarray}

At this point, we would like to note that the pure state (\ref{new3}) of
    the outgoing $np$ pair contains components with all possible neutron
    and proton spin projections, which are induced by the transition
    matrix elements $\langle m_n m_p | T | m_n' m_p' \rangle$:
\begin{eqnarray}
\frac {T |\psi\rangle } {[\text{Tr} (T |\psi\rangle
\langle \psi | T^{\dagger} )]^{\frac {1} {2} }} &=& \sum_{ m_n m_p }
\frac { \langle m_n m_p| T |\psi\rangle }
{[\text{Tr} (T |\psi\rangle
\langle \psi | T^{\dagger} )]^{\frac {1} {2} }}~
|m_n m_p\rangle
\label{new6} ~.
\end{eqnarray}
Therefore, this state will, in general, be unentangled, even if the
initial state $|\psi\rangle$ is entangled.
However, there are interesting exceptions to this rule, arising mainly
from the fact that the largest matrix elements
$\langle m_n m_p| T | m_{n'} m_{p'} \rangle$ correspond to
transitions that preserve the sign of the nucleon spin projection.
As a result, the number of significant matrix elements, which
primarily determine the amplitudes of the final state (\ref{new6})
for different outgoing spin projections $(m_n, m_p)$, is reduced
from 16 to just 4:
$\langle +\frac{1}{2} +\frac{1}{2}| T |+\frac{1}{2} +\frac{1}{2}\rangle$,
$\langle +\frac{1}{2} -\frac{1}{2}| T |+\frac{1}{2} -\frac{1}{2}\rangle$,
$\langle -\frac{1}{2} +\frac{1}{2}| T |-\frac{1}{2} +\frac{1}{2}\rangle$, and
$\langle -\frac{1}{2} -\frac{1}{2}| T |-\frac{1}{2} -\frac{1}{2}\rangle$.
At the end of this section, we will demonstrate that this, combined with
the energy and angular dependence of the transition matrix, enables
the identification of strongly entangled Bell-like states at higher energies
and specific angles, with only a small admixture of entanglement-spoiling
contributions.

\begin{figure}
\includegraphics[scale=0.8]{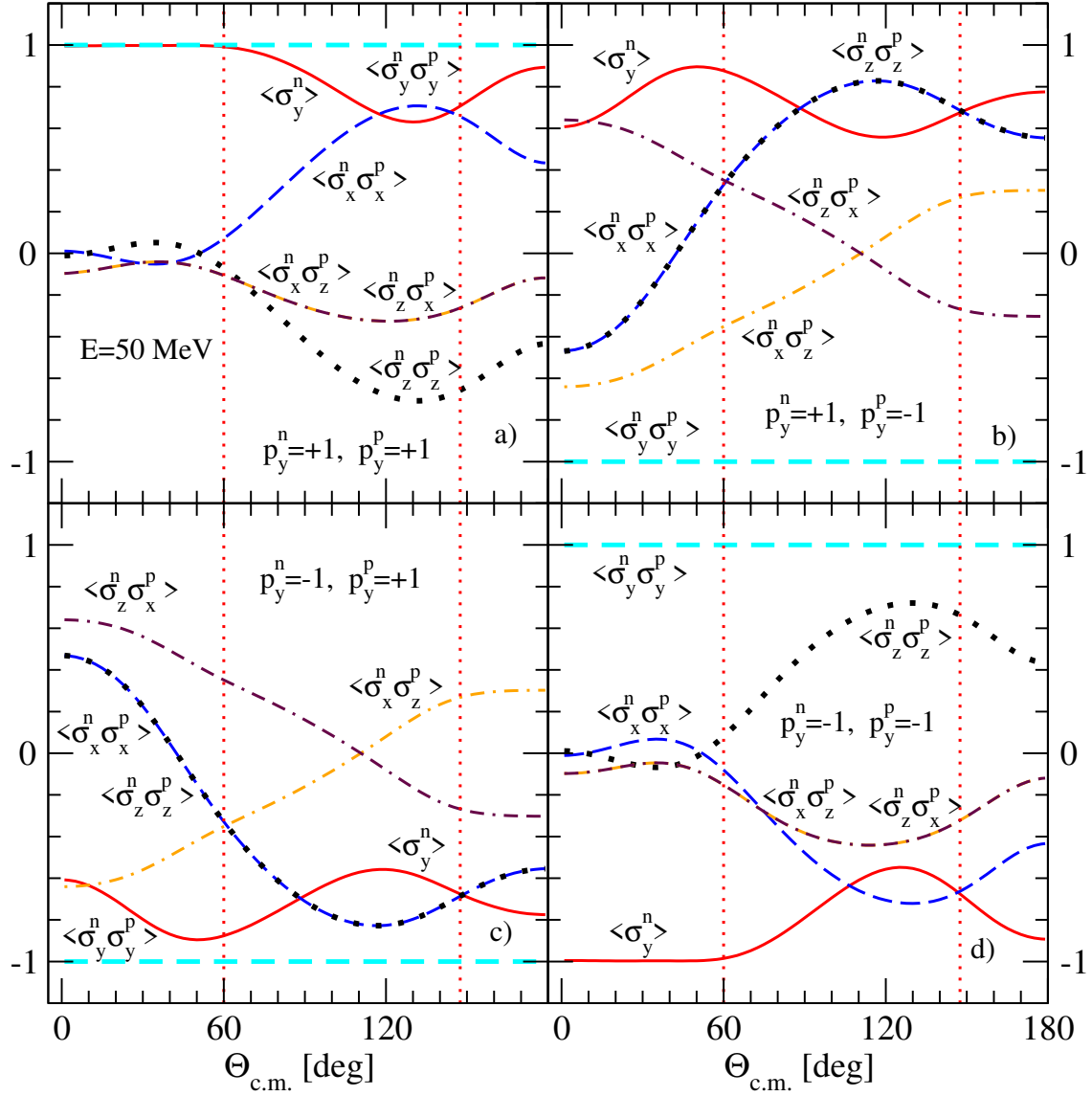}  
% Here is how to import EPS art
\caption{(color online) The 
  neutron final polarization    
  $\langle \sigma_y^n\rangle $ ((red) solid line)
  and nonvanishing spin correlations
  $\langle \sigma_x^n\sigma_x^p\rangle $ ((blue) long dashed line) ,
  $\langle \sigma_y^n\sigma_y^p\rangle $ ((cyan) short dashed line),
  $\langle \sigma_z^n\sigma_z^p\rangle $ ((black) dotted line),
  $\langle \sigma_x^n\sigma_z^p\rangle $ ((orange) dashed-dotted line),
  $\langle \sigma_z^n\sigma_x^p\rangle $ ((maroon) double-dashed-dotted line), 
    in the $\vec p (\vec n,\vec n)\vec p$ reaction at incoming neutron
    lab. energy $E=50$~MeV, predicted with the AV18 potential,
    shown in function of the c.m. scattering angle.
    The incoming neutron
    and proton polarizations are: a) $p_y^n=+1, p_y^p=+1$, 
    b) $p_y^n=+1$, $p_y^p=-1$, c) $p_y^n=-1$, $p_y^p=+1$,
    d) $p_y^n=-1$, $p_y^p=-1$. 
}
 \label{fig10}
\end{figure}

In Fig.~\ref{fig10} we show at $E=50$~MeV, for the four combinations of
 the neutron and proton polarizations
$p_y^n=p_y^p=\pm 1$,  the neutron final polarization
 $\langle \sigma_y^n\rangle $ of Eq.~(\ref{eqqq_1}) and
 all nonvanishing spin correlations
 of Eqs.~(\ref{eqqq_2})-(\ref{eqqq_6}), calculated using Eq.~(\ref{new5}).
 The absolute value of the spin correlation
 $\langle \sigma_y^n \sigma_y^p\rangle =1$
across the entire angular range. Also the neutron polarization is
far from zero, being close to $1$ for $\Theta_{c.m.} \le 60^o$.
The values of all these observables further
%provide support
 suggest that these states are not entangled. 
Moreover, the angular behaviour of these observables explains
the ranges of values for polarization 
 and spin correlations, shown for two c.m. scattering angles
 $\Theta_{c.m.}=60^o$ and $147.5^o$ in maps of
 Figs.~\ref{fig2}-\ref{fig8}. To facilitate the comparison, the
 position of these two angles is
 indicated in Figs.~\ref{fig10}a-\ref{fig10}d by vertical dotted lines.

\begin{table}[ht!]
\centering  
\begin{tabular}{ |c|c|c|c|c| }
\hline
$p_y^n$ and $p_y^p$ combinations & $\alpha_{+\frac {1} {2} +\frac {1} {2}  }$
& $\alpha_{+\frac {1} {2} -\frac {1} {2} } $  & $\alpha_{-\frac {1} {2} +\frac {1} {2}  } $ 
& $\alpha_{-\frac {1} {2} -\frac {1} {2} }$ \\
\hline
a) $p_y^n=+1~p_y^p=+1$  & $+\frac {1} {2}$
& $+\frac {i} {2} $
& $+\frac {i} {2} $  & $-\frac {1} {2}$  \\
\hline
b) $p_y^n=+1~p_y^p=-1$  & $-\frac {1} {2}$
& $+\frac {i} {2}  $
& $-\frac {i} {2}  $  &  $-\frac {1} {2}$ \\
\hline
c) $p_y^n=-1~p_y^p=+1$  & $-\frac {1} {2}$
&  $-\frac {i} {2}  $
&  $+\frac {i} {2}  $  &  $-\frac {1} {2}$ \\
\hline
d) $p_y^n=-1~p_y^p=-1$  & $+\frac {1} {2}$
& $-\frac {i} {2} $
& $-\frac {i} {2} $  & $-\frac {1} {2}$  \\
\hline
\end{tabular}
\caption{Values of the complex coefficients $\alpha_{mm'}$  for the
four pure initial states  (\ref{eq13_a0}), with  different
combinations of the neutron and proton polarizations shown in the first
column.}
\label{tab2}
\end{table}

\begin{figure}
\includegraphics[scale=0.8]{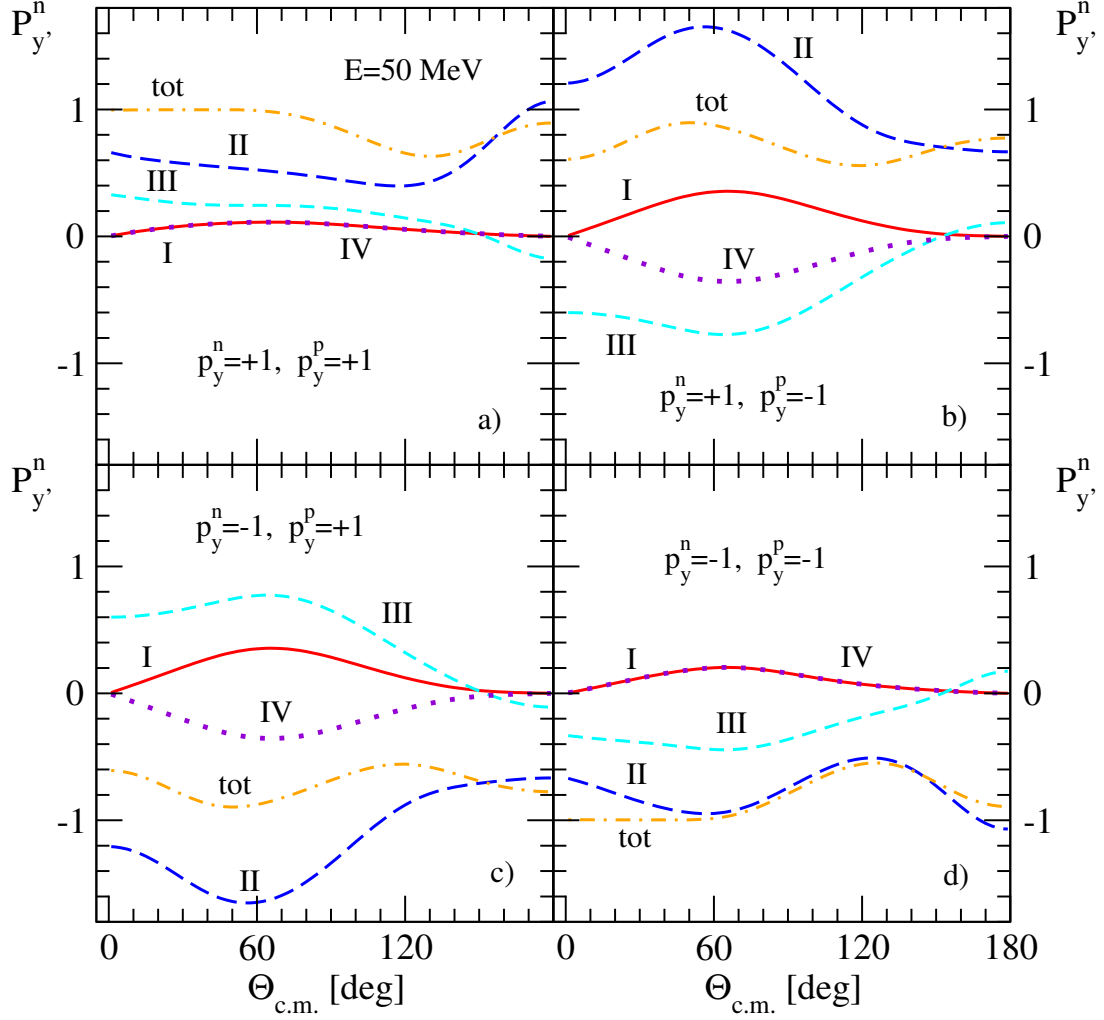}  
% Here is how to import EPS art
\caption{(color online) The y-component of the
  neutron final polarization   
  $\langle \sigma_y^n\rangle $ ((orange) dashed-dotted line)
  together with its building terms 
    in the $\vec p (\vec n,\vec n)p$ reaction at the incoming neutron
    lab. energy $E=50$~MeV with incoming neutron
    and proton polarizations: a) $p_y^n=+1, p_y^p=+1$, 
    b) $p_y^n=+1$, $p_y^p=-1$, c) $p_y^n=-1$, $p_y^p=+1$,
    d) $p_y^n=-1$, $p_y^p=-1$. Contributions of each of the four terms in
    (\ref{eqqq_1}) are denoted by:
    I - the first term with $P_{y'}^{(0)}$ ((red) solid line),
    II - the second term with $K_{0,y}^{y',0}$ ((blue) long dashed line),
    III - the third term with
    $K_{y,0}^{y',0}$ ((cyan) short dashed line),
    IV - the fourth term with $K_{y,y}^{y',0}$ ((violet) dotted line).
}
 \label{fig11}
\end{figure}

\begin{figure}
\includegraphics[scale=0.8]{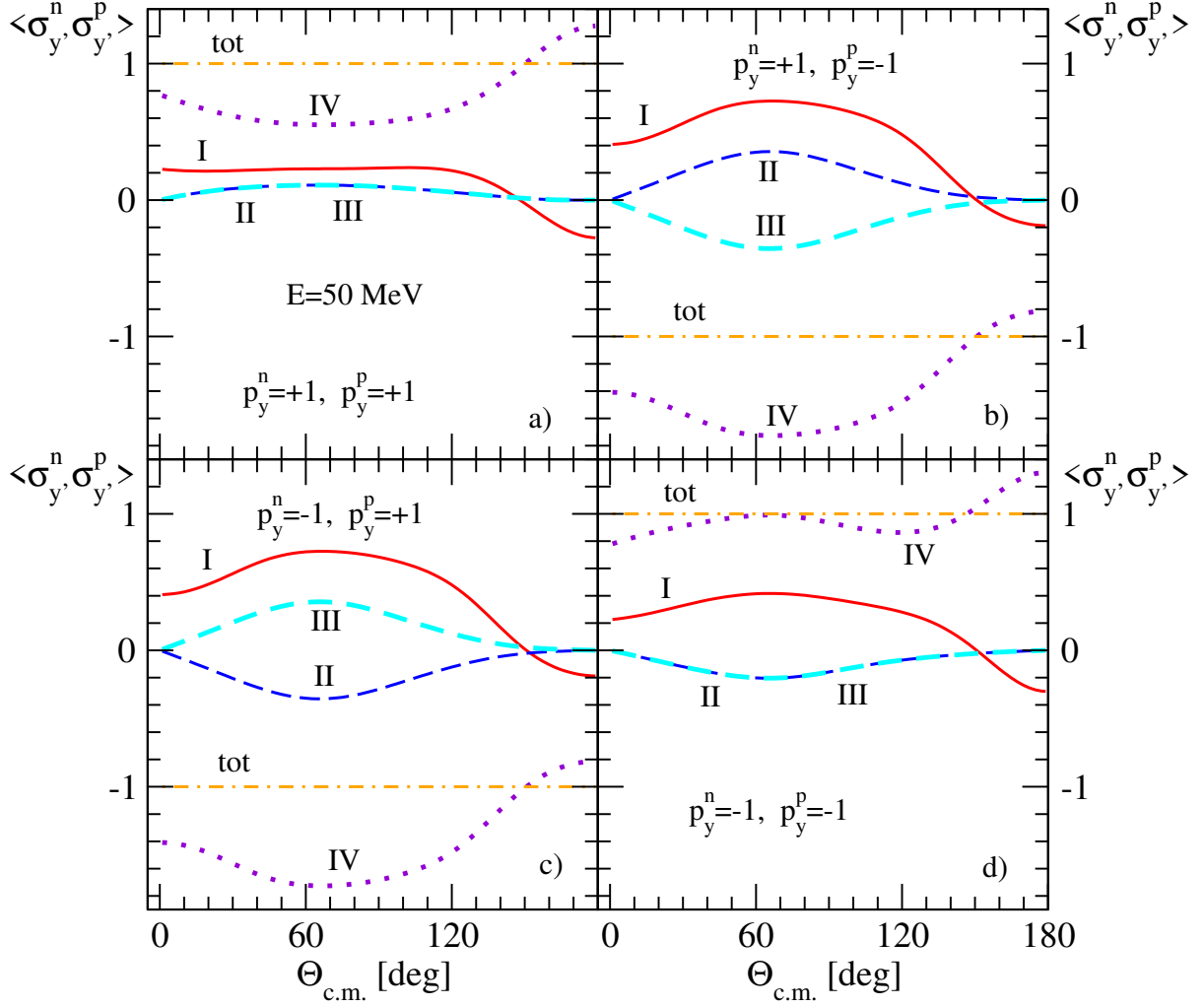}  
% Here is how to import EPS art
\caption{(color online) The correlation between 
    y-components of the outgoing
    neutron and proton spins
 $\langle \sigma_{y'}^n \sigma_{y'}^p\rangle $ ((orange) dashed-dotted line) 
    together with its forming terms 
    in the $\vec p (\vec n,\vec n)\vec p$ reaction at the incoming neutron
    lab. energy $E=50$~MeV with  incoming neutron
    and proton polarizations: a) $p_y^n=+1, p_y^p=+1$, 
    b) $p_y^n=+1$, $p_y^p=-1$, c) $p_y^n=-1$, $p_y^p=+1$,
    d) $p_y^n=-1$, $p_y^p=-1$. Contributions of each of the four terms in
    (\ref{eqqq_2}) are denoted as: I - the first term with
    $\langle \sigma_{y'}^n \sigma_{y'}^p\rangle ^{(0)}$ ((red) solid line),
    II - the second term with $K_{0,y}^{y',y'}$ ((blue) long dashed line),
    III - the third term with
    $K_{y,0}^{y',y'}$ ((cyan) short dashed line),
    IV - the fourth term with $K_{y,y}^{y',y'}$ ((violet) dotted line).
}
 \label{fig12}
\end{figure}

For the special case of four combinations of
 the neutron and proton polarizations
 $p_y^n=p_y^p=\pm 1$ we illustrate the importance of each of the
  four contributing terms
 in forming the final neutron polarization
$\langle \sigma_y^n\rangle $ of (\ref{eqqq_1}) (Fig.~\ref{fig11}) and 
 the final neutron-proton
 spin correlation $\langle \sigma_y^n \sigma_y^p\rangle $ of (\ref{eqqq_2})
 (Fig.~\ref{fig12}). 
 The most important contribution to the creation of
 $\langle \sigma_y^n\rangle $ seems
to be the polarization transfer from the incoming neutron to the final neutron. 
The final neutron-proton
spin correlation $\langle \sigma_y^n \sigma_y^p\rangle $ stems predominantly
from
the spin correlation transfer from doubly polarized initial state.
However, this changes and the picture becomes more complicated
when the values of the initial neutron and proton polarizations deviate
from the maximal values of $\pm 1$.

\begin{figure}
\begin{center}
\begin{tabular}{c}
%\resizebox{92mm}{!}{\includegraphics[angle=270]{gnu_entangl_power_e10p0_eps_fig13a.eps}} \\
%\resizebox{92mm}{!}{\includegraphics[angle=270]{gnu_entangl_power_e10p0_conc_fig13b.eps}} \\
%\resizebox{92mm}{!}{\includegraphics[angle=270]{gnu_entangl_power_e10p0_conc2_fig13c.eps}} \\
\resizebox{86mm}{!}{\includegraphics[angle=270]{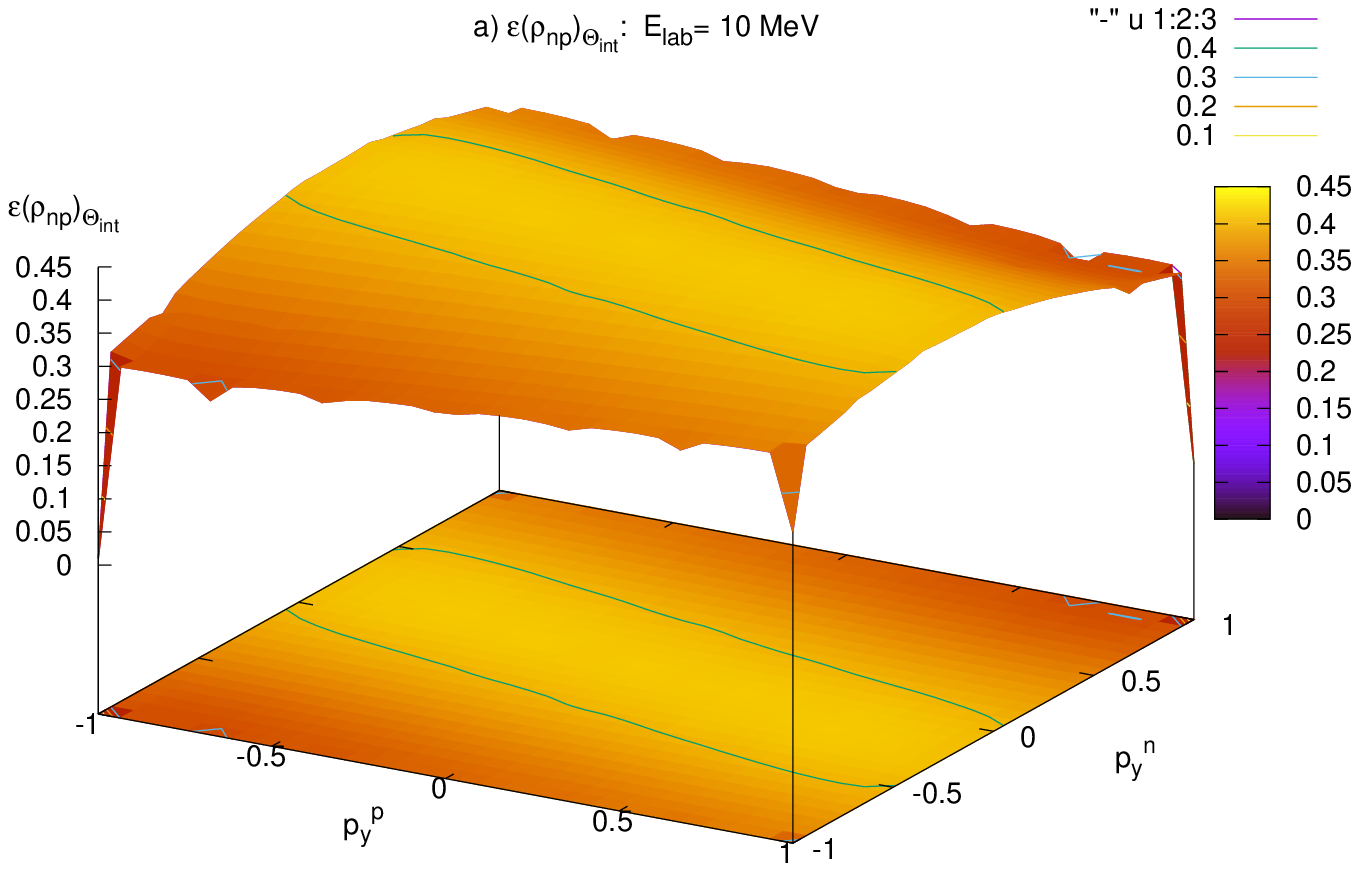}} \\
\resizebox{86mm}{!}{\includegraphics[angle=270]{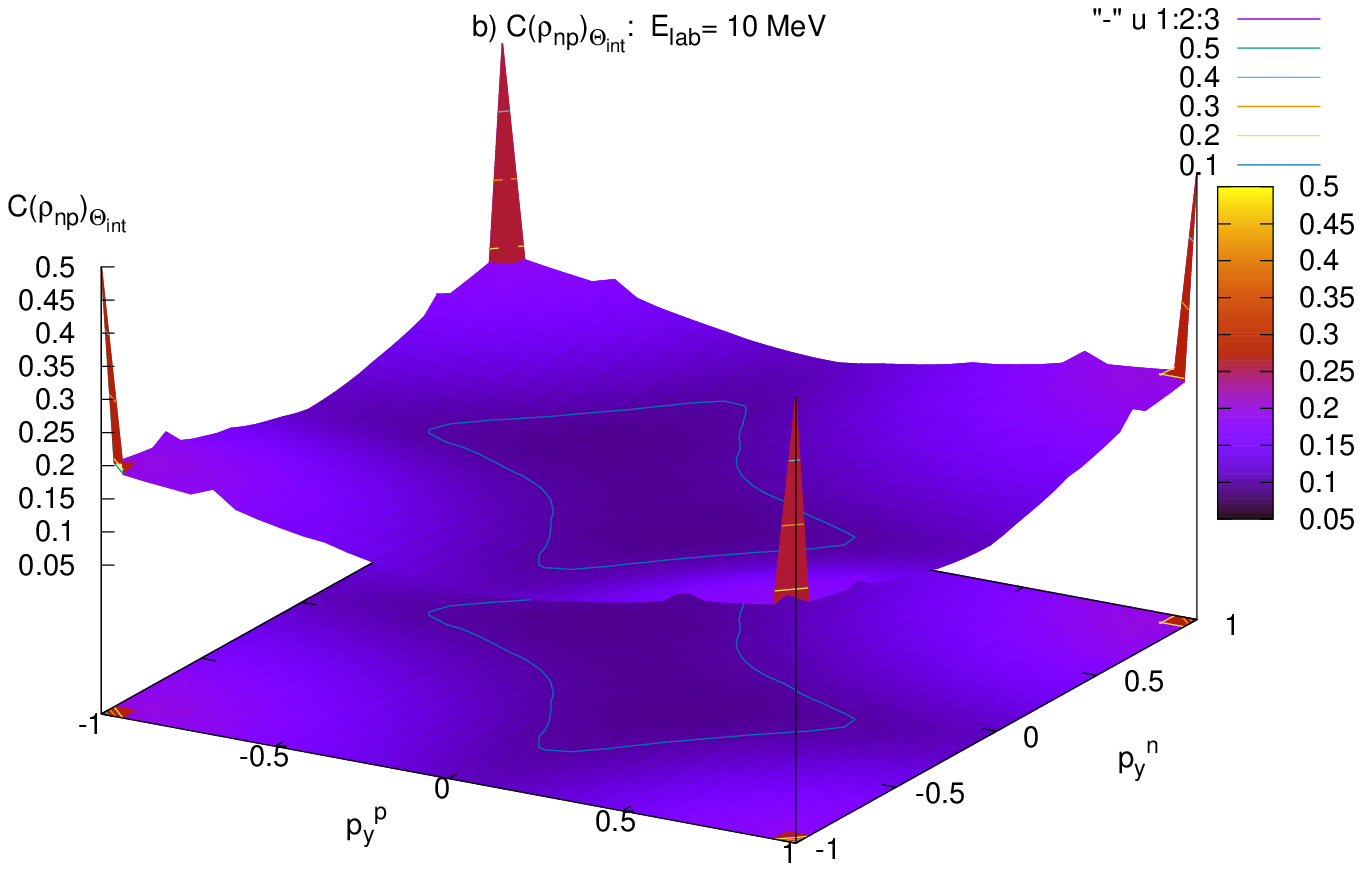}} \\
\resizebox{86mm}{!}{\includegraphics[angle=270]{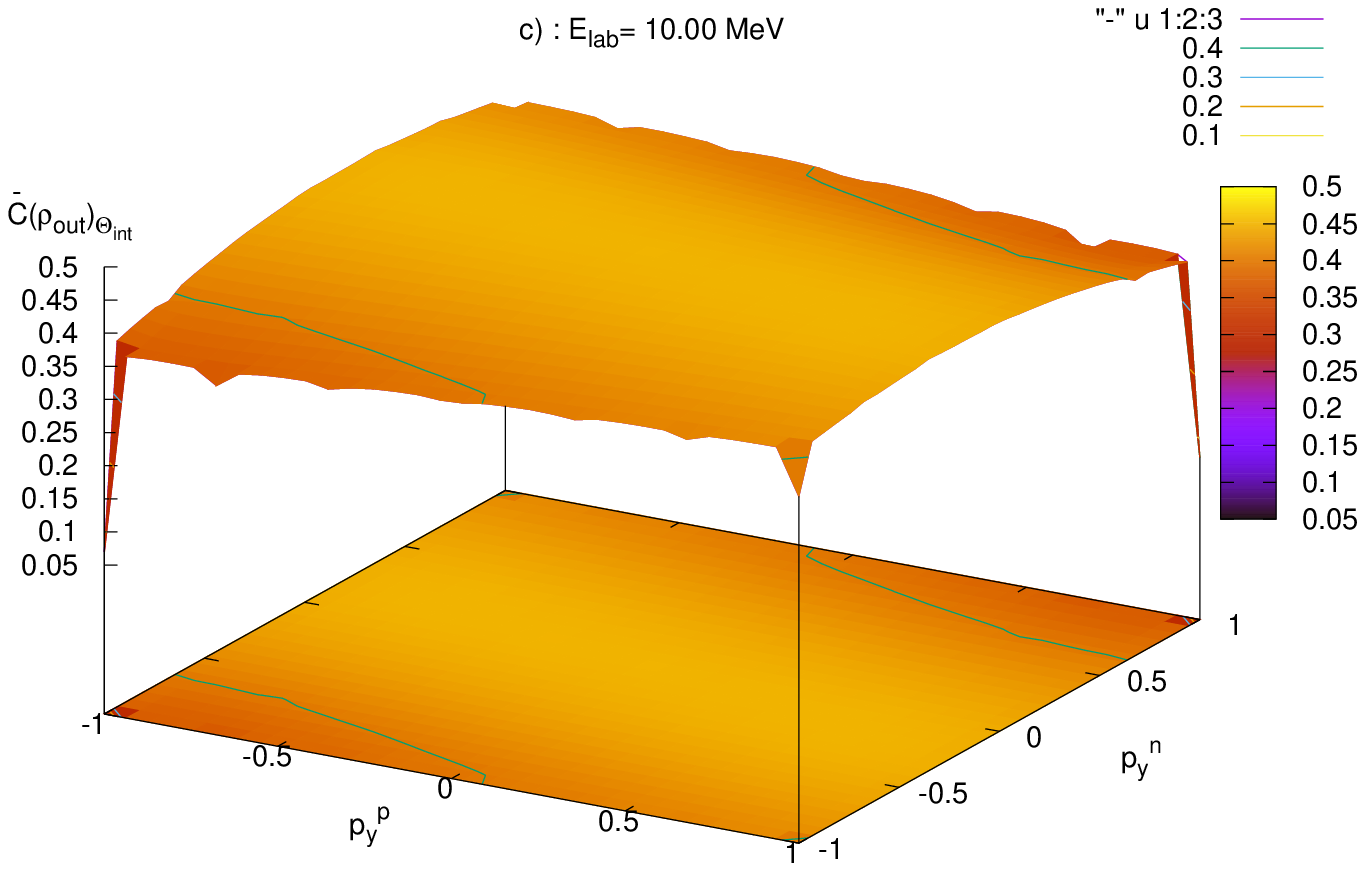}} \\
\end{tabular}%
\caption{
  (color online)
  The entanglement power $\epsilon(\rho_{np})_{\Theta_{int}}$ a), 
  concurrence $C(\rho_{np})_{\Theta_{int}}$ b) and
  ${\bar C}(\rho_{np})_{\Theta_{int}}$ c),
  averaged over the c.m. angle
  $\Theta_{c.m.}$, are shown for the final spin states
  in $\vec p(\vec n,\vec n)\vec p$
  scattering  for $E_{lab}=10$~MeV. All quantities are presented as 
  functions of the polarizations of the incoming neutron $p_y^n$ and
  proton $p_y^p$ and were calculated using the AV18 potential.}
\label{fig13}
\end{center}
\end{figure}

\begin{figure}
\begin{center}
\begin{tabular}{c}
%\resizebox{100mm}{!}{\includegraphics[angle=270]{gnu_entangl_power_e50p0_eps_fig14a.eps}} \\
%\resizebox{100mm}{!}{\includegraphics[angle=270]{gnu_entangl_power_e50p0_conc_fig14b.eps}} \\
%\resizebox{100mm}{!}{\includegraphics[angle=270]{gnu_entangl_power_e50p0_conc2_fig14c.eps}} \\
\resizebox{97mm}{!}{\includegraphics[angle=270]{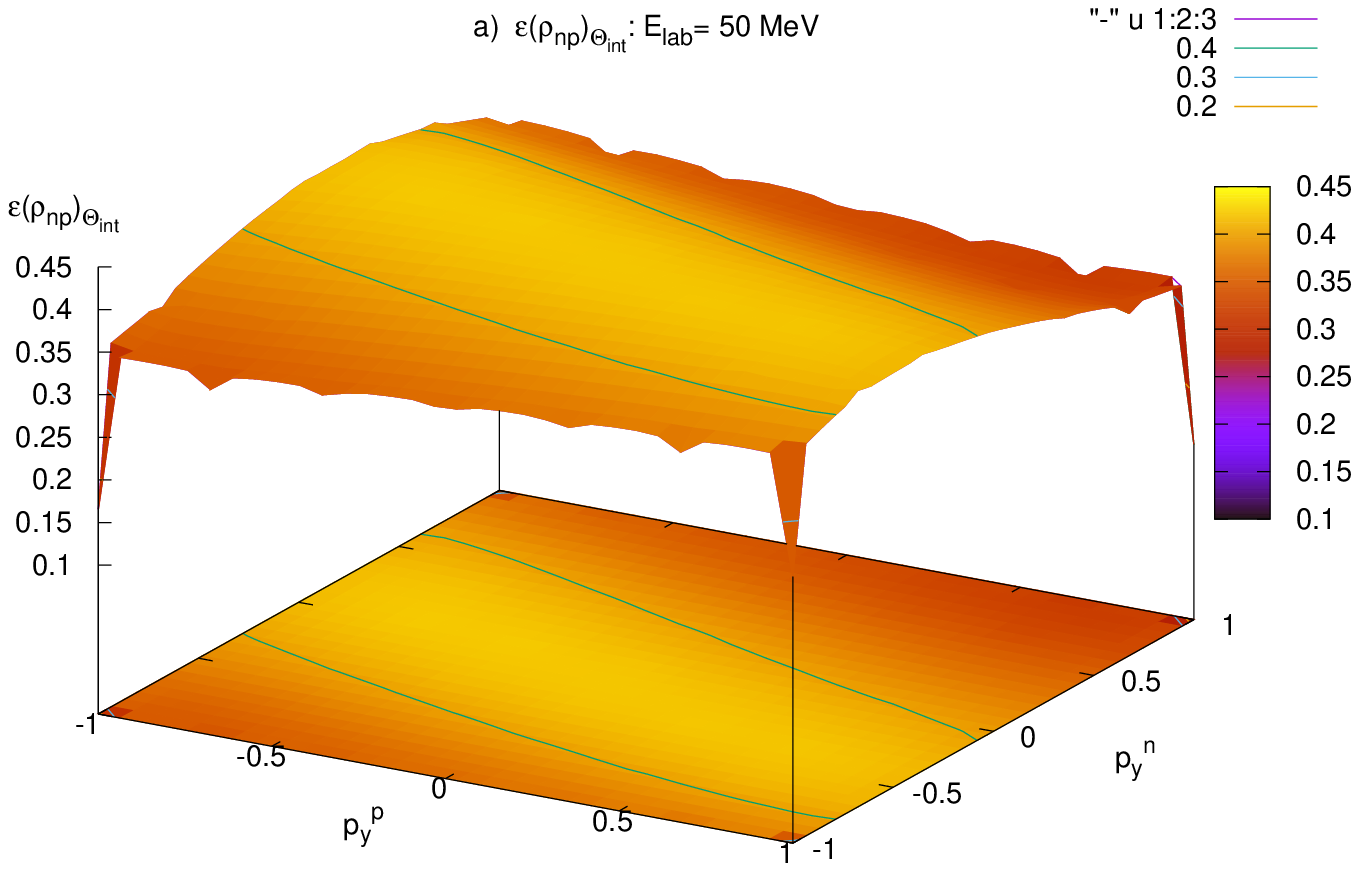}} \\
\resizebox{97mm}{!}{\includegraphics[angle=270]{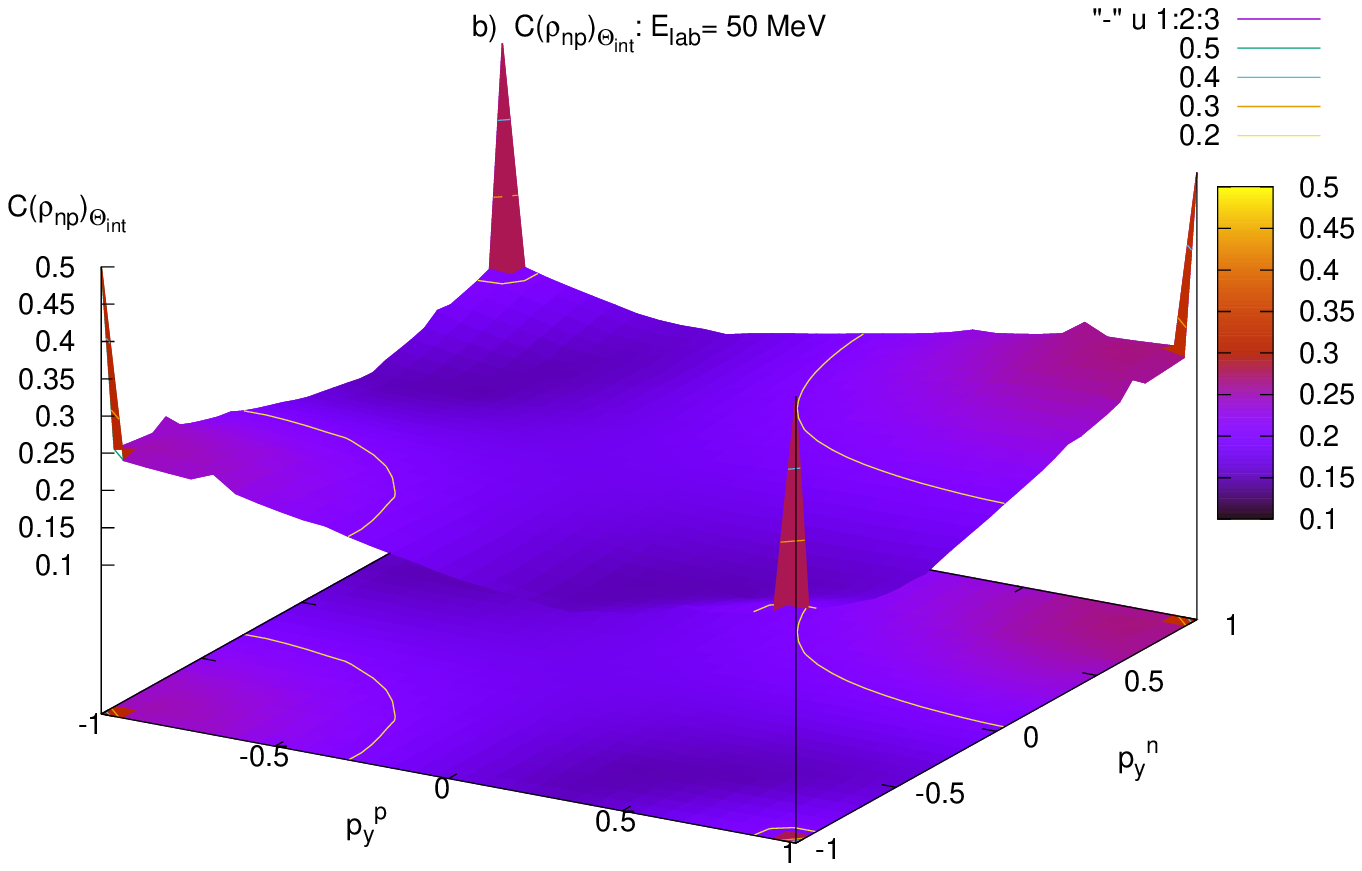}} \\
\resizebox{97mm}{!}{\includegraphics[angle=270]{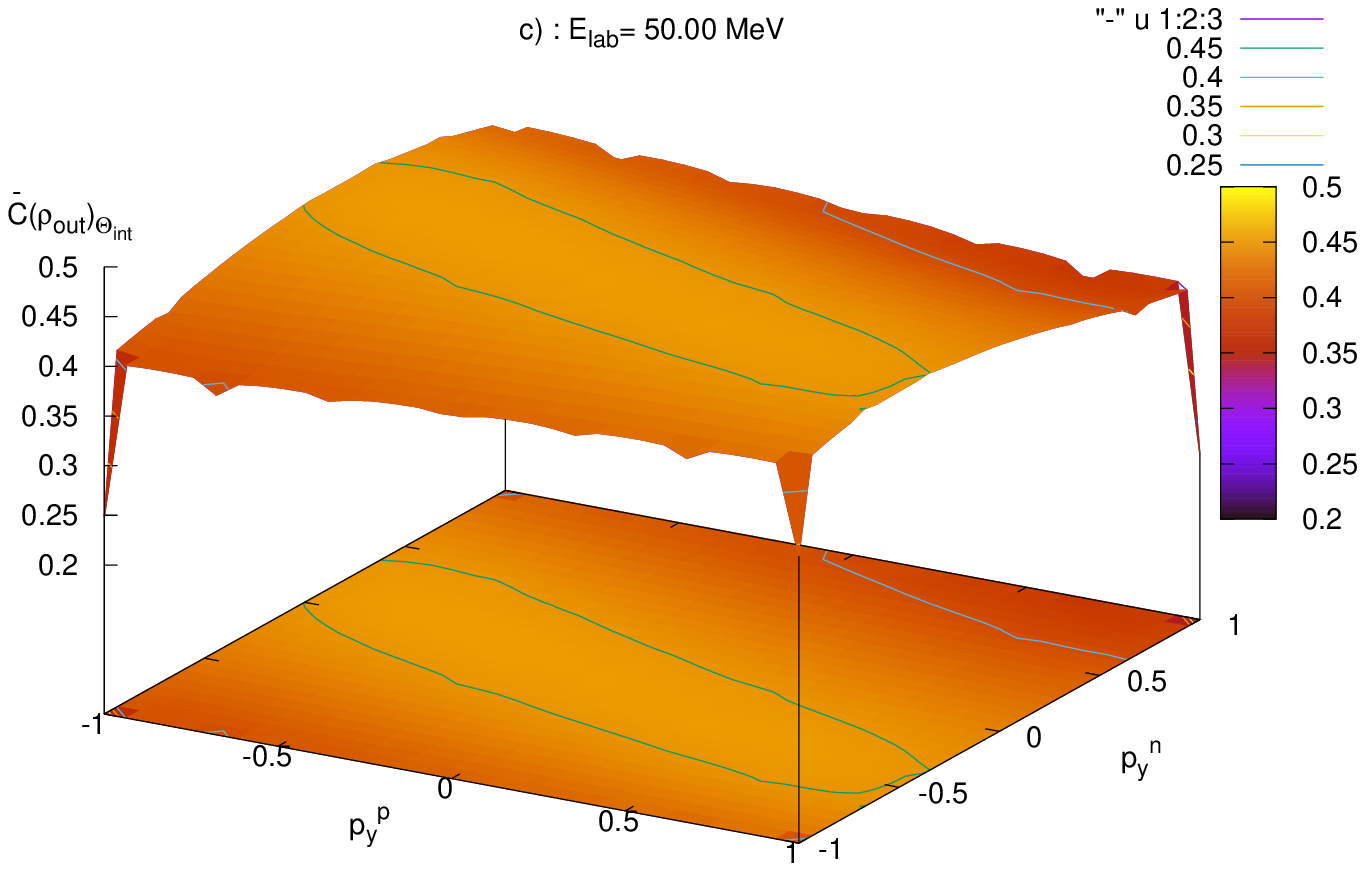}} \\
\end{tabular}%
\caption{
  (color online)
  Same as in Fig.\ref{fig13}, but for $E_{lab}=50$~MeV.}
\label{fig14}
\end{center}
\end{figure}

\begin{figure}
\begin{center}
\begin{tabular}{c}
%\resizebox{100mm}{!}{\includegraphics[angle=270]{gnu_entangl_power_e100p0_eps_fig15a.eps}} \\
%\resizebox{100mm}{!}{\includegraphics[angle=270]{gnu_entangl_power_e100p0_conc_fig15b.eps}} \\
%\resizebox{100mm}{!}{\includegraphics[angle=270]{gnu_entangl_power_e100p0_conc2_fig15c.eps}} \\
\resizebox{97mm}{!}{\includegraphics[angle=270]{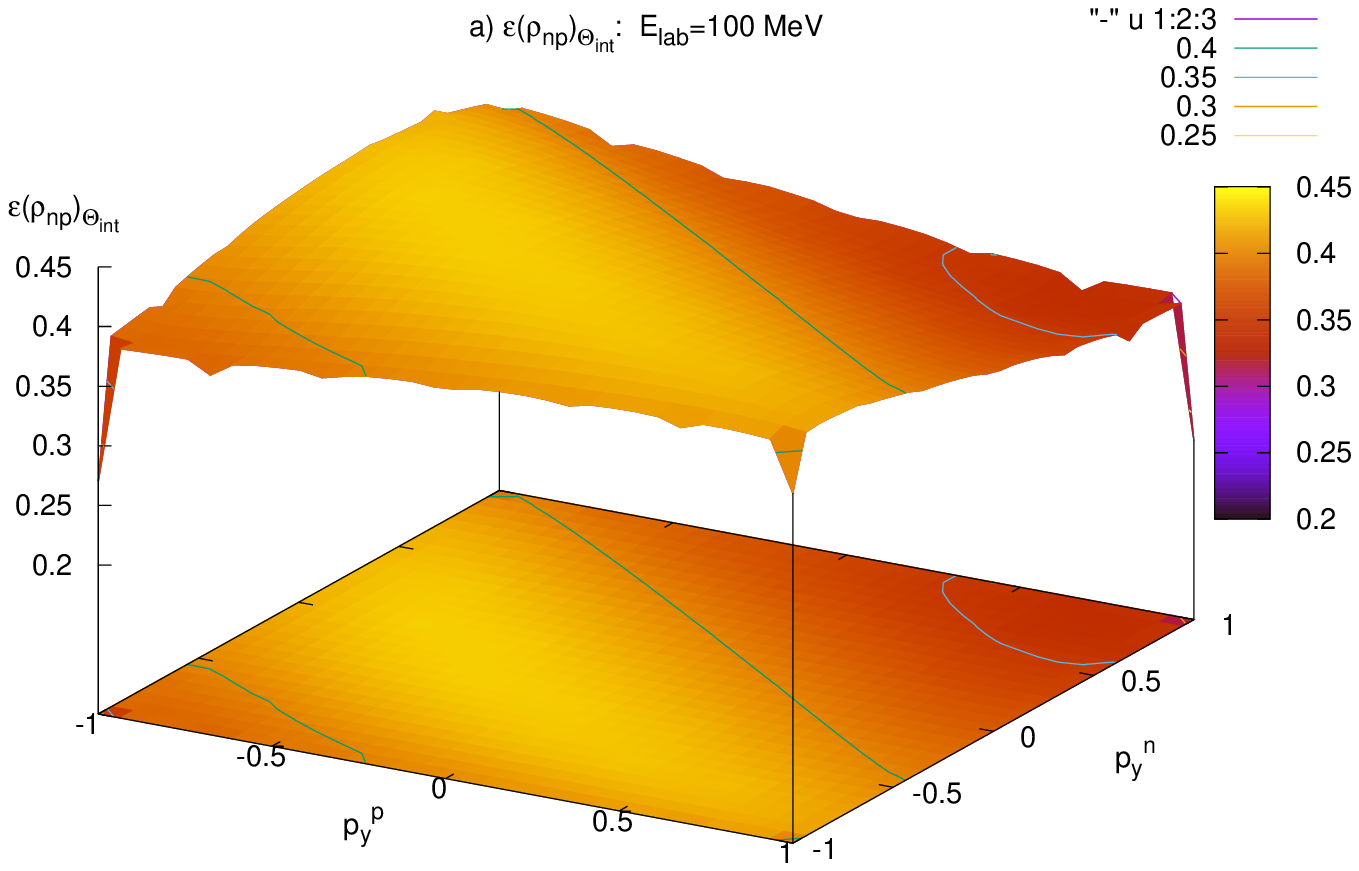}} \\
\resizebox{97mm}{!}{\includegraphics[angle=270]{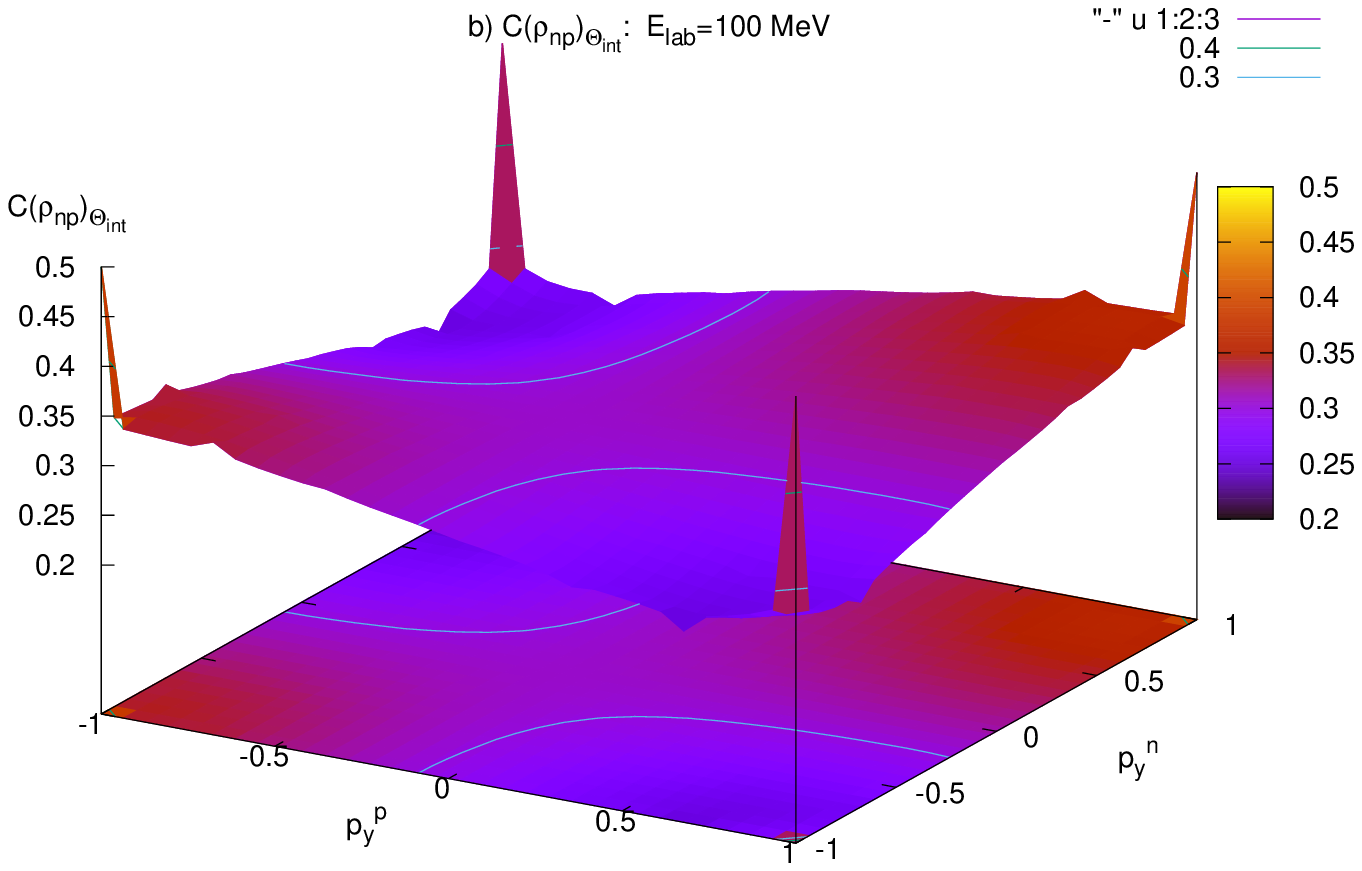}} \\
\resizebox{97mm}{!}{\includegraphics[angle=270]{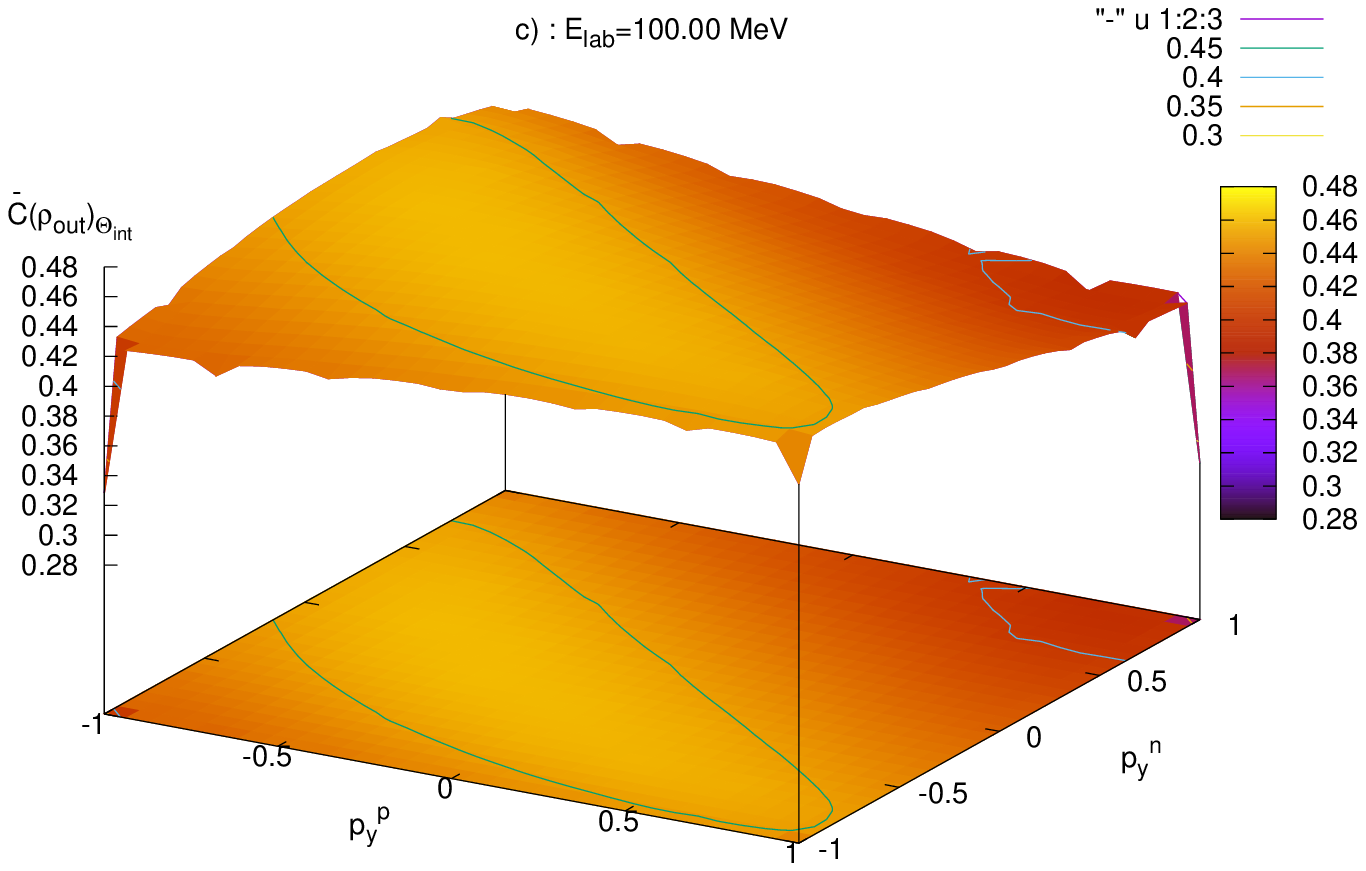}} \\
\end{tabular}%
\caption{
  (color online)
  Same as in Fig.\ref{fig13}, but for $E_{lab}=100$~MeV.}
\label{fig15}
\end{center}
\end{figure}

In spite of suggestions about the lack of entanglement in impure
    final $np$ spin states, stemming from direct comparisons of their
    spin polarizations and correlations to those of pure entangled states,
    we checked their quantitative degree of entanglement.
    To that end, we calculated, according to Eqs.~(\ref{a_eq1})-(\ref{a_eq3}),
    their entanglement power
    and concurrences, averaged over scattering angles.
    The results at $E_{lab} = 10$, $50$, and $100$~MeV are shown in
    Figs.\ref{fig13}, \ref{fig14}, and \ref{fig15}, respectively.

    The entanglement power reaches a value of
    $\epsilon(\rho_{np})_{\Theta_{int}} = 0.45$ at all energies
    (see Figs.\ref{fig13}a--\ref{fig15}a), which is close to
    the maximal possible value of $0.5$.
    At $E_{lab} = 10$~MeV (see Fig.\ref{fig13}a), it occurs in a relatively
    narrow region of incoming neutron polarizations, which clearly broadens
    with increasing energy (see Figs.\ref{fig14}a and \ref{fig15}a).
    It is interesting to note that the minimal values of the entanglement
    power for the impure final states also grow with energy, reaching
    approximately $0.3$, $0.35$, and $0.4$ at $E_{lab} = 10$, $50$,
    and $100$~MeV, respectively.
    Both facts indicate a growing entanglement of impure final states
    with increasing energy.
 
Similar conclusions can be drawn from the behavior of the concurrences
$C(\rho_{np})_{\Theta_{int}}$ and ${\bar C}(\rho_{np})_{\Theta_{int}}$
(Figs.~\ref{fig13}b,c--\ref{fig15}b,c). However, for
$C(\rho_{np})_{\Theta_{int}}$,
the maximal values achieved are still farther from the maximal
value of $0.5$ typical for pure entangled states.

At $E_{lab} = 10$~MeV (see Fig.\ref{fig13}b), the concurrence
$C(\rho_{np})_{\Theta_{int}}$ lies approximately in the range $0.05 - 0.2$,
increasing at $E_{lab} = 50$~MeV (see Fig.\ref{fig14}b) to about $0.1 - 0.25$,
and reaching at $E_{lab} = 100$~MeV (see Fig.\ref{fig15}b) roughly $0.2 - 0.35$.
This increase in the concurrence $C(\rho_{np})_{\Theta_{int}}$ again
indicates growing entanglement for statistical mixtures of $np$ final
states with increasing energy.

The results for ${\bar C}(\rho_{np})_{\Theta_{int}}$ correlate well with
those for the entanglement power.

%\begin{sidewaystable}[htbp]
\begin{table}[ht!]
\centering
\caption{Values of the amplitudes $\alpha_{m_nm_p}$  and their contributions to
  the norm, $|\alpha_{m_nm_p}|^2$, as defined in Eq.~(\ref{eq13_a0}),
  for the outgoing $np$ pure states at energy $E$ and
  c.m. angle $\Theta_{c.m.}$ corresponding to the angular regions where maxima
  of the entanglement power and concurrence are observed in Fig.~\ref{fig17}.
  These final states originate from pure initial states with maximal neutron
  and proton polarizations. 
  The specific polarization combination $(p_y^n, p_y^p)$ is indicated by
  the superscript on the energy value in the first column:
  a) $(+1,+1)$, b) $(+1,-1)$, c) $(-1,+1)$, and d) $(-1,-1)$.
  The corresponding type of Bell state [as defined in Eq.~(\ref{eq17_a0})]
  is listed in the last column. 
}
\fontsize{9.0pt}{11pt}
\selectfont
\label{tab3}
\begin{tabular}{ |c|c|c|c|c|c|c|c|c|c| }
\hline
No & E$~^{a)}$  & $\Theta_{c.m.}$ & $\alpha_{+\frac {1} {2} +\frac {1} {2}  }$
& $\alpha_{+\frac {1} {2} -\frac {1} {2} } $
& $\alpha_{-\frac {1} {2} +\frac {1} {2}  } $
& $\alpha_{-\frac {1} {2} -\frac {1} {2} }$
& $|\alpha_{+\frac {1} {2} +\frac {1} {2} }|^2$
& $|\alpha_{+\frac {1} {2} -\frac {1} {2} }|^2$
&  Bell  \\ 
& MeV & deg & & &   &
& $|\alpha_{-\frac {1} {2} -\frac {1} {2} }|^2$  &
$|\alpha_{-\frac {1} {2} +\frac {1} {2} }|^2$ &type \\
\hline
1 & 100$~^{a)} $ & $ 98.5  $ & $ +0.325 +i 0.126 $  & $ -0.561 +i 0.251 $
& $ -0.561 +i 0.251 $
&  $ -0.325 -i 0.126 $ & $ 0.122 $  &$ 0.378 $ &  $| \psi_{III} \rangle$ \\
\hline
2 & 100$~^{b)} $ &  $17.0$ & $-0.475 -i 0.459$  & 
$ -0.221 - i 0.120 $ &  $ +0.221 + i 0.120 $  & 
$ -0.475 -i 0.459 $  & $ 0.437 $ & $ 0.063  $  & $| \psi_{I} \rangle$ \\
\hline
3 & 100$~^{b)} $ &  $138.5 $ & $ -0.053 -i 0.256 $  & 
$ -0.044 +i 0.656 $ &  $ +0.044 -i 0.656 $  & 
$ -0.053 -i 0.256 $  & $ 0.068 $   & $ 0.432 $ &  $| \psi_{IV} \rangle$ \\
\hline
4 & 100$~^{c)} $ &  $17.0$ & $-0.475 -i 0.459$  & 
$ +0.221 + i 0.120 $ &  $ -0.221 - i 0.120 $  & 
$ -0.475 -i 0.459 $  & $ 0.437 $ & $ 0.063  $ & $| \psi_{I} \rangle$ \\
\hline
5 & 100$~^{c)} $ &  $ 138.5 $ & $ -0.053 -i 0.256 $  & 
$ +0.044 -i 0.656 $ &  $ -0.044 +i 0.656 $  & 
$ -0.053 -i 0.256 $  & $ 0.068 $   &$ 0.432 $  & $| \psi_{IV} \rangle$ \\
\hline
6 & 100$~^{d)} $ & $ 143.0  $ & $ +0.602 +i 0.360 $ &
$ -0.012 -i 0.091  $ &  $ -0.012 -i 0.091  $
&  $ -0.602 -i 0.360  $ & $ 0.492 $
& $  0.009 $ & $| \psi_{II} \rangle$ \\
\hline
7 & 50$~^{a)} $ & $ 106.5  $ & $ -0.268 -i 0.242 $ &
$ -0.584 +i 0.167 $ &  $ -0.584 +i 0.167 $
&  $ +0.268 +i 0.242 $ & $ 0.131 $  & $ 0.370 $ & $| \psi_{III} \rangle$ \\
\hline
8 & 50$~^{b)} $ &  $0.0$ & $-0.378 -i 0.517 $  & 
$ -0.300 - i 0.008 $ &  $ +0.300 + i 0.008 $
& 
$ -0.378 -i 0.517 $  & $ 0.410 $   &$ 0.090  $ &  $| \psi_{I} \rangle$ \\
\hline
9 & 50$~^{b)} $ &  -  &   &   &  &  & $ 0.175 $
& $ 0.325 $ & \\
\hline
10 & 50$~^{c)} $ &  $0.0$ & $ -0.378 -i 0.517 $  & 
$ +0.300 + i 0.008 $ &  $ -0.300 - i 0.008 $
& 
$ -0.378 -i 0.517 $  & $ 0.410 $   & $ 0.090  $  & $| \psi_{I} \rangle$ \\
\hline
11 & 50$~^{c)} $ &  - &    &  &   & 
  & $ 0.175 $   & $ 0.325 $  &  \\
\hline
12 & 50$~^{d)} $ & $ 123.5  $ & $ +0.379 +i 0.561 $ &
$ +0.175 -i 0.103  $ &  $ +0.175 -i 0.103  $
&  $ -0.379 -i 0.561  $ & $ 0.459 $
&$  0.041 $ &  $| \psi_{II} \rangle$ \\
\hline
\end{tabular}
\end{table}
%\end{sidewaystable}

\begin{figure}
\includegraphics[scale=0.79]{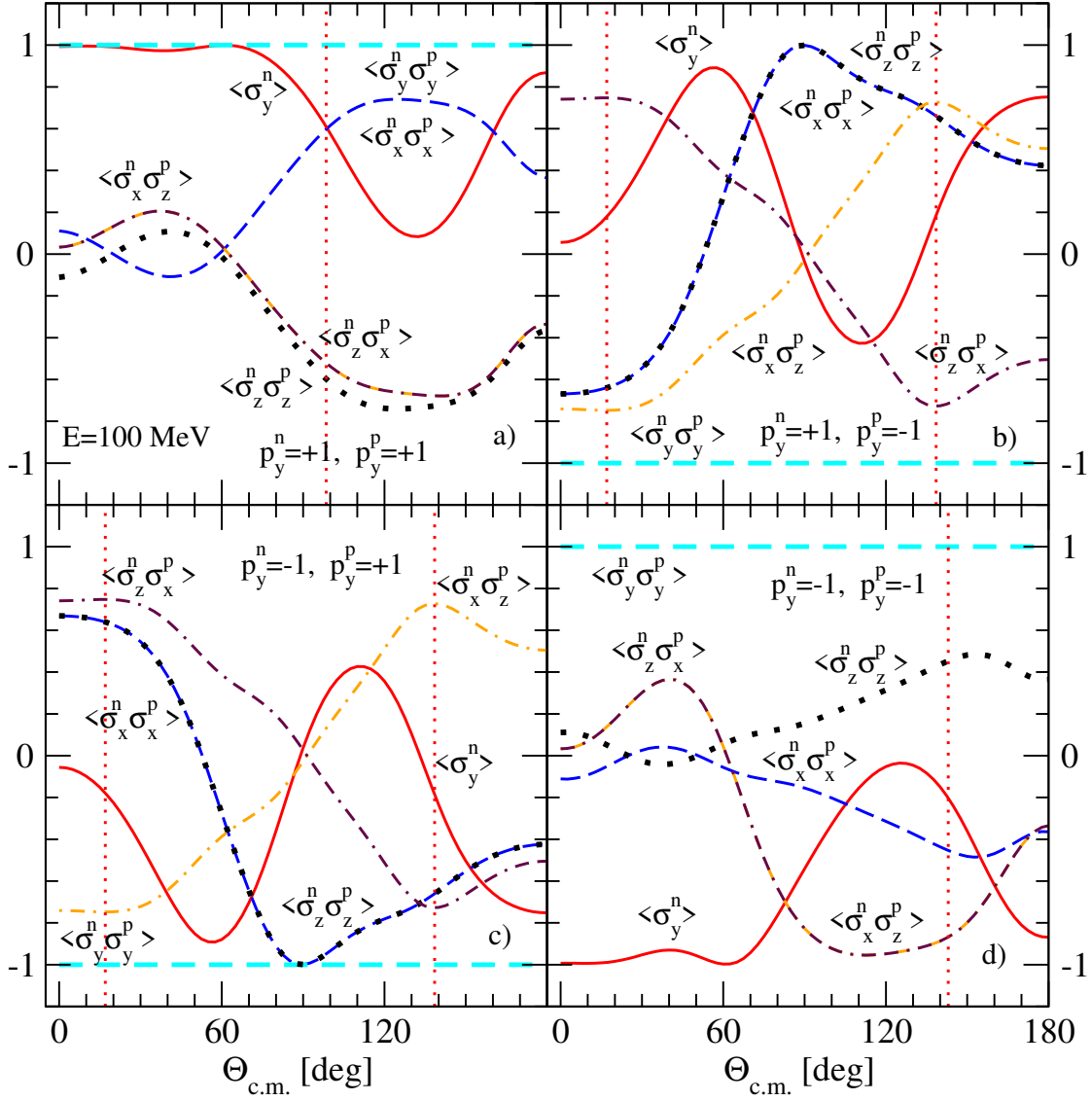}  
% Here is how to import EPS art
\caption{(color online)
  Same as in Fig.\ref{fig10}, but for $E_{lab}=100$~MeV.
  Vertical (red) dotted lines indicate the positions of entangled states
  listed in Table~\ref{tab3}.
 }
 \label{fig16}
\end{figure}

It is remarkable that for the only pure final states arising from pure
cross product incoming $np$ states with maximal neutron and proton
polarizations $p_y^n, p_y^p = \pm 1$, the angle-integrated concurrence
$C(\rho_{np})_{\Theta_{int}}$ reaches the maximal value of $0.5$ at
each energy (see Figs.\ref{fig13}b--\ref{fig15}b).
This is a consequence of the fact that the spin correlation
$\langle \sigma_y^n \sigma_y^p \rangle$ for these states is independent
of both energy and angle, and takes a constant value of $\pm 1$
(see Figs.\ref{fig10} and \ref{fig16} for $E = 50$ and $100$~MeV, respectively).
 It is worth noting here that this behavior of the spin correlation
$\langle \sigma_y^n \sigma_y^p \rangle$ results from the interplay
between different components in Eq.~(\ref{eqqq_2}), which defines
this correlation.
Namely, direct algebraic calculation yields:
$K_{y,y}^{y',y'} = 1$,
$\langle \sigma_{y'}^n \sigma_{y'}^p \rangle^{(0)} = C_{y,y}$,
$K_{0,y}^{y',y'} = A_y^n$, and
$K_{y,0}^{y',y'} = A_y^p$.
Therefore, for any values of $p_y^n$ and $p_y^p$, one obtains:
\begin{eqnarray}
\langle \sigma_{y'}^n \sigma_{y'}^p\rangle &=& \frac
{ C_{y,y}
+ p_y^n A_y^n
+ p_y^p A_y^p
+ p_y^n p_y^p } {1 + p_y^n A_y^n
+ p_y^p A_y^p
+ p_y^n p_y^p C_{y,y} } ~.
\label{a_eq4}
\end{eqnarray}
It follows that for our pure final states,
$\langle \sigma_{y'}^n \sigma_{y'}^p \rangle = +1$
for $p_y^n = p_y^p = \pm 1$, and
$\langle \sigma_{y'}^n \sigma_{y'}^p \rangle = -1$
for $p_y^n = -p_y^p = +1$ or $-1$, independently of energy and angle. 
In contrast, the angle-integrated entanglement power
 $\epsilon(\rho_{np})_{\Theta_{int}}$ and concurrence
 ${\bar C}(\rho_{np})_{\Theta_{int}}$ for these four pure states vary with
the state and increase with rising energy
(see Figs.~\ref{fig13}a,c--\ref{fig15}a,c).
This suggests that these quantities depend on angle, or that the
entanglement in these states increases with energy.

\begin{figure}
\includegraphics[scale=0.65]{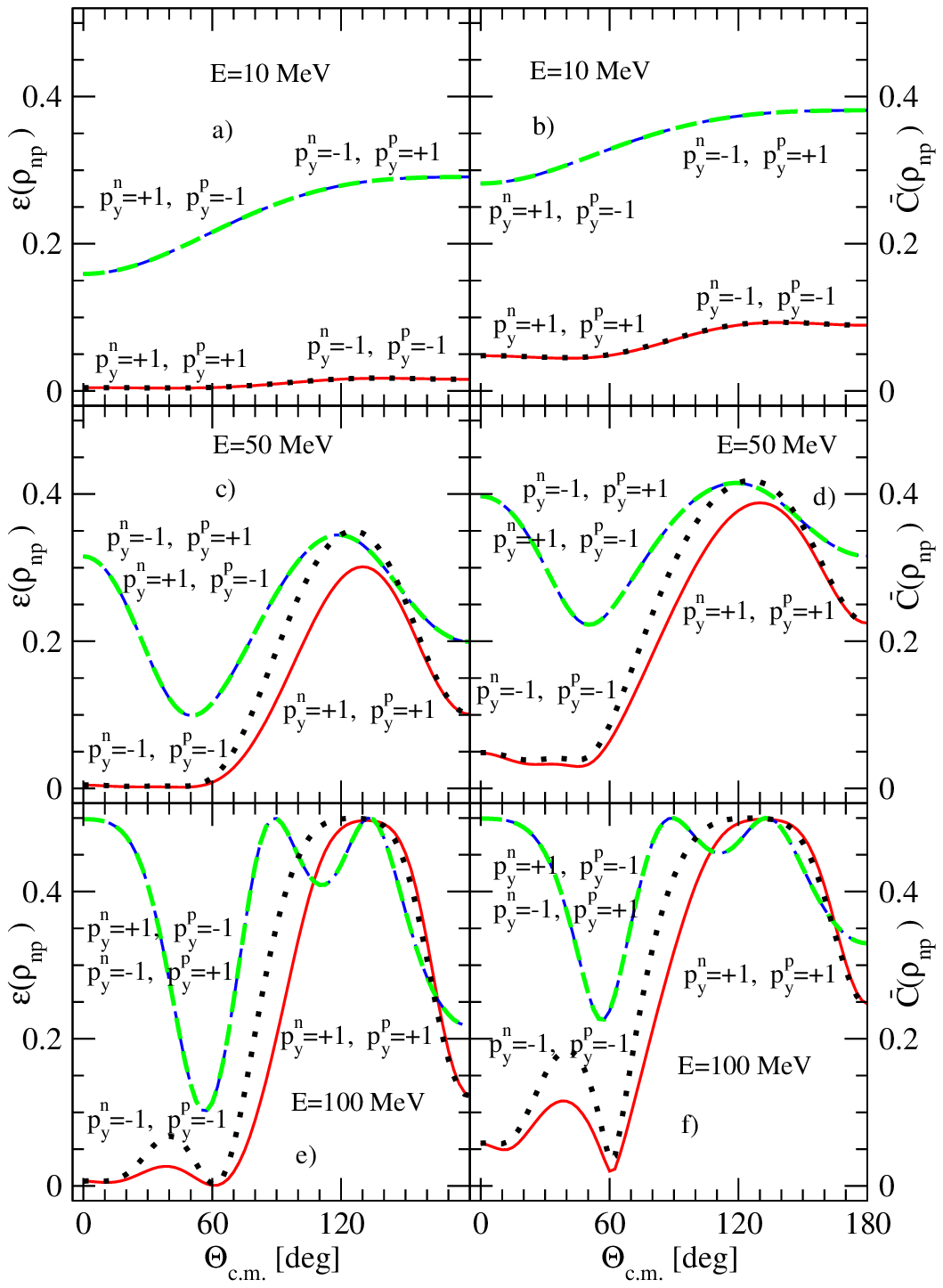}  
% Here is how to import EPS art
\caption{(color online)
  Angular distributions of the entanglement power $\epsilon(\rho_{np})$
   and the concurrence ${\bar C}(\rho_{np})$   
    in the $\vec p (\vec n,\vec n)\vec p$ scattering at incoming neutron
    lab. energies $E=10$~MeV a), b),  $E=50$~MeV c), d), and $E=100$~MeV e), f),
    for the pure final $np$ states which originate from pure cross product
    initial states with maximal neutron
    and proton polarizations:
    $p_y^n=+1, p_y^p=+1$ (red) solid line, 
    $p_y^n=+1$, $p_y^p=-1$ (blue) long dashed line, 
    $p_y^n=-1$, $p_y^p=+1$, (green) short dashed line, 
    and $p_y^n=-1$, $p_y^p=-1$ (black) dotted line. 
    The predictions were obtained using the AV18 potential.
}
 \label{fig17}
\end{figure}

In Fig.~\ref{fig17}, we show the angular distributions of the
entanglement power $\epsilon(\rho_{np})$ and concurrence ${\bar C}(\rho_{np})$
at three energies for each of
the final $np$ pure states.
At $E = 10$~MeV, both the entanglement power and concurrence are smooth
and close to zero for the states arising from incoming configurations
with the same signs of neutron and proton polarizations
(see Figs.\ref{fig17}a) and b)).
For the remaining two states, both quantities are larger but still smooth,
remaining below $0.3$ and $0.4$, respectively.

At higher energies, $E = 50$~MeV (Figs.\ref{fig17}c) and d))
and $E = 100$~MeV (Figs.\ref{fig17}e) and f)), the entanglement power
and concurrence strongly depend on scattering angle and exhibit clear
 maxima and minima at specific c.m. angles
for each combination of neutron and proton polarizations.
At $E = 50$~MeV, the maxima reach approximately $0.3 - 0.4$,
while at $E = 100$~MeV, they reach the maximal allowed value of $0.5$.

To check whether the states at the angles corresponding to these maxima
indeed exhibit strong entanglement, we calculated, according
to Eq.(\ref{new6}), the four coefficients $\alpha_{m_n m_p}$ specifying
the state, as well as their contributions to the norm $|\alpha_{m_n m_p}|^2$,
for each neutron and proton spin projection $(m_n, m_p)$.
We searched for angles at which the smaller contributions
$|\alpha_{m_n m_p}|^2$ were minimized.
The results are summarized in Table~\ref{tab3}.

Indeed, at $E = 100$~MeV, we found strongly entangled pure states in
the angular regions where maxima of the entanglement power and concurrence
occur, for each combination of neutron and proton spin projections.
These states cover all types of Bell states specified in (\ref{eq17_a0})
and contain only small contributions from entanglement-destroying components.

For the combination of incoming polarizations $(p_y^n, p_y^p) = (+1, +1)$,
we identified only one such state at $\Theta_{c.m.} = 98.5^{\circ}$,
resembling the Bell $| \psi_{III} \rangle$ state, though with a rather
large admixture of entanglement-spoiling components, contributing
approximately $24\%$ to the norm (Table~\ref{tab3}, state $No~1$).

Similarly, for $(p_y^n, p_y^p) = (-1, -1)$, we found one state
at $\Theta_{c.m.} = 143.0^{\circ}$ of the Bell $| \psi_{II} \rangle$
type, with small entanglement-breaking contributions below
approximately $2\%$ (Table~\ref{tab3}, state $No~6$).
 That is the most entangled state we have found. 

For $(p_y^n, p_y^p) = (+1, -1)$ and $(-1, +1)$, we encountered two states:
one at $\Theta_{c.m.} = 17.0^{\circ}$ (states $No~2$ and $4$ in Table~\ref{tab3}),
and another at $\Theta_{c.m.} = 138.5^{\circ}$ (states $No~3$ and $5$),
corresponding to the Bell $| \psi_{I} \rangle$ and $| \psi_{IV} \rangle$
types, respectively.
Both exhibit entanglement-breaking contributions below approximately $13\%$.

The difference between the entangled states for $(p_y^n, p_y^p) = (+1, -1)$
and $(-1, +1)$ lies in the opposite signs of their entanglement-spoiling
components.

As the energy decreases, both the entanglement power and concurrence
diminish (see Fig.\ref{fig17}), which is reflected by an increase in
the entanglement-destroying contributions.
At $E = 50$~MeV, despite a similar structure of maxima and minima in
the entanglement power and concurrence as at $E = 100$~MeV, we found
only four states out of the six identified at $100$~MeV, with shifted
angular positions (states $No~7$, $8$, $10$, and $12$ in Table~\ref{tab3}).
It was no longer possible to locate two states ($No~9$ and $11$,
counterparts of states $No~3$ and $5$ at $100$~MeV), due to
the excessively large contributions (approximately $35\%$) from
the smaller amplitudes in this angular region.

At $E = 10$~MeV, the different coefficients $\alpha_{m_n m_p}$ contribute
nearly equally at all angles (around $25\%$ each), making it impossible
to identify any trace of entanglement in the outgoing pure final states.

\section{Summary and conclusions}
\label{sumary}

We investigated  polarization states of the outgoing
neutron-proton pair in elastic $np$ scattering, looking for signatures
of entanglement. Since these spin states are specified unequivocally
by the values
of the final neutron and proton polarizations and their spin correlations, 
we computed  all contributions to these quantities, starting
from the component induced
in unpolarized $np$ scattering, followed by two pieces
due to the single transfers either from a polarized neutron or proton,
and concluding
with the part originating from the transfer from a doubly polarized
initial spin state. This last term, to the best of our knowledge, has not been
considered up to now in previous studies. This enabled us to trace
the formation of
final polarizations and spin correlations.

We found that at each scattering angle and
for all the allowed  values ($-1 \le p^{n(p)}_y \le +1$) of the initial
neutron and proton polarizations
pointing at the $y$-direction
(with the only exception being the case when both polarizations take on the 
 maximal values of $\pm 1$) the final spin states are statistical
 mixtures of states. 
 An investigation of the outgoing neutron and proton polarizations, as well as
their spin correlations in these predominantly impure final states, revealed
variations not only with energy and scattering angle but also with
the polarizations of the incoming neutron and proton.
By calculating the angle-integrated entanglement power and concurrence,
we found these states to be entangled to some extent, with indications
of increasing entanglement at higher energies.

The dominance of impure states among the final spin states is a consequence of
the standard procedure used to prepare polarization states in nuclear
scattering experiments.
In this procedure, the polarization state of each incoming particle is
prepared independently, which typically results in only partially
polarized particles.
As a consequence, the initial polarized state is usually not a pure
quantum state but rather a statistical mixture of spin states.
Only when the polarizations of the incoming neutron and proton reach $\pm 1$,
such that all neutrons and protons occupy the same spin state, does
the initial state become pure.
As a result, the final state of the $np$ pair is also pure.
    However, it should be emphasized that even if the initial state
    were entangled—such as one of the Bell states—the final state, while
    still pure, would not necessarily remain entangled.
    This is because the presence of components with different neutron and
    proton spin projections, induced by the transition matrix
    $T_{{ m_{\text{in}} }}^{{ m_{\text{out}} }}$, can destroy the initial
    entanglement.

    The selectivity of the transition matrix—specifically, the fact that its
    largest elements correspond to transitions that preserve the sign
    of the nucleon spin projection—can also act in the opposite direction:
    by suppressing undesired components, it may induce entanglement
    in the final pure states.

    By calculating the angular distributions of the entanglement power
    and the concurrence $\bar C$ for the final pure states,
    we found that at $E = 100$~MeV and specific scattering angles, some
    final states exhibit entanglement resembling different types of Bell states.

    As the energy decreases, however, the increasing contribution of
    components that spoil entanglement leads to the disappearance of
    such states.

The question remains open if the presence of additional nucleons would help
create  entanglement 
 in the spin-state of two outgoing nucleons produced in reactions
with many-nucleon systems. 
The nucleon-induced deuteron breakup reaction 
 seems to be a natural candidate to investigate this problem.
It would be also interesting to look for signatures of entanglement in
 two-nucleon spin states produced via a decay of complex nuclear systems.

\clearpage

\acknowledgements

This work was supported by the National Science Centre,
Poland under Grant
IMPRESS-U 2024/06/Y/ST2/00135.   
 It was also supported in part by the Excellence
Initiative – Research University Program at the Jagiellonian
University in Krak\'ow. 
The numerical calculations were partly performed on the supercomputers of
the JSC, J\"ulich, Germany.

\clearpage
                                
\appendix

\section{Final polarization (spin-correlation) transfer coefficients
  from single or doubly spin-polarized initial state
  to the outgoing neutron (neutron-proton pair) in elastic $np$ scattering:
  Cartesian observables expressed in terms of the
  spherical ones}
\label{a1}

For elastic $np$ scattering with a polarized neutron or/and proton 
in the initial state and polarization of the outgoing neutron
or/and proton measured,  
$\vec p(\vec n,\vec{n'})\vec {p'}$, we denote the    
spin correlation transfer spherical tensors by
 $t_{k_pq_p,k_nq_n}^{k_{n'}q_{n'},k_{p'}q_{p'}}$
(defined in (\ref{eq4_a0})) 
 and the corresponding 
  spin correlation transfer Cartesian tensors
  by $K_{p,n}^{n',p'}$, defined analogously
  as in Eq.~(5.34) of ~\cite{ohlsen1972}:
 \begin{eqnarray}
	 K_{j,i}^{l',m'}  &=& \frac{\text{Tr} ( T {\mathcal{\sigma}}_j \sigma_i T^{\dagger}
     \sigma_{l'} \sigma_{m'}) }
	 { \text{Tr} ( T T^{\dagger}) } ~.
\label{eqqq_oh}
\end{eqnarray}   
 Using definitions
 of Ref.~\cite{ohlsen1972}, specifically relations between final
 polarization, induced polarization and polarization transfers
  (Eq.~(5.35)  in ~\cite{ohlsen1972}), 
 the following relations between Cartesian single,  $K_{0,n}^{n',p'}$,
 $K_{p,0}^{n',p'}$ (double $K_{p,n}^{n',p'}$) spin correlation transfers,  
  and  spherical single, $t_{00,k_nq_n}^{k_{n'}q_{n'},k_{p'}q_{p'}}$ 
 $t_{k_pq_p,00}^{k_{n'}q_{n'},k_{p'}q_{p'}}$,
 (double $t_{k_pq_p,k_nq_n}^{k_{n'}q_{n'},k_{p'}q_{p'}}$),  spin correlation transfer 
 tensors, as well as 
   between Cartesian  single, $K_{0,n}^{n',0}$, $K_{p,0}^{n',0}$ 
 (double $K_{p,n}^{n',0}$) spin polarizations   
 and  spherical single, $t_{00,k_nq_n}^{k_{n'}q_{n'},00}$,
 $t_{k_pq_p,00}^{k_{n'}q_{n'},00}$ (double $t_{k_pq_p,k_nq_n}^{k_{n'}q_{n'},00}$)   
  spin polarization  tensors, were derived 
 (the underlined Cartesian tensors vanish for elastic scattering
 due to parity conservation ~\cite{ohlsen1972}):

\vskip15pt
 
{\bf{I. Single polarization transfers $K_{0,n}^{n',0}$}}

\vskip15pt

1. $K_{0,x}^{z',0}  =  \frac {\sqrt{2}} {2} 
             ( t_{00,1-1}^{10,00} - t_{00,1+1}^{10,00} ) $

2. $\underline{K_{0,y}^{z',0} } =  - \frac {i\sqrt{2}} {2} 
             ( t_{00,1-1}^{10,00} + t_{00,1+1}^{10.00} ) $

3. $K_{0,z}^{z',0}  = t_{00,10}^{10,00}$

%#---------------------------------------------------------

4. $K_{0,x}^{x',0} = \frac {1} {2}
                            ( t_{00,1-1}^{1-1,00}
                             - t_{00,1-1}^{1+1,00}
                             - t_{00,1+1}^{1-1,00} + t_{00,1+1}^{1+1,00} ) $

5. $\underline{ K_{0,y}^{x',0} } = - \frac {i} { 2 }
                   ( t_{00,1-1}^{1-1,00} - t_{00,1-1}^{1+1,00}
                                  + t_{00,1+1}^{1-1,00} - t_{00,1+1}^{1+1,00} ) $

6. $K_{0,z}^{x',0} = \frac {\sqrt{2}} {2} 
                      ( t_{00,10}^{1-1,00} - t_{00,10}^{1+1,00} ) $

%#------------------------------------------------------------------

7. $\underline{ K_{0,x}^{y',0} } = \frac {i} { 2 }
                               ( t_{00,1+1}^{1+1,00} + t_{00,1-1}^{1+1,00}
                               - t_{00,1+1}^{1-1,00} - t_{00,1+1}^{1+1,00} ) $
		     
8. $K_{0,y}^{y',0} = \frac {1} { 2 }
                     ( t_{00,1-1}^{1-1,00} + t_{00,1-1}^{1+1,00}
                             + t_{00,1+1}^{1-1,00} + t_{00,1+1}^{1+1,00} ) $

9. $\underline{ K_{0,z}^{y',0} } = \frac {i\sqrt{2}} {2} 
                    ( t_{00,10}^{1-1,00} + t_{00,10}^{1+1,00} ) $

%#------------------------------------------------------------------

\vskip15pt
                      
{\bf{II. Double polarization transfers $K_{p,n}^{n',0}$}}

\vskip15pt

1. $\underline{ K_{x,x}^{z',0} } =  \frac {1} {2} 
             ( t_{1-1,1-1}^{10,00} - t_{1-1,1+1}^{10,00}
             - t_{1+1,1-1}^{10,00} + t_{1+1,1+1}^{10,00}  )$

2. $K_{y,x}^{z',0} =  - \frac {i} {2} 
             ( t_{1-1,1-1}^{10,00} - t_{1-1,1+1}^{10,00}
             + t_{1+1,1-1}^{10,00} - t_{1+1,1+1}^{10,00}  )$

3. $\underline{ K_{z,x}^{z',0} } = \frac {\sqrt{ 2 }} {2}
             (t_{10,1-1}^{10,00} - t_{10,1+1}^{10,00} )$

4. $K_{x,y}^{z',0} =  - \frac {i} {2} 
             ( t_{1-1,1-1}^{10,00} + t_{1-1,1+1}^{10,00} 
             - t_{1+1,1-1}^{10,00} - t_{1+1,1+1}^{10,00}  )$

5. $\underline{ K_{y,y}^{z',0} } = - \frac {1} {2} 
             (t_{1-1,1-1}^{10,00} + t_{1-1,1+1}^{10,00}
             + t_{1+1,1-1}^{10,00} + t_{1+1,1+1}^{10,00}  )$

6. $K_{z,y}^{z',0} =  - \frac {i \sqrt{2} } {2}
                  (t_{10,1-1}^{10,00} + t_{10,1+1}^{10,00} )$

7. $\underline{ K_{x,z}^{z',0} } = \frac {\sqrt{2}} {2}
                  (t_{1-1,10}^{10,00} - t_{1+1,10}^{10,00} )$

8. $K_{y,z}^{z',0} = - \frac {i \sqrt{2} } {2}
                  (t_{1-1,10}^{10,00} + t_{1+1,10}^{10,00} )$

9. $\underline{ K_{z,z}^{z',0} }=  t_{10,10}^{10,00} $

%#---------------------------------------------------------

10. $\underline{ K_{x,x}^{x',0} }=  \frac {\sqrt{2}  } {4}
                             ( t_{1-1,1-1}^{1-1,00}
                             - t_{1-1,1-1}^{1+1,00}
                             - t_{1-1,1+1}^{1-1,00} + t_{1-1,1+1}^{1+1,00}
                             - t_{1+1,1-1}^{1-1,00} + t_{1+1,1-1}^{1+1,00}
                             + t_{1+1,1+1}^{1-1,00} - t_{1+1,1+1}^{1+1,00} )$

11. $K_{y,x}^{x',0} = - \frac {i\sqrt{2}} { 4 }
                             ( t_{1-1,1-1}^{1-1,00} - t_{1-1,1-1}^{1+1,00}
                              -t_{1-1,1+1}^{1-1,00} + t_{1-1,1+1}^{1+1,00}
                              +t_{1+1,1-1}^{1-1,00} - t_{1+1,1-1}^{1+1,00}
                              -t_{1+1,1+1}^{1-1,00} + t_{1+1,1+1}^{1+1,00} )$

12. $\underline{ K_{z,x}^{x',0} }=  \frac {1} {2} 
                   ( t_{10,1-1}^{1-1,00} - t_{10,1-1}^{1+1,00}
                    -t_{10,1+1}^{1-1,00} + t_{10,1+1}^{1+1,00} )$

13. $K_{x,y}^{x',0} = - \frac {i \sqrt{2}} { 4 }
                   ( t_{1-1,1-1}^{1-1,00} - t_{1-1,1-1}^{1+1,00}
                    +t_{1-1,1+1}^{1-1,00} - t_{1-1,1+1}^{1+1,00}
                    -t_{1+1,1-1}^{1-1,00} + t_{1+1,1-1}^{1+1,00}
                    -t_{1+1,1+1}^{1-1,00} + t_{1+1,1+1}^{1+1,00} )$

14. $\underline{ K_{y,y}^{x',0} }= \frac {\sqrt{2}} { 4  }
                                  ( - t_{1-1,1-1}^{1-1,00} + t_{1-1,1-1}^{1+1,00}
                                    - t_{1-1,1+1}^{1-1,00} + t_{1-1,1+1}^{1+1,00}
                                    - t_{1+1,1-1}^{1-1,00} + t_{1+1,1-1}^{1+1,00}
                                    - t_{1+1,1+1}^{1-1,00} + t_{1+1,1+1}^{1+1,00} )$

15. $K_{z,y}^{x',0} = - \frac {i} {2} 
                                 ( t_{10,1-1}^{1-1,00} - t_{10,1-1}^{1+1,00}
                                  +t_{10,1+1}^{1-1,00} - t_{10,1+1}^{1+1,00} )$

16. $\underline{ K_{x,z}^{x',0} }=  \frac {1} {2} 
                      ( t_{1-1,10}^{1-1,00} - t_{1-1,10}^{1+1,00}
                       -t_{1+1,10}^{1-1,00} + t_{1+1,10}^{1+1,00} )$

17. $K_{y,z}^{x',0} = - \frac {i} {2} 
                    ( t_{1-1,10}^{1-1,00} - t_{1-1,10}^{1+1,00}
                     +t_{1+1,10}^{1-1,00} - t_{1+1,10}^{1+1,00} )$

18. $\underline{ K_{z,z}^{x',0} }= \frac {\sqrt{2} } { 2 }
                      ( t_{10,10}^{1-1,00} - t_{10,10}^{1+1,00} )$

%#------------------------------------------------------------------

19. $K_{x,x}^{y',0} = \frac {i\sqrt{2} } { 4 }
                               ( t_{1-1,1-1}^{1-1,00} + t_{1-1,1-1}^{1+1,00}
                               - t_{1-1,1+1}^{1-1,00} - t_{1-1,1+1}^{1+1,00}
                               - t_{1+1,1-1}^{1-1,00} - t_{1+1,1-1}^{1+1,00}
                               + t_{1+1,1+1}^{1-1,00} + t_{1+1,1+1}^{1+1,00} )$

20. $\underline{ K_{y,x}^{y',0} }= \frac {\sqrt{2}} { 4  }
                              ( t_{1-1,1-1}^{1-1,00} + t_{1-1,1-1}^{1+1,00}
                              - t_{1-1,1+1}^{1-1,00} - t_{1-1,1+1}^{1+1,00}
                              + t_{1+1,1-1}^{1-1,00} + t_{1+1,1-1}^{1+1,00}
                              - t_{1+1,1+1}^{1-1,00} - t_{1+1,1+1}^{1+1,00} )$

21. $K_{z,x}^{y',0} = \frac {i} {2} 
                     ( t_{10,1-1}^{1-1,00} + t_{10,1-1}^{1+1,00}
                     - t_{10,1+1}^{1-1,00} - t_{10,1+1}^{1+1,00} )$
		     
22. $\underline{ K_{x,y}^{y',0} }= \frac {\sqrt{2}} { 4 }
                     ( t_{1-1,1-1}^{1-1,00} + t_{1-1,1-1}^{1+1,00}
                     + t_{1-1,1+1}^{1-1,00} + t_{1-1,1+1}^{1+1,00}
                     - t_{1+1,1-1}^{1-1,00} - t_{1+1,1-1}^{1+1,00}
                     - t_{1+1,1+1}^{1-1,00} - t_{1+1,1+1}^{1+1,00} )$

23. $K_{y,y}^{y',0} = - \frac {i\sqrt{2}} { 4 }
                             (  t_{1-1,1-1}^{1-1,00} + t_{1-1,1-1}^{1+1,00}
                              + t_{1-1,1+1}^{1-1,00} + t_{1-1,1+1}^{1+1,00}
                              + t_{1+1,1-1}^{1-1,00} + t_{1+1,1-1}^{1+1,00}
                              + t_{1+1,1+1}^{1-1,00} + t_{1+1,1+1}^{1+1,00} )$

24. $\underline{ K_{z,y}^{y',0} }=  \frac {1} {2} 
                    ( t_{10,1-1}^{1-1,00} + t_{10,1-1}^{1+1,00}
                    + t_{10,1+1}^{1-1,00} + t_{10,1+1}^{1+1,00} )$

25. $K_{x,z}^{y',0} =  \frac {i} {2} 
                    ( t_{1-1,10}^{1-1,00} + t_{1-1,10}^{1+1,00}
                    - t_{1+1,10}^{1-1,00} - t_{1+1,10}^{1+1,00} )$

26. $\underline{ K_{y,z}^{y',0} }=  \frac {1} {2} 
                    ( t_{1-1,10}^{1-1,00} + t_{1-1,10}^{1+1,00}
                    + t_{1+1,10}^{1-1,00} + t_{1+1,10}^{1+1,00} )$

27. $K_{z,z}^{y',0} = \frac {i \sqrt{2} } { 2 }
                    ( t_{10,10}^{1-1,00} + t_{10,10}^{1+1,00} )$

%#------------------------------------------------------------------

\vskip15pt
 
{\bf{III. Single spin correlation transfers $K_{0,n}^{n',p'}$}}

\vskip15pt

1. $\underline{K_{0,x}^{z',x'}  } =  \frac {1} {2} 
( t_{00,1+1}^{10,1+1} - t_{00,1-1}^{10,1+1}
 - t_{00,1+1}^{10,1-1} + t_{00,1-1}^{10,1-1}  )$

2. $K_{0,x}^{z',y'}  =  \frac {i} {2} 
( -t_{00,1+1}^{10,1+1} + t_{00,1-1}^{10,1+1}
 - t_{00,1+1}^{10,1-1} + t_{00,1-1}^{10,1-1}  )$

3. $\underline{ K_{0,x}^{z',z'}  } =  \frac {\sqrt{2}} {2} 
( - t_{00,1+1}^{10,10} + t_{00,1-1}^{10,10} ) $

4. $K_{0,y}^{z',x'} =   \frac {i} {2} 
             ( t_{00,1+1}^{10,1+1} + t_{00,1-1}^{10,1+1} 
             - t_{00,1+1}^{10,1-1} - t_{00,1-1}^{10,1-1}  )$

5. $\underline{K_{0,y}^{z',y'} } =  \frac {1} {2} 
             ( t_{00,1+1}^{10,1+1} + t_{00,1-1}^{10,1+1} 
             + t_{00,1+1}^{10,1-1} + t_{00,1-1}^{10,1-1}  )$

6. $K_{0,y}^{z',z'} =  - \frac {i\sqrt{2}} {2} 
             ( t_{00,1+1}^{10,10} + t_{00,1-1}^{10,10} ) $ 
             
7. $\underline{ K_{0,z}^{z',x'}  } = \frac {\sqrt{2}} {2}
             (-t_{00,10}^{10,1+1} + t_{00,10}^{10,1-1} )$

8. $K_{0,z}^{z',y'}  = \frac {i\sqrt{2}} {2}
             (t_{00,10}^{10,1+1} + t_{00,10}^{10,1-1} )$

9. $\underline{ K_{0,z}^{z',z'}  } = t_{00,10}^{10,10}$
             
%#---------------------------------------------------------

10. $\underline{K_{0,x}^{x',x'} } =  \frac {\sqrt{2}  } {4}
                            (- t_{00,1+1}^{1+1,1+1}
                             + t_{00,1+1}^{1-1,1+1}
                             + t_{00,1-1}^{1+1,1+1} - t_{00,1-1}^{1-1,1+1}
                             + t_{00,1+1}^{1+1,1-1} - t_{00,1+1}^{1-1,1-1}
                             - t_{00,1-1}^{1+1,1-1} + t_{00,1-1}^{1-1,1-1} )$

11. $K_{0,x}^{x',y'} = \frac {i\sqrt{2}  } {4}
                            (  t_{00,1+1}^{1+1,1+1}
                             - t_{00,1+1}^{1-1,1+1}
                             - t_{00,1-1}^{1+1,1+1} + t_{00,1-1}^{1-1,1+1}
                             + t_{00,1+1}^{1+1,1-1} - t_{00,1+1}^{1-1,1-1}
                             - t_{00,1-1}^{1+1,1-1} + t_{00,1-1}^{1-1,1-1} )$

12. $\underline{K_{0,x}^{x',z'} } = \frac {1} {2} 
                            ( t_{00,1+1}^{1+1,10}
                             - t_{00,1+1}^{1-1,10}
                             - t_{00,1-1}^{1+1,10} + t_{00,1-1}^{1-1,10} ) $
                             
13. $K_{0,y}^{x',x'} = - \frac {i \sqrt{2}} { 4 }
                   (- t_{00,1+1}^{1+1,1+1} + t_{00,1+1}^{1-1,1+1}
                    - t_{00,1-1}^{1+1,1+1} + t_{00,1-1}^{1-1,1+1}
                    + t_{00,1+1}^{1+1,1-1} - t_{00,1+1}^{1-1,1-1}
                    + t_{00,1-1}^{1+1,1-1} - t_{00,1-1}^{1-1,1-1} )$

14. $\underline{ K_{0,y}^{x',y'} } =  \frac {\sqrt{2}} { 4 }
                  (- t_{00,1+1}^{1+1,1+1} + t_{00,1+1}^{1-1,1+1}
                   - t_{00,1-1}^{1+1,1+1} + t_{00,1-1}^{1-1,1+1}
                   - t_{00,1+1}^{1+1,1-1} + t_{00,1+1}^{1-1,1-1}
                   - t_{00,1-1}^{1+1,1-1} + t_{00,1-1}^{1-1,1-1} )$

15. $K_{0,y}^{x',z'} = - \frac {i} { 2 }
                   ( t_{00,1+1}^{1+1,10} - t_{00,1+1}^{1-1,10}
                   + t_{00,1-1}^{1+1,10} - t_{00,1-1}^{1-1,10} ) $
                                  
16. $\underline{K_{0,z}^{x',x'} } = \frac {1} {2} 
                    ( t_{00,10}^{1+1,1+1} - t_{00,10}^{1-1,1+1}
                    - t_{00,10}^{1+1,1-1} + t_{00,10}^{1-1,1-1} )$

17. $K_{0,z}^{x',y'} = - \frac {i} {2} 
                      ( t_{00,10}^{1+1,1+1} - t_{00,10}^{1-1,1+1}
                      + t_{00,10}^{1+1,1-1} - t_{00,10}^{1-1,1-1} )$

18. $\underline{K_{0,z}^{x',z'} } = \frac {\sqrt{2}} {2} 
                      ( - t_{00,10}^{1+1,10} + t_{00,10}^{1-1,10} ) $

%#------------------------------------------------------------------

19. $K_{0,x}^{y',x'} = \frac {i\sqrt{2} } { 4 }
                               ( t_{00,1+1}^{1+1,1+1} + t_{00,1+1}^{1-1,1+1}
                               - t_{00,1-1}^{1+1,1+1} - t_{00,1-1}^{1-1,1+1}
                               - t_{00,1+1}^{1+1,1-1} - t_{00,1+1}^{1-1,1-1}
                               + t_{00,1-1}^{1+1,1-1} + t_{00,1-1}^{1-1,1-1} )$

20. $\underline{ K_{0,x}^{y',y'} } = \frac {\sqrt{2} } { 4 }
                               ( t_{00,1+1}^{1+1,1+1} + t_{00,1+1}^{1-1,1+1}
                               - t_{00,1-1}^{1+1,1+1} - t_{00,1-1}^{1-1,1+1}
                               + t_{00,1+1}^{1+1,1-1} + t_{00,1+1}^{1-1,1-1}
                               - t_{00,1-1}^{1+1,1-1} - t_{00,1-1}^{1-1,1-1} )$

21. $K_{0,x}^{y',z'} = - \frac {i} { 2 }
                               ( t_{00,1+1}^{1+1,10} + t_{00,1+1}^{1-1,10}
                               - t_{00,1-1}^{1+1,10} - t_{00,1-1}^{1-1,10} ) $
		     
22. $\underline{K_{0,y}^{y',x'} } = \frac {\sqrt{2}} { 4 }
                    (- t_{00,1+1}^{1+1,1+1} - t_{00,1+1}^{1-1,1+1}
                     - t_{00,1-1}^{1+1,1+1} - t_{00,1-1}^{1-1,1+1}
                     + t_{00,1+1}^{1+1,1-1} + t_{00,1+1}^{1-1,1-1}
                     + t_{00,1-1}^{1+1,1-1} + t_{00,1-1}^{1-1,1-1} )$

23. $K_{0,y}^{y',y'} = \frac {i\sqrt{2}} { 4 }
                     ( t_{00,1+1}^{1+1,1+1} + t_{00,1+1}^{1-1,1+1}
                     + t_{00,1-1}^{1+1,1+1} + t_{00,1-1}^{1-1,1+1}
                     + t_{00,1+1}^{1+1,1-1} + t_{00,1+1}^{1-1,1-1}
                     + t_{00,1-1}^{1+1,1-1} + t_{00,1-1}^{1-1,1-1} )$
                             
24. $\underline{K_{0,y}^{y',z'} } = \frac {1} { 2 }
                     ( t_{00,1+1}^{1+1,10} + t_{00,1+1}^{1-1,10}
                     + t_{00,1-1}^{1+1,10} + t_{00,1-1}^{1-1,10} ) $
                             
25. $K_{0,z}^{y',x'} = - \frac {i} {2} 
                    ( t_{00,10}^{1+1,1+1} + t_{00,10}^{1-1,1+1}
                    - t_{00,10}^{1+1,1-1} - t_{00,10}^{1-1,1-1} )$

26. $\underline{ K_{0,z}^{y',y'} } =  \frac {1} {2} 
                   (- t_{00,10}^{1+1,1+1} - t_{00,10}^{1-1,1+1}
                    - t_{00,10}^{1+1,1-1} - t_{00,10}^{1-1,1-1} )$

27. $K_{0,z}^{y',z'} =  \frac {i\sqrt(2)} {2} 
                    ( t_{00,10}^{1+1,10} + t_{00,10}^{1-1,10} ) $
                             
%#------------------------------------------------------------------

\vskip15pt
                      
{\bf{IV. Double spin correlation transfers $K_{p,n}^{n',p'}$}}

\vskip15pt

1. $K_{x,x}^{z',x'} =  \frac {\sqrt{2}} {4} 
                                (-t_{1+1,1+1}^{10,1+1} + t_{1+1,1-1}^{10,1+1}
                                 +t_{1+1,1+1}^{10,1-1} - t_{1+1,1-1}^{10,1-1}  
                                 +t_{1-1,1+1}^{10,1+1} - t_{1-1,1-1}^{10,1+1}
                                 -t_{1-1,1+1}^{10,1-1} + t_{1-1,1-1}^{10,1-1}  )$

2. $\underline{ K_{x,x}^{z',y'} } =  \frac {i\sqrt{2}} {4} 
                                ( t_{1+1,1+1}^{10,1+1} - t_{1+1,1-1}^{10,1+1}
                                + t_{1+1,1+1}^{10,1-1} - t_{1+1,1-1}^{10,1-1}  
                                - t_{1-1,1+1}^{10,1+1} + t_{1-1,1-1}^{10,1+1}
                                - t_{1-1,1+1}^{10,1-1} + t_{1-1,1-1}^{10,1-1}  )$
                                 
3. $K_{x,x}^{z',z'} =   \frac {1} {2} 
                                ( t_{1+1,1+1}^{10,10} - t_{1+1,1-1}^{10,10}
                                 -t_{1-1,1+1}^{10,10} + t_{1-1,1-1}^{10,10}  )$ 

4. $\underline{ K_{y,x}^{z',x'} } =  \frac {i\sqrt{2}} {4} 
            (- t_{1+1,1+1}^{10,1+1} + t_{1+1,1-1}^{10,1+1}
             + t_{1+1,1+1}^{10,1-1} - t_{1+1,1-1}^{10,1-1}  
             - t_{1-1,1+1}^{10,1+1} + t_{1-1,1-1}^{10,1+1}
             + t_{1-1,1+1}^{10,1-1} - t_{1-1,1-1}^{10,1-1}  )$
             
5. $K_{y,x}^{z',y'} = \frac {\sqrt{2}} {4}
            (- t_{1+1,1+1}^{10,1+1} + t_{1+1,1-1}^{10,1+1} 
             - t_{1+1,1+1}^{10,1-1} + t_{1+1,1-1}^{10,1-1}  
             - t_{1-1,1+1}^{10,1+1} + t_{1-1,1-1}^{10,1+1} 
             - t_{1-1,1+1}^{10,1-1} + t_{1-1,1-1}^{10,1-1}  )$
             
6. $\underline{ K_{y,x}^{z',z'} } =  \frac {i} {2} 
              (t_{1+1,1+1}^{10,10} - t_{1+1,1-1}^{10,10}
             + t_{1-1,1+1}^{10,10} - t_{1-1,1-1}^{10,10}  )$
             
7. $K_{z,x}^{z',x'} =  \frac {1} {2}  (t_{10,1+1}^{10,1+1} - t_{10,1-1}^{10,1+1} 
                                  - t_{10,1+1}^{10,1-1} + t_{10,1-1}^{10,1-1} )$

8. $\underline{ K_{z,x}^{z',y'} } = \frac {i} {2}
              (-t_{10,1+1}^{10,1+1} + t_{10,1-1}^{10,1+1} 
               -t_{10,1+1}^{10,1-1} + t_{10,1-1}^{10,1-1} )$

9. $K_{z,x}^{z',z'} =  \frac {\sqrt{ 2 }} {2}
              (- t_{10,1+1}^{10,10} + t_{10,1-1}^{10,10} )$
             
10. $\underline{ K_{x,y}^{z',x'} } =  \frac {i\sqrt{2}} {4} 
            (- t_{1+1,1+1}^{10,1+1} - t_{1+1,1-1}^{10,1+1} 
             + t_{1+1,1+1}^{10,1-1} + t_{1+1,1-1}^{10,1-1}  
             + t_{1-1,1+1}^{10,1+1} + t_{1-1,1-1}^{10,1+1} 
             - t_{1-1,1+1}^{10,1-1} - t_{1-1,1-1}^{10,1-1}  )$
             
11. $K_{x,y}^{z',y'} =  \frac {\sqrt{2}} {4}
            (- t_{1+1,1+1}^{10,1+1} - t_{1+1,1-1}^{10,1+1}
             - t_{1+1,1+1}^{10,1-1} - t_{1+1,1-1}^{10,1-1}  
             + t_{1-1,1+1}^{10,1+1} + t_{1-1,1-1}^{10,1+1}
             + t_{1-1,1+1}^{10,1-1} + t_{1-1,1-1}^{10,1-1}  )$

12. $\underline{ K_{x,y}^{z',z'} } =  \frac {i} {2} 
             ( t_{1+1,1+1}^{10,10} + t_{1+1,1-1}^{10,10} 
             - t_{1-1,1+1}^{10,10} - t_{1-1,1-1}^{10,10}  )$
             
13. $K_{y,y}^{z',x'} =  \frac {\sqrt{2}} {4} 
              (t_{1+1,1+1}^{10,1+1} + t_{1+1,1-1}^{10,1+1}
             - t_{1+1,1+1}^{10,1-1} - t_{1+1,1-1}^{10,1-1}  
             + t_{1-1,1+1}^{10,1+1} + t_{1-1,1-1}^{10,1+1}
             - t_{1-1,1+1}^{10,1-1} - t_{1-1,1-1}^{10,1-1}  )$
             
14. $\underline{ K_{y,y}^{z',y'} } = - \frac {i\sqrt{2}} {4} 
            (+ t_{1+1,1+1}^{10,1+1} + t_{1+1,1-1}^{10,1+1}
             + t_{1+1,1+1}^{10,1-1} + t_{1+1,1-1}^{10,1-1}  
             + t_{1-1,1+1}^{10,1+1} + t_{1-1,1-1}^{10,1+1}
             + t_{1-1,1+1}^{10,1-1} + t_{1-1,1-1}^{10,1-1}  )$
             
15. $K_{y,y}^{z',z'} = - \frac {1} {2} 
              (t_{1+1,1+1}^{10,10} + t_{1+1,1-1}^{10,10}
             + t_{1-1,1+1}^{10,10} + t_{1-1,1-1}^{10,10}  )$
             
16. $\underline{K_{z,y}^{z',x'} } =  \frac {i} {2}
             ( t_{10,1+1}^{10,1+1} + t_{10,1-1}^{10,1+1} 
              -t_{10,1+1}^{10,1-1} - t_{10,1-1}^{10,1-1} )$

17. $K_{z,y}^{z',y'} =  \frac {1} {2}
              ( t_{10,1+1}^{10,1+1} + t_{10,1-1}^{10,1+1} 
               +t_{10,1+1}^{10,1-1} + t_{10,1-1}^{10,1-1} )$

18. $\underline{K_{z,y}^{z',z'} } =  - \frac {i \sqrt{2} } {2}
              (t_{10,1+1}^{10,10} + t_{10,1-1}^{10,10} )$
             
19. $K_{x,z}^{z',x'} = \frac {1} {2}
              ( t_{1+1,10}^{10,1+1} - t_{1+1,10}^{10,1-1} 
              - t_{1-1,10}^{10,1+1} + t_{1-1,10}^{10,1-1} )$

20. $\underline{ K_{x,z}^{z',y'} } = \frac {i} {2}
              (- t_{1+1,10}^{10,1+1} - t_{1+1,10}^{10,1-1} 
               + t_{1-1,10}^{10,1+1} + t_{1-1,10}^{10,1-1} )$

21. $K_{x,z}^{z',z'} = \frac {\sqrt{2}} {2}
              (-t_{1+1,10}^{10,10} + t_{1-1,10}^{10,10} )$
             
22. $\underline{K_{y,z}^{z',x'} } = \frac {i} {2}
             ( t_{1+1,10}^{10,1+1} - t_{1+1,10}^{10,1-1} 
              +t_{1-1,10}^{10,1+1} - t_{1-1,10}^{10,1-1} )$

23. $K_{y,z}^{z',y'} =  \frac {1} {2}
             ( t_{1+1,10}^{10,1+1} + t_{1+1,10}^{10,1-1} 
             + t_{1-1,10}^{10,1+1} + t_{1-1,10}^{10,1-1} )$

24. $\underline{K_{y,z}^{z',z'} } = - \frac {i \sqrt{2} } {2}
              (t_{1+1,10}^{10,10} + t_{1-1,10}^{10,10} )$
             
25. $K_{z,z}^{z',x'} =  \frac {\sqrt{2} } {2}
              (-t_{10,10}^{10,1+1} + t_{10,10}^{10,1-1} )$

26. $\underline{ K_{z,z}^{z',y'} }= \frac {i\sqrt{2}} {2} 
              ( t_{10,10}^{10,1+1} + t_{10,10}^{10,1-1} )$
             
27. $K_{z,z}^{z',z'} =   t_{10,10}^{10,10} $
             
%#---------------------------------------------------------

28. $K_{x,x}^{x',x'} = \frac {1} {4}
                             ( t_{1+1,1+1}^{1+1,1+1}
                             - t_{1+1,1+1}^{1-1,1+1}
                             - t_{1+1,1-1}^{1+1,1+1} + t_{1+1,1-1}^{1-1,1+1}
                             - t_{1+1,1+1}^{1+1,1-1} + t_{1+1,1+1}^{1-1,1-1}
                             + t_{1+1,1-1}^{1+1,1-1} - t_{1+1,1-1}^{1-1,1-1}
                             - t_{1-1,1+1}^{1+1,1+1} + t_{1-1,1+1}^{1-1,1+1}
                             + t_{1-1,1-1}^{1+1,1+1} - t_{1-1,1-1}^{1-1,1+1}
                             + t_{1-1,1+1}^{1+1,1-1} - t_{1-1,1+1}^{1-1,1-1}
                             - t_{1-1,1-1}^{1+1,1-1} + t_{1-1,1-1}^{1-1,1-1} )$

29. $\underline{ K_{x,x}^{x',y'} }= \frac {i} {4}
                            (- t_{1+1,1+1}^{1+1,1+1} + t_{1+1,1+1}^{1-1,1+1}
                             + t_{1+1,1-1}^{1+1,1+1} - t_{1+1,1-1}^{1-1,1+1}
                             - t_{1+1,1+1}^{1+1,1-1} + t_{1+1,1+1}^{1-1,1-1}
                             + t_{1+1,1-1}^{1+1,1-1} - t_{1+1,1-1}^{1-1,1-1} 
                             + t_{1-1,1+1}^{1+1,1+1} - t_{1-1,1+1}^{1-1,1+1}
                             - t_{1-1,1-1}^{1+1,1+1} + t_{1-1,1-1}^{1-1,1+1}
                             + t_{1-1,1+1}^{1+1,1-1} - t_{1-1,1+1}^{1-1,1-1}
                             - t_{1-1,1-1}^{1+1,1-1} + t_{1-1,1-1}^{1-1,1-1} )$
                             
30. $K_{x,x}^{x',z'} = \frac {\sqrt{2}  } {4}
                            (- t_{1+1,1+1}^{1+1,10}
                             + t_{1+1,1+1}^{1-1,10}
                             + t_{1+1,1-1}^{1+1,10} - t_{1+1,1-1}^{1-1,10}
                             + t_{1-1,1+1}^{1+1,10} - t_{1-1,1+1}^{1-1,10}
                             - t_{1-1,1-1}^{1+1,10} + t_{1-1,1-1}^{1-1,10} )$

31. $\underline{K_{y,x}^{x',x'} } =  \frac {i} { 4 }
                             ( t_{1+1,1+1}^{1+1,1+1} - t_{1+1,1+1}^{1-1,1+1}
                             - t_{1+1,1-1}^{1+1,1+1} + t_{1+1,1-1}^{1-1,1+1}
                             - t_{1+1,1+1}^{1+1,1-1} + t_{1+1,1+1}^{1-1,1-1}
                             + t_{1+1,1-1}^{1+1,1-1} - t_{1+1,1-1}^{1-1,1-1}
                             + t_{1-1,1+1}^{1+1,1+1} - t_{1-1,1+1}^{1-1,1+1}
                             - t_{1-1,1-1}^{1+1,1+1} + t_{1-1,1-1}^{1-1,1+1}
                             - t_{1-1,1+1}^{1+1,1-1} + t_{1-1,1+1}^{1-1,1-1}
                             + t_{1-1,1-1}^{1+1,1-1} - t_{1-1,1-1}^{1-1,1-1} )$
                             
32. $K_{y,x}^{x',y'} = \frac {1} { 4 }
                             ( t_{1+1,1+1}^{1+1,1+1} - t_{1+1,1+1}^{1-1,1+1}
                             - t_{1+1,1-1}^{1+1,1+1} + t_{1+1,1-1}^{1-1,1+1}
                             + t_{1+1,1+1}^{1+1,1-1} - t_{1+1,1+1}^{1-1,1-1}
                             - t_{1+1,1-1}^{1+1,1-1} + t_{1+1,1-1}^{1-1,1-1} 
                             + t_{1-1,1+1}^{1+1,1+1} - t_{1-1,1+1}^{1-1,1+1}
                             - t_{1-1,1-1}^{1+1,1+1} + t_{1-1,1-1}^{1-1,1+1}
                             + t_{1-1,1+1}^{1+1,1-1} - t_{1-1,1+1}^{1-1,1-1}
                             - t_{1-1,1-1}^{1+1,1-1} + t_{1-1,1-1}^{1-1,1-1} )$
                             
33. $\underline{K_{y,x}^{x',z'} } = \frac {i\sqrt{2}} { 4 }
                            (- t_{1+1,1+1}^{1+1,10} + t_{1+1,1+1}^{1-1,10}
                             + t_{1+1,1-1}^{1+1,10} - t_{1+1,1-1}^{1-1,10}
                             - t_{1-1,1+1}^{1+1,10} + t_{1-1,1+1}^{1-1,10}
                             + t_{1-1,1-1}^{1+1,10} - t_{1-1,1-1}^{1-1,10} )$
                                
34. $K_{z,x}^{x',x'} =  \frac {\sqrt{2}} {4} 
                  (- t_{10,1+1}^{1+1,1+1} + t_{10,1+1}^{1-1,1+1}
                   + t_{10,1-1}^{1+1,1+1} - t_{10,1-1}^{1-1,1+1} 
                   + t_{10,1+1}^{1+1,1-1} - t_{10,1+1}^{1-1,1-1}
                   - t_{10,1-1}^{1+1,1-1} + t_{10,1-1}^{1-1,1-1} )$
                   
35. $\underline{ K_{z,x}^{x',y'} }=  \frac {i\sqrt{2}} {4} 
                   ( t_{10,1+1}^{1+1,1+1} - t_{10,1+1}^{1-1,1+1}
                   - t_{10,1-1}^{1+1,1+1} + t_{10,1-1}^{1-1,1+1} 
                   + t_{10,1+1}^{1+1,1-1} - t_{10,1+1}^{1-1,1-1}
                   - t_{10,1-1}^{1+1,1-1} + t_{10,1-1}^{1-1,1-1} )$
                   
36. $K_{z,x}^{x',z'} =  \frac {1} {2} 
                   ( t_{10,1+1}^{1+1,10} - t_{10,1+1}^{1-1,10}
                   - t_{10,1-1}^{1+1,10} + t_{10,1-1}^{1-1,10} )$
                   
37. $\underline{K_{x,y}^{x',x'} }= \frac {i} { 4 }
                   ( t_{1+1,1+1}^{1+1,1+1} - t_{1+1,1+1}^{1-1,1+1}
                   + t_{1+1,1-1}^{1-1,1+1} - t_{1+1,1+1}^{1-1,1+1}
                   - t_{1+1,1+1}^{1+1,1-1} + t_{1+1,1+1}^{1-1,1-1}
                   - t_{1+1,1-1}^{1+1,1-1} + t_{1+1,1-1}^{1-1,1-1} 
                   - t_{1-1,1+1}^{1+1,1+1} + t_{1-1,1+1}^{1-1,1+1}
                   - t_{1-1,1-1}^{1+1,1+1} + t_{1-1,1-1}^{1-1,1+1}
                   + t_{1-1,1+1}^{1+1,1-1} - t_{1-1,1+1}^{1-1,1-1}
                   + t_{1-1,1-1}^{1+1,1-1} - t_{1-1,1-1}^{1-1,1-1} )$
                   
38. $K_{x,y}^{x',y'} =  \frac {1} { 4 }
                   ( t_{1+1,1+1}^{1+1,1+1} - t_{1+1,1+1}^{1-1,1+1}
                   + t_{1+1,1-1}^{1+1,1+1} - t_{1+1,1-1}^{1-1,1+1}
                   + t_{1+1,1+1}^{1+1,1-1} - t_{1+1,1+1}^{1-1,1-1}
                   + t_{1+1,1-1}^{1+1,1-1} - t_{1+1,1-1}^{1-1,1-1}
                   - t_{1-1,1+1}^{1+1,1+1} + t_{1-1,1+1}^{1-1,1+1}
                   - t_{1-1,1-1}^{1+1,1+1} + t_{1-1,1-1}^{1-1,1+1}
                   - t_{1-1,1+1}^{1+1,1-1} + t_{1-1,1+1}^{1-1,1-1}
                   - t_{1-1,1-1}^{1+1,1-1} + t_{1-1,1-1}^{1-1,1-1} )$

39. $\underline{K_{x,y}^{x',z'} }= \frac {i \sqrt{2}} { 4 }
                  (- t_{1+1,1+1}^{1+1,10} + t_{1+1,1+1}^{1-1,10}
                   - t_{1+1,1-1}^{1+1,10} + t_{1+1,1-1}^{1-1,10}
                   + t_{1-1,1+1}^{1+1,10} - t_{1-1,1+1}^{1-1,10}
                   + t_{1-1,1-1}^{1+1,10} - t_{1-1,1-1}^{1-1,10} )$
                                  
40. $K_{y,y}^{x',x'} =  \frac {1} { 4  }
                 ( - t_{1+1,1+1}^{1+1,1+1} + t_{1+1,1+1}^{1-1,1+1}
                   - t_{1+1,1-1}^{1+1,1+1} + t_{1+1,1-1}^{1-1,1+1}
                   + t_{1+1,1+1}^{1+1,1-1} - t_{1+1,1+1}^{1-1,1-1}
                   + t_{1+1,1-1}^{1+1,1-1} - t_{1+1,1-1}^{1-1,1-1} 
                   - t_{1-1,1+1}^{1+1,1+1} + t_{1-1,1+1}^{1-1,1+1}
                   - t_{1-1,1-1}^{1+1,1+1} + t_{1-1,1-1}^{1-1,1+1}
                   + t_{1-1,1+1}^{1+1,1-1} - t_{1-1,1+1}^{1-1,1-1}
                   + t_{1-1,1-1}^{1+1,1-1} - t_{1-1,1-1}^{1-1,1-1} )$
                   
41. $\underline{ K_{y,y}^{x',y'} }= \frac {i} { 4  }
                 ( + t_{1+1,1+1}^{1+1,1+1} - t_{1+1,1+1}^{1-1,1+1}
                   + t_{1+1,1-1}^{1+1,1+1} - t_{1+1,1-1}^{1-1,1+1}
                   + t_{1+1,1+1}^{1+1,1-1} - t_{1+1,1+1}^{1-1,1-1}
                   + t_{1+1,1-1}^{1+1,1-1} - t_{1+1,1-1}^{1-1,1-1} 
                   + t_{1-1,1+1}^{1+1,1+1} - t_{1-1,1+1}^{1-1,1+1}
                   + t_{1-1,1-1}^{1+1,1+1} - t_{1-1,1-1}^{1-1,1+1}
                   + t_{1-1,1+1}^{1+1,1-1} - t_{1-1,1+1}^{1-1,1-1}
                   + t_{1-1,1-1}^{1+1,1-1} - t_{1-1,1-1}^{1-1,1-1} )$
                   
42. $K_{y,y}^{x',z'} =  \frac {\sqrt{2}} { 4  }
                 (   t_{1+1,1+1}^{1+1,10} - t_{1+1,1+1}^{1-1,10}
                   + t_{1+1,1-1}^{1+1,10} - t_{1+1,1-1}^{1-1,10}
                   + t_{1-1,1+1}^{1+1,10} - t_{1-1,1+1}^{1-1,10}
                   + t_{1-1,1-1}^{1+1,10} - t_{1-1,1-1}^{1-1,10} )$
                                 
43. $\underline{K_{z,y}^{x',x'} } = \frac {i\sqrt{2}} {4} 
                  (- t_{10,1+1}^{1+1,1+1} + t_{10,1+1}^{1-1,1+1}
                   - t_{10,1-1}^{1+1,1+1} + t_{10,1-1}^{1-1,1+1} 
                   + t_{10,1+1}^{1+1,1-1} - t_{10,1+1}^{1-1,1-1}
                   + t_{10,1-1}^{1+1,1-1} - t_{10,1-1}^{1-1,1-1} )$
                   
44. $K_{z,y}^{x',y'} =  \frac {\sqrt{2}} {4} 
                  (- t_{10,1+1}^{1+1,1+1} + t_{10,1+1}^{1-1,1+1}
                   - t_{10,1-1}^{1+1,1+1} + t_{10,1-1}^{1-1,1+1} 
                   - t_{10,1+1}^{1+1,1-1} + t_{10,1+1}^{1-1,1-1}
                   - t_{10,1-1}^{1+1,1-1} + t_{10,1-1}^{1-1,1-1} )$
                   
45. $\underline{K_{z,y}^{x',z'} }= \frac {i} {2} 
                   ( t_{10,1+1}^{1+1,10} - t_{10,1+1}^{1-1,10}
                   + t_{10,1-1}^{1+1,10} - t_{10,1-1}^{1-1,10} )$
                      
46. $K_{x,z}^{x',x'} =  \frac {\sqrt{2}} {4} 
                   (- t_{1+1,10}^{1+1,1+1} + t_{1+1,10}^{1-1,1+1}
                    + t_{1+1,10}^{1+1,1-1} - t_{1+1,10}^{1-1,1-1} 
                    + t_{1-1,10}^{1+1,1+1} - t_{1-1,10}^{1-1,1+1}
                    - t_{1-1,10}^{1+1,1-1} + t_{1-1,10}^{1-1,1-1} )$

47. $\underline{ K_{x,z}^{x',y'} }= \frac {i\sqrt{2}} {4} 
                    ( t_{1+1,10}^{1+1,1+1} - t_{1+1,10}^{1-1,1+1}
                    + t_{1+1,10}^{1+1,1-1} - t_{1+1,10}^{1-1,1-1} 
                    - t_{1-1,10}^{1+1,1+1} + t_{1-1,10}^{1-1,1+1}
                    - t_{1-1,10}^{1+1,1-1} + t_{1-1,10}^{1-1,1-1} )$
                    
48. $K_{x,z}^{x',z'} =  \frac {1} {2} 
                    ( t_{1+1,10}^{1+1,10} - t_{1+1,10}^{1-1,10}
                    - t_{1-1,10}^{1+1,10} + t_{1-1,10}^{1-1,10} )$
                    
49. $\underline{K_{y,z}^{x',x'} }= \frac {i\sqrt{2}} {4} 
                     (- t_{1+1,10}^{1+1,1+1} + t_{1+1,10}^{1-1,1+1}
                      + t_{1+1,10}^{1+1,1-1} - t_{1+1,10}^{1-1,1-1} 
                      - t_{1-1,10}^{1+1,1+1} + t_{1-1,10}^{1-1,1+1}
                      + t_{1-1,10}^{1+1,1-1} - t_{1-1,10}^{1-1,1-1} )$
                      
50. $K_{y,z}^{x',y'} =  \frac {\sqrt{2}} {4} 
                   (- t_{1+1,10}^{1+1,1+1} + t_{1+1,10}^{1-1,1+1}
                    - t_{1+1,10}^{1+1,1-1} + t_{1+1,10}^{1-1,1-1} 
                    - t_{1-1,10}^{1+1,1+1} + t_{1-1,10}^{1-1,1+1}
                    - t_{1-1,10}^{1+1,1-1} + t_{1-1,10}^{1-1,1-1} )$
                    
51. $\underline{K_{y,z}^{x',z'} }= \frac {i} {2} 
                      ( t_{1+1,10}^{1+1,10} - t_{1+1,10}^{1-1,10}
                      + t_{1-1,10}^{1+1,10} - t_{1-1,10}^{1-1,10} )$
                      
52. $K_{z,z}^{x',x'}=  \frac {1} { 2 }
                      ( t_{10,10}^{1+1,1+1} - t_{10,10}^{1-1,1+1} 
                      - t_{10,10}^{1+1,1-1} + t_{10,10}^{1-1,1-1} )$

53. $\underline{ K_{z,z}^{x',y'} }= \frac {i} {2}
                      (- t_{10,10}^{1+1,1+1} + t_{10,10}^{1-1,1+1} 
                       - t_{10,10}^{1+1,1-1} + t_{10,10}^{1-1,1-1} )$

54. $K_{z,z}^{x',z'} =  \frac {\sqrt{2} } { 2 }
                      (- t_{10,10}^{1+1,10} + t_{10,10}^{1-1,10} )$

%#------------------------------------------------------------------

55. $\underline{K_{x,x}^{y',x'} }= \frac {i} { 4 }
                              (- t_{1+1,1+1}^{1+1,1+1} - t_{1+1,1+1}^{1-1,1+1}
                               + t_{1+1,1-1}^{1+1,1+1} + t_{1+1,1-1}^{1-1,1+1}
                               + t_{1+1,1+1}^{1+1,1-1} + t_{1+1,1+1}^{1-1,1-1}
                               - t_{1+1,1-1}^{1+1,1-1} - t_{1+1,1-1}^{1-1,1-1} 
                               + t_{1-1,1+1}^{1+1,1+1} + t_{1-1,1+1}^{1-1,1+1}
                               - t_{1-1,1-1}^{1+1,1+1} - t_{1-1,1-1}^{1-1,1+1}
                               - t_{1-1,1+1}^{1+1,1-1} - t_{1-1,1+1}^{1-1,1-1}
                               + t_{1-1,1-1}^{1+1,1-1} + t_{1-1,1-1}^{1-1,1-1} )$
                               
56. $K_{x,x}^{y',y'} =  \frac {1} { 4 }
                              (- t_{1+1,1+1}^{1+1,1+1} - t_{1+1,1+1}^{1-1,1+1}
                               + t_{1+1,1-1}^{1+1,1+1} + t_{1+1,1-1}^{1-1,1+1}
                               - t_{1+1,1+1}^{1+1,1-1} - t_{1+1,1+1}^{1-1,1-1}
                               + t_{1+1,1-1}^{1+1,1-1} + t_{1+1,1-1}^{1-1,1-1}
                               + t_{1-1,1+1}^{1+1,1+1} + t_{1-1,1+1}^{1-1,1+1}
                               - t_{1-1,1-1}^{1+1,1+1} - t_{1-1,1-1}^{1-1,1+1}
                               + t_{1-1,1+1}^{1+1,1-1} + t_{1-1,1+1}^{1-1,1-1}
                               - t_{1-1,1-1}^{1+1,1-1} - t_{1-1,1-1}^{1-1,1-1} )$

57. $\underline{K_{x,x}^{y',z'} }=  \frac {i\sqrt{2} } { 4 }
                               ( t_{1+1,1+1}^{1+1,10} + t_{1+1,1+1}^{1-1,10}
                               - t_{1+1,1-1}^{1+1,10} - t_{1+1,1-1}^{1-1,10}
                               - t_{1-1,1+1}^{1+1,10} - t_{1-1,1+1}^{1-1,10}
                               + t_{1-1,1-1}^{1+1,10} + t_{1-1,1-1}^{1-1,10} )$
                               
58. $K_{y,x}^{y',x'} =  \frac {1} { 4  }
                              ( t_{1+1,1+1}^{1+1,1+1} + t_{1+1,1+1}^{1-1,1+1}
                              - t_{1+1,1-1}^{1+1,1+1} - t_{1+1,1-1}^{1-1,1+1}
                              - t_{1+1,1+1}^{1+1,1-1} - t_{1+1,1+1}^{1-1,1-1}
                              + t_{1+1,1-1}^{1+1,1-1} + t_{1+1,1-1}^{1-1,1-1} 
                              + t_{1-1,1+1}^{1+1,1+1} + t_{1-1,1+1}^{1-1,1+1}
                              - t_{1-1,1-1}^{1+1,1+1} - t_{1-1,1-1}^{1-1,1+1}
                              - t_{1-1,1+1}^{1+1,1-1} - t_{1-1,1+1}^{1-1,1-1}
                              + t_{1-1,1-1}^{1+1,1-1} + t_{1-1,1-1}^{1-1,1-1} )$
                              
59. $\underline{ K_{y,x}^{y',y'} }= \frac {i} { 4  }
                             (- t_{1+1,1+1}^{1+1,1+1} - t_{1+1,1+1}^{1-1,1+1}
                              + t_{1+1,1-1}^{1+1,1+1} + t_{1+1,1-1}^{1-1,1+1}
                              - t_{1+1,1+1}^{1+1,1-1} - t_{1+1,1+1}^{1-1,1-1}
                              + t_{1+1,1-1}^{1+1,1-1} + t_{1+1,1-1}^{1-1,1-1} 
                              - t_{1-1,1+1}^{1+1,1+1} - t_{1-1,1+1}^{1-1,1+1}
                              + t_{1-1,1-1}^{1+1,1+1} + t_{1-1,1-1}^{1-1,1+1}
                              - t_{1-1,1+1}^{1+1,1-1} - t_{1-1,1+1}^{1-1,1-1}
                              + t_{1-1,1-1}^{1+1,1-1} + t_{1-1,1-1}^{1-1,1-1} )$
                              
60. $K_{y,x}^{y',z'} =  \frac {\sqrt{2}} { 4  }
                             (- t_{1+1,1+1}^{1+1,10} - t_{1+1,1+1}^{1-1,10}
                              + t_{1+1,1-1}^{1+1,10} + t_{1+1,1-1}^{1-1,10}
                              - t_{1-1,1+1}^{1+1,10} - t_{1-1,1+1}^{1-1,10}
                              + t_{1-1,1-1}^{1+1,10} + t_{1-1,1-1}^{1-1,10} )$
                              
61. $\underline{K_{z,x}^{y',x'} }= \frac {i\sqrt{2}} {4} 
                     ( t_{10,1+1}^{1+1,1+1} + t_{10,1+1}^{1-1,1+1}
                     - t_{10,1-1}^{1+1,1+1} - t_{10,1-1}^{1-1,1+1} 
                     - t_{10,1+1}^{1+1,1-1} - t_{10,1+1}^{1-1,1-1}
                     + t_{10,1-1}^{1+1,1-1} + t_{10,1-1}^{1-1,1-1} )$
                     
62. $K_{z,x}^{y',y'} = \frac {\sqrt{2}} {4} 
                     ( t_{10,1+1}^{1+1,1+1} + t_{10,1+1}^{1-1,1+1}
                     - t_{10,1-1}^{1+1,1+1} - t_{10,1-1}^{1-1,1+1} 
                     + t_{10,1+1}^{1+1,1-1} + t_{10,1+1}^{1-1,1-1}
                     - t_{10,1-1}^{1+1,1-1} - t_{10,1-1}^{1-1,1-1} )$
                     
63. $\underline{K_{z,x}^{y',z'} }= \frac {i} {2} 
                    (- t_{10,1+1}^{1+1,10} - t_{10,1+1}^{1-1,10}
                     + t_{10,1-1}^{1+1,10} + t_{10,1-1}^{1-1,10} )$
                     
64. $K_{x,y}^{y',x'} = \frac {1} { 4 }
                     ( t_{1+1,1+1}^{1+1,1+1} + t_{1+1,1+1}^{1-1,1+1}
                     + t_{1+1,1-1}^{1+1,1+1} + t_{1+1,1-1}^{1-1,1+1}
                     - t_{1+1,1+1}^{1+1,1-1} - t_{1+1,1+1}^{1-1,1-1}
                     - t_{1+1,1-1}^{1+1,1-1} - t_{1+1,1-1}^{1-1,1-1} 
                     - t_{1-1,1+1}^{1+1,1+1} - t_{1-1,1+1}^{1-1,1+1}
                     - t_{1-1,1-1}^{1+1,1+1} - t_{1-1,1-1}^{1-1,1+1}
                     + t_{1-1,1+1}^{1+1,1-1} + t_{1-1,1+1}^{1-1,1-1}
                     + t_{1-1,1-1}^{1+1,1-1} + t_{1-1,1-1}^{1-1,1-1} )$

65. $\underline{ K_{x,y}^{y',y'} }= \frac {i} { 4 }
                    (- t_{1+1,1+1}^{1+1,1+1} - t_{1+1,1+1}^{1-1,1+1}
                     - t_{1+1,1-1}^{1+1,1+1} - t_{1+1,1-1}^{1-1,1+1}
                     - t_{1+1,1+1}^{1+1,1-1} - t_{1+1,1+1}^{1-1,1-1}
                     - t_{1+1,1-1}^{1+1,1-1} - t_{1+1,1-1}^{1-1,1-1} 
                     + t_{1-1,1+1}^{1+1,1+1} + t_{1-1,1+1}^{1-1,1+1}
                     + t_{1-1,1-1}^{1+1,1+1} + t_{1-1,1-1}^{1-1,1+1}
                     + t_{1-1,1+1}^{1+1,1-1} + t_{1-1,1+1}^{1-1,1-1}
                     + t_{1-1,1-1}^{1+1,1-1} + t_{1-1,1-1}^{1-1,1-1} )$
                     
66. $K_{x,y}^{y',z'} =  \frac {\sqrt{2}} { 4 }
                    (- t_{1+1,1+1}^{1+1,10} - t_{1+1,1+1}^{1-1,10}
                     - t_{1+1,1-1}^{1+1,10} - t_{1+1,1-1}^{1-1,10}
                     + t_{1-1,1+1}^{1+1,10} + t_{1-1,1+1}^{1-1,10}
                     + t_{1-1,1-1}^{1+1,10} + t_{1-1,1-1}^{1-1,10} )$
                             
67. $K_{y,y}^{y',x'} = \frac {i} { 4 }
                             (  t_{1+1,1+1}^{1+1,1+1} + t_{1+1,1+1}^{1-1,1+1}
                              + t_{1+1,1-1}^{1+1,1+1} + t_{1+1,1-1}^{1-1,1+1}
                              - t_{1+1,1+1}^{1+1,1-1} - t_{1+1,1+1}^{1-1,1-1}
                              - t_{1+1,1-1}^{1+1,1-1} - t_{1+1,1-1}^{1-1,1-1} 
                              + t_{1-1,1+1}^{1+1,1+1} + t_{1-1,1+1}^{1-1,1+1}
                              + t_{1-1,1-1}^{1+1,1+1} + t_{1-1,1-1}^{1-1,1+1}
                              - t_{1-1,1+1}^{1+1,1-1} - t_{1-1,1+1}^{1-1,1-1}
                              - t_{1-1,1-1}^{1+1,1-1} - t_{1-1,1-1}^{1-1,1-1} )$
                              
68. $K_{y,y}^{y',y'} =  \frac {1} { 4 }
                             (  t_{1+1,1+1}^{1+1,1+1} + t_{1+1,1+1}^{1-1,1+1}
                              + t_{1+1,1-1}^{1+1,1+1} + t_{1+1,1-1}^{1-1,1+1}
                              + t_{1+1,1+1}^{1+1,1-1} + t_{1+1,1+1}^{1-1,1-1}
                              + t_{1+1,1-1}^{1+1,1-1} + t_{1+1,1-1}^{1-1,1-1} 
                              + t_{1-1,1+1}^{1+1,1+1} + t_{1-1,1+1}^{1-1,1+1}
                              + t_{1-1,1-1}^{1+1,1+1} + t_{1-1,1-1}^{1-1,1+1}
                              + t_{1-1,1+1}^{1+1,1-1} + t_{1-1,1+1}^{1-1,1-1}
                              + t_{1-1,1-1}^{1+1,1-1} + t_{1-1,1-1}^{1-1,1-1} )$
                              
69. $\underline{K_{y,y}^{y',z'} }= - \frac {i\sqrt{2}} { 4 }
                                (  t_{1+1,1+1}^{1+1,10} + t_{1+1,1+1}^{1-1,10}
                                 + t_{1+1,1-1}^{1+1,10} + t_{1+1,1-1}^{1-1,10}
                                 + t_{1-1,1+1}^{1+1,10} + t_{1-1,1+1}^{1-1,10}
                                 + t_{1-1,1-1}^{1+1,10} + t_{1-1,1-1}^{1-1,10} )$
                                 
70. $K_{z,y}^{y',x'} = \frac {\sqrt{2}} {4} 
                   (- t_{10,1+1}^{1+1,1+1} - t_{10,1+1}^{1-1,1+1}
                    - t_{10,1-1}^{1+1,1+1} - t_{10,1-1}^{1-1,1+1} 
                    + t_{10,1+1}^{1+1,1-1} + t_{10,1+1}^{1-1,1-1}
                    + t_{10,1-1}^{1+1,1-1} + t_{10,1-1}^{1-1,1-1} )$
                    
71. $\underline{ K_{z,y}^{y',y'} }=  \frac {i\sqrt{2}} {4} 
                    ( t_{10,1+1}^{1+1,1+1} + t_{10,1+1}^{1-1,1+1}
                    + t_{10,1-1}^{1+1,1+1} + t_{10,1-1}^{1-1,1+1} 
                    + t_{10,1+1}^{1+1,1-1} + t_{10,1+1}^{1-1,1-1}
                    + t_{10,1-1}^{1+1,1-1} + t_{10,1-1}^{1-1,1-1} )$
                    
72. $K_{z,y}^{y',z'} =  \frac {1} {2} 
                    ( t_{10,1+1}^{1+1,10} + t_{10,1+1}^{1-1,10}
                    + t_{10,1-1}^{1+1,10} + t_{10,1-1}^{1-1,10} )$
                    
73. $K\underline{_{x,z}^{y',x'} }=  \frac {i\sqrt{2}} {4} 
                      ( t_{1+1,10}^{1+1,1+1} + t_{1+1,10}^{1-1,1+1}
                      - t_{1+1,10}^{1+1,1-1} - t_{1+1,10}^{1-1,1-1} 
                      - t_{1-1,10}^{1+1,1+1} - t_{1-1,10}^{1-1,1+1}
                      + t_{1-1,10}^{1+1,1-1} + t_{1-1,10}^{1-1,1-1} )$
                      
74. $K_{x,z}^{y',y'} = \frac {\sqrt{2}} {4} 
                    ( t_{1+1,10}^{1+1,1+1} + t_{1+1,10}^{1-1,1+1}
                    + t_{1+1,10}^{1+1,1-1} + t_{1+1,10}^{1-1,1-1} 
                    - t_{1-1,10}^{1+1,1+1} - t_{1-1,10}^{1-1,1+1}
                    - t_{1-1,10}^{1+1,1-1} - t_{1-1,10}^{1-1,1-1} )$

75. $\underline{K_{x,z}^{y',z'} }=  \frac {i} {2} 
                     (- t_{1+1,10}^{1+1,10} - t_{1+1,10}^{1-1,10}
                      + t_{1-1,10}^{1+1,10} + t_{1-1,10}^{1-1,10} )$
                      
76. $K_{y,z}^{y',x'} =  \frac {\sqrt{2}} {4} 
                   (- t_{1+1,10}^{1+1,1+1} - t_{1+1,10}^{1-1,1+1}
                    + t_{1+1,10}^{1+1,1-1} + t_{1+1,10}^{1-1,1-1} 
                    - t_{1-1,10}^{1+1,1+1} - t_{1-1,10}^{1-1,1+1}
                    + t_{1-1,10}^{1+1,1-1} + t_{1-1,10}^{1-1,1-1} )$
                    
77. $\underline{ K_{y,z}^{y',y'} }=  \frac {i\sqrt{2}} {4} 
                    ( t_{1+1,10}^{1+1,1+1} + t_{1+1,10}^{1-1,1+1}
                    + t_{1+1,10}^{1+1,1-1} + t_{1+1,10}^{1-1,1-1} 
                    + t_{1-1,10}^{1+1,1+1} + t_{1-1,10}^{1-1,1+1}
                    + t_{1-1,10}^{1+1,1-1} + t_{1-1,10}^{1-1,1-1} )$
                    
78. $K_{y,z}^{y',z'} =  \frac {1} {2} 
                    ( t_{1+1,10}^{1+1,10} + t_{1+1,10}^{1-1,10}
                    + t_{1-1,10}^{1+1,10} + t_{1-1,10}^{1-1,10} )$
                    
79. $\underline{K_{z,z}^{y',x'} }=  \frac {i} { 2 }
                    (- t_{10,10}^{1+1,1+1} - t_{10,10}^{1-1,1+1} 
                     + t_{10,10}^{1+1,1-1} + t_{10,10}^{1-1,1-1} )$

80. $K_{z,z}^{y',y'} = \frac {1} { 2 }
                    (- t_{10,10}^{1+1,1+1} - t_{10,10}^{1-1,1+1} 
                     - t_{10,10}^{1+1,1-1} - t_{10,10}^{1-1,1-1} )$

81. $\underline{K_{z,z}^{y',z'} }=  \frac {i \sqrt{2} } { 2 }
                    ( t_{10,10}^{1+1,10} + t_{10,10}^{1-1,10} )$

%#------------------------------------------------------------------

\clearpage

\end{document}